%% file: notes.tex
\documentclass{memoir}
\usepackage{amsmath,amsfonts,amssymb,amsthm,bm}
\usepackage{bbm,nccmath}
\usepackage[framemethod=tikz]{mdframed} 
\usepackage{stmaryrd,mathrsfs,mathtools}
\usepackage{array} 
\usepackage{multirow}
\usepackage{wasysym}
\usepackage{enumitem}
\usepackage[most]{tcolorbox}
\usepackage{hyperref}
\usepackage{breakurl}
\usepackage{color}
\usepackage{tikz}
\usetikzlibrary{cd,calc}
\usepackage{titlesec}
\usepackage[leqno,fleqn,intlimits]{empheq}  
\usepackage{atbegshi}

\tikzset{
  symbol/.style={
    draw=none,
    every to/.append style={
      edge node={node [sloped, allow upside down, auto=false]{$#1$}}}
  }
}

\setstocksize{11in}{8.5in}
\settrimmedsize{11in}{8.5in}{*}
\settrims{0in}{0in}
\settypeblocksize{8.5in}{5.75in}{*}
\setlrmargins{1.3in}{*}{*}
\setulmargins{1.0in}{*}{*}
\checkandfixthelayout

\numberwithin{equation}{section}
\input{macros}
\newcommand{\cp}{\mathsf{cp}}

\input{theorems}

\newcounter{step}

\let\origmaketitle\maketitle
\def\maketitle{
  \begingroup
  \def\uppercasenonmath##1{} 
  \let\MakeUppercase\relax 
  \origmaketitle
  \endgroup
}
\makeatletter
\newcommand*{\textoverline}[1]{$\overline{\hbox{#1}}\m@th$}
\makeatother
\newcolumntype{L}{>{$}l<{$}}
\newcolumntype{R}{>{$}r<{$}}
\newcolumntype{C}{>{$}c<{$}}

\newsavebox{\boxt}
\savebox{\boxt}{${\btheta}$}
\newsavebox{\boxp}
\savebox{\boxp}{${\bpi}$}
\newsavebox{\boxa}
\savebox{\boxa}{$\dot\cE(t)=\Lambda(t)\cE(t)$}

  \renewcommand{\thesection}{\arabic{section}}
    \titleformat{\section}[hang]%
                {\normalfont\bfseries\MakeUppercase}{\thesection}{1em}{}
  \titleformat{\subsection}[block]%
              {\filcenter\normalfont\bfseries}{\thesubsection}{1em}{}
  \titleformat{\subsubsection}[block]%
              {\normalfont\bfseries}{\thesubsubsection}{1em}{}
  \renewcommand{\thesubsection}{\thesection\Alph{subsection}}

\begin{document}

\AtBeginShipoutNext{\AtBeginShipoutNext{\AtBeginShipoutDiscard}}
\begin{titlingpage}
  \title{Notes on completely positive maps
    and continuous-time Markovian CP evolution. 
  A geometry-flavored perspective
  }
  \author{Paul~E.~Lammert \\
  \href{mailto:lammert@psu.edu}{lammert@psu.edu}}
\date{June 16, 2026}
\maketitle
\begin{abstract}
  These notes provide a detailed and self-contained
  exposition of basic theory of CP maps and continuous-time
  Markovian evolution.
  The infinite-dimensional (separable) setting is handled
  as an extension of the finite-dimensional one.
  The treatment stands on two legs.
  For the finite-dimensional part, a basis-free version of
  the Choi-Jamio{\l}kowski isomorphism
  called simply \textit{Jamio{\l}kowski transform}.
  And, for the extension, the \textit{ground matrix element topology} (GMET),
  which does for the superoperators on trace-class operators what the
  weak-operator topology does for bounded operators on a Hilbert space.
  Background in open quantum systems or quantum information theory is
  not assumed.
\end{abstract}
\end{titlingpage}

\renewcommand{\chapternumberline}[1]{}
\setcounter{tocdepth}{2}
\tableofcontents
\setcounter{secnumdepth}{-2}
\section{Introduction}


\subsubsection*{Aims and scope}
These notes cover some central foundational topics in the theory of
open quantum systems.
Specifically, we deal with basic theory of completely positive ($\CP$)
maps and its application to continuous-time Markov $\CP$ evolutions.
Both the finite-dimensional and infinite-dimensional (separable)
settings are dealt with. The latter builds on the former and the
development is as self-contained as practically possible.

There are two major \textit{can-be-written-in-the-form}
type results in this subject ---
on for $\CP$ maps (Kraus decomposition), and one for generators of
continuous-time Markovian $\CP$ evolution (Lindblad representation).
I aim to develop a perspective and corresponding tools with which
these emerge naturally and easily.

Exposition of the basic topics here covered seems to have ossified significantly,
and a fresh perspective may be welcome.
The traditional approaches in infinite dimension are not, I think, really
comfortable for people who don't eat $C^*$-flakes for breakfast.
I, myself, do not really understand Lindblad's paper\cite{Lindblad-76}.

\subsubsection*{Look elsewhere for breadth}
So, viewed against the full sweep of the theory of open quantum systems
as it exists today, the scope is narrow.
There are many excellent sources the reader can, and should, go to
for more breadth, or for topics at the growth front of the field, such as
\cite{Alicki+Lendi,Breuer+Petruccione,Rivas+Huelga,Vacchini,Lidar20},
or \cite{Nielsen+Chuang} for quantum computation and information.
Besides, I am competent to expound almost none of that.
If, dear reader, you find certain parts of those sources baffling as I did,
perhaps these notes are for you.

\subsubsection*{Preview}

The question with which we begin is, what plays the r\^{o}le for
mixed states that ordinary Hilbert space operators do for pure states?
The answer is the completely positive ($\CP$) maps.
(Or \textit{operators}, or \textit{superoperators}, but \textit{maps}
has the virtue of being a single syllable.)
In fact, these are the morphisms of a category, as explained in the
stage-setting \S~\ref{sec:CP cat}.

The definition of $\CP$, straightforward though it may appear, is rather
opaque in practice. This motivates the development of the
\textit{Jamio{\l}kowski transform}, $\jam$, in \S\S~\ref{sec:Jamiolkowski},
\ref{sec:correspondence}.
This is a basis-free, hence well-typed, version of the Choi-Jamio{\l}kowski
isomorphism, which provides an isometry
\begin{equation}\nonumber
  \CP(\cH,\cK) \bijam \Pos(\Blin(\cH,\cK))
\end{equation}
of the $\CP$ maps over Hilbert spaces $\cH$ and $\cK$ with the
positive operators on the Hilbert space $\Blin(\cH,\cK)$.
This tool is limited to the finite-dimensional setting since
only then is $\Blin(\cH,\cK)$ a Hilbert space in the way we need.

In \S~\ref{sec:Kraus},
$\jam$ is used to establish a correspondence between decomposition of
mixed states (positive operators) into pure states (rank-1) and
a decomposition
\begin{equation}\nonumber
  \Lambda = \sum \btheta(A_i)
\end{equation}
of a $\CP$ map $\Lambda$ into a sum of $\CP$ maps which themselves cannot
be so decomposed except into multiples of themselves.
The geometrical and algebraic significance of this \textit{Kraus decomposition}
is thereby established.

The Jamio{\l}kowski transform thus provides a method of constructing
Kraus decompositions. However, it will not work in infinite dimensions and
is maybe not very practical for large finite dimensions, either.
So, we develop an algorithm for partial Kraus decomposition which 
turns out later, to work very well in the separable setting
(\S~\ref{sec:CP separable}).

Extending things to the separable setting presents a problem because
the Jamio{\l}kowski transform does not work for the infinite-dimensional
spaces we are interested in. On the other hand, it should apply to
finite-dimensional ``approximations'', depending on what we can make of
that concept. This is a matter of topology, and norm topology is woefully
inadequate. Thus, \S~\ref{sec:GMET} develops the
\textit{ground matrix element topology} (GMET).
It does for $\Blin(\TC(\cH),\TC(\cK))$ (superoperators) what the weak-operator
topology does for a space of bounded Hilbert space operators.
Specifically, closed bounded sets are GMET-compact.
\S~\ref{sec:seminorms and topology} covers relevant parts of basic
point-set topology.

Another batch of questions concern an analog of the Schr\"{o}dinger equation,
namely a so-called master equation
\begin{equation}\nonumber
\frac{d\rho(t)}{dt} = \Lambda(t)\rho(t)  
\end{equation}
for a density operator. The generator $\Lambda(t)$ here is a,
possibly time-dependent,
linear operator on $\rho$. We require the evolution generated by this to
be $\CP$. To bring that to the surface, it makes more sense to just write
the equation for the evolution operator $\cE(t,s)$ in $\CP(\cH)$ as
\begin{equation}\label{eq:dE/dt}
  \frac{d}{dt}\cE(t,s)
  = \Lambda(t)\cE(t,s).
\end{equation}
How must the generator be restricted to ensure that $\cE(t,s)$
is $\CP$? How is it conveniently parametrized?
How do we map the generator to a parametrization?
It is clear that the generator must be in the tangent cone
to $\CP(\cH)$ at the identity, designated $\cp(\cH)$.
It is also pretty easy to motivate and demonstrate that
$\CP$ evolutions are generated by operators
$L(\Psi,K)$ parametrized by $\Psi\colon\CP(\cH)$ and $K\colon\Blin(\cH)$,
and given by
\begin{equation}\nonumber\label{eq:intro Lindblad}
  \rho \mapsto \Psi\rho + K\rho + \rho K^\dag.
\end{equation}
The hard(er) part is to show that the entire tangent cone $\cp(\cH)$ is
in the range of $L$.
Jamio{\l}kowski transform comes to our aid again in finite dimension,
converting this to an easier question about the tangent cone to
$\Pos(\Blin(\cH))$ at projection onto the identity.
Using GMET and other tools of finite-dimensional approximation, this
is extended to the separable setting. Unfortunately, that extension is
nonconstructive, but perhaps that defect can be overcome.
All this is done in \S~\ref{sec:generators CP evolution}.

The preceding paragraph slid by the question of restictions on the time dependence
of the generator. In fact, it is no more severe than for a general linear evolution,
and for this reason, \S~\ref{sec:evolution} gives a careful treatment of this
less-constrained situation. Neither continuity nor local boundedness of the
generator is necessary, or even helpful. But, dropping those means interpreting
the time derivatives in a weak sense or, better, reformulating as an integral
equation. That is probably a good thing to do anyway, as it provides a
more flexible starting point to move beyond the extreme Markovian
idealization represented by (\ref{eq:dE/dt}).

We've touched on the synthesis of a generator of $\CP$ evolution from
a pair $(\Psi,K)$ of a $\CP$ map and a bounded operator on $\cH$.
Can we go the other way? The parametrization (\ref{eq:intro Lindblad})
is (generally) far from unique, so that amounts to a selection procedure.
\S~\ref{sec:parametrizers} shows that linear such selections exist and
amount to a choice of a certain kind of subspace of hermiticity-preserving
superoperators. Of course, in the separable case this is also nonconstructive,
but it does yield a little more insight into the structure of $\cp(\cH)$.


\subsubsection*{The finite-dimensional track}

It is expected that some readers will be interested only in the
finite-dimensional setting, or at least want to stick to that for
a first reading. Up through \S\ref{sec:miscellany}, finite-dimensionality
is an implicit assumption, so those sections are completely safe.
\S\ref{sec:things fall apart} -- \S\ref{sec:CP separable} should be skipped.
\S\ref{sec:evolution} is essentially dimension-agnostic.
The finite-dimensional and separable cases are mixed together in
the remaining sections, but in distinct organizational units below
the section level. 

%
%
%
\subsubsection*{Revisions are to be expected}

Or, possibly, hoped for.
At this point, I am reasonably satisfied with the organization
and execution of the main threads. (At least until somebody points out
a mistake!)
Much of the supporting material needs improvement, though, which
I hope to achieve without much increasing the length, since these notes
are already too long.
Suggestions and comments are welcome.
\newline

\vspace*{0.8cm}
\centerline{\rule{0.5\textwidth}{0.8pt}}
\vspace*{15pt}
\centerline{\huge{$\CP$ maps}}
\vspace*{10pt}
\centerline{\rule{0.5\textwidth}{0.8pt}}
\vspace*{0.8cm}
\addcontentsline{toc}{part}{$\CP$ maps}

\setcounter{secnumdepth}{4}
\section{$\CP$: a category for open quantum systems}
\label{sec:CP cat}

Shall we begin with the question, what is a $\CP$ map?
A better question is, what \textit{are} the $\CP$ maps?
This section develops the concept of \textit{completely positive} (\CP) maps
from the ground up; the definition of \CP\ will not even appear until \S~\ref{sec:CP safe}.
The general viewpoint is that $\CP$ maps (whatever those turn out to be) are to general
(a.k.a. \textit{mixed}) quantum states as Hilbert space operators are to state vectors.
We might describe our quest as one for an appropriate category of quantum operations
on open systems, as long as we avoid reading too much into the word ``operation''.
For instance, just letting things evolve for a while under their proper dynamics
should count.
``Appropriate'' means that the class of quantum operations satisfies these
operationally-grounded requirements:
(i) members map positive operators (representing states) into positive operators
in a way respecting probabilistic mixture,
(ii) contain identities (``nothing happens''),
(iii) be closed under combination in series (composition) when this makes sense
at the level of set functions,
(iv) be closed under combination in parallel (tensor product).
The last of these has, famously, surprising bite.

\subsection{The world of closed quantum  systems}\label{sec:closed}

Usually, we represent the state of a closed quantum system by a
state vector $h$ in a Hilbert space $\cH$ associated with the system.
There was an example of a notational convention which will be largely adhered to.
Because we use vectors in various Hilbert spaces, it is helpful
to use names for vectors (e.g., `$h$') which show this membership
(here, `$\cH$') on their faces. 
Also, we do not use the Dirac notation, except on rare occasion;
this has advantages in working with superoperators.

In the ``world'' of closed quantum systems, we have vectors and operators
that can be applied to them. 
To each ordered pair $(\cH,\cK)$ of Hilbert spaces
there corresponds the collection $\Blin(\cH,\cK)$ of linear maps from $\cH$ to $\cK$.
There are no general grounds for banning any of these from quantum theoretical
discourse.
If $A\colon \Blin(\cH,\cK)$ and $B\colon \Blin(\cK,\cN)$, they can be composed
as $BA$ to yield a map in $\Blin(\cH,\cN)$.
Further, for each $\cH$, there is an identity $\Id_\cH$ in $\Blin(\cH,\cH)$
such that $A\Id_\cH = A$ or $\Id_\cH B = B$ whenever these compositions are well-typed.
These simple structural properties license $\Hilb$ as a \textit{category}.
The Hilbert spaces are the \textit{objects} of the category, and linear maps between
them are the \textit{morphisms}. We call this category $\Hilb$.

\subsection{\dots is a part of the world of open quantum systems}
\label{sec:OQS world}

State vectors are completely inadequate for representing states of open quantum
systems; such states are instead represented by positive operators.
$\Pos(\cH)$, the class of positive operators on $\cH$ can be defined by
\begin{equation}\nonumber
A \in \Pos(\cH) \;\equiv\; \forall h\colon\cH,\; \inpr{h}{A h} \ge 0.
\end{equation}
There is a tradition of calling these more general kind of states
``mixed states'', but this gets things backwards, in my view.
They are not a special fancy kind of state; rather the states representable by
state vectors are a special and important restricted class, namely the pure states.
Occasionally, we use the traditional term \textit{density matrix} for positive
operator.
In Dirac notation, the positive operator representing the state corresponding
to $\Dket{\psi}$ is written $\Dket{\psi}\Dbra{\psi}$.
We use a couple of different notations for this.

There is a natural conjugate-linear map $\psi \mapsto \pam{\psi}$, defined
by $\pam{\psi}(\phi) = \inpr{\psi}{\phi}$,
which carries $\cH$ onto the space of $\Cmplx$-linear maps
\mbox{$\cH \rightarrow \Cmplx$}, i.e., the \textit{dual space} $\cH^*$.
The associated Hilbert space structure on $\cH^*$ is
\begin{equation}\label{eq:dual inner product}
\inpr{\pam{\psi}}{\pam{\phi}} = \inpr{{\phi}}{{\psi}}.
\end{equation}
This is a good definition because every vector in $\cH^*$ has the form $\pam{h}$ for
some $h$ in $\cH$ (Riesz representation theorem, though that's unnecessary in
finite dimensions). 

With this notation, we write the pure state corresponding to a state vector
$h$ as $h\pam{h}$. We also use the following more concise notation.
\begin{definition}[$\bpi$]\label{def:pi}
For $h$ in Hilbert space $\cH$, 
\begin{equation}
\bpi(h) \defeq h\pam{h}
\end{equation}
is a member of $\Blin(\cH)$ proportional to the orthoprojector
onto $h$, but not equal to it unless $h$ is normalized. 
\end{definition}
$\bpi$ carries a state vector into the corresponding state.
Any positive operator $\rho$ can be written as a sum of the form
$\sum \bpi(h_i)$. The most obvious, but not only, way
to do this is to use a set of orthogonal eigenvectors of $\rho$.

Now, we begin to set up a category --- $\CP$ --- appropriate for open quantum
systems, and analogous to $\Hilb$. 
The main question is, what are the morphisms?
We do not answer this all at once. But, at least some are readily at hand.
For, $T\colon\Blin(\cH,\cK)$ acts as $h\mapsto Th$, and this translates to
$\bpi(h) \mapsto \bpi(Th) = T(h\pam{h})T^\dag = T(\bpi(h))T^\dag$.
Extending this over linear combinations, we are
led to a new notation embodying an important idea.
\begin{definition}[$\blank$]\label{def:box}
  If $S\colon\Blin(\cH,\cK)$ and $T\colon\Blin(\cM,\cN)$, then    
\begin{equation}\nonumber
 S\blank T
\end{equation}
is the superoperator
$\Blin(\cN,\cH) \rightarrow \Blin(\cM,\cK)$ given by
$X \mapsto SXT$.
\end{definition}
The box shows where the argument goes; dummy arguments are thereby avoided.
\begin{definition}[$\btheta$, $\btheta$-map]\label{def:theta-map}
For $A\colon\Blin(\cH,\cK)$,
$\btheta(A)\in\Blin(\Blin(\cH),\Blin(\cK))$ is given by  
\begin{equation}\label{eq:theta}
\btheta(A) \defeq A\blank A^\dag.
\end{equation}
A superoperator of the form $\btheta(A)$ is called a \textit{$\btheta$-map}.
\end{definition}
With this, we can now say that, just as the denizen $h$ of the ``world of closed
quantum systems'' is translated to the ``world of open quantum systems'' as
$\bpi(h)$, $T$ is translated as $\btheta(T)$. A question suggested by this is,
is it the case that --- just as all legitimate states are combinations
like $\sum\bpi(h_i)$ --- all legitimate operations on such states have the
form $\sum \btheta(A_i)$? We shall see that this is indeed the case;
in the literature, this goes under the moniker \textit{Kraus decomposition}.
In fact, \S~\ref{sec:Kraus and extreme} shows how this can be viewed
as an instance of decomposition
of a positive operator. In the other direction,
\S\ref{sec:states as morphisms} shows a sense in which
$\bpi$ is an instance of $\btheta$.
But that is getting ahead of the story.

\subsection{Why  a category?}

The kind of structure referred to by the term \textit{category}
is by no means exotic, but well-grounded in simple operational ideas.
For a system identified by the
Hilbert space $\cH$, absolutely nothing happening to it should count as an operation.
This is the identity on $\Pos(\cH)$. The only other requirement is that operations
can be composed is ``the types match''.
Operations taking a state of the $\cH$-system to one of the $\cK$-system can be
followed by one going from $\cK$ to $\cN$ to obtain one going from $\cH$ to $\cN$.
That is, operations can be combined in \textit{series}.
That is all (although there is more terminology attached to this.)

\subsection{Why  \textit{linear} superoperators?}

The quantum operations we seek should take positive operators to positive operators,
but linearity should not be taken for granted. 
We have nested subsets of $\Blin(\cH)$ as follows:
\begin{equation}\nonumber
\Pos_1(\cH) \subset
\Pos(\cH) \subset
   \SA(\cH) \subset
   \Blin(\cH). 
 \end{equation}
 Here, $\SA(\cH)$ denotes the hermitian (`self-adjoint') operators and
 $\Pos_1(\cH)$ the positive operators with trace $1$.
 The credentials of members of $\Pos_1(\cH)$
 as representing proper states are unimpeachable.
 It seems all we can require of a putative \textit{quantum operation} $\Gamma$
 \textit{a priori} is that it map $\Pos_1(\cH)$ into $\Pos_1(\cK)$ ($\cK$ potentially
 a different Hilbert space) respecting convex combinations:
 \begin{equation}
   \Gamma(s \rho + (1-s)\sigma) = s \Gamma(\rho) + (1-s)\Gamma(\sigma),
   \qquad 0 \le s \le 1.
 \end{equation}
 This says no more than that the action of $\Gamma$ is compatible with
 probabilistic mixing of states. Now we show that this has a unique
 extension to a map $\Pos(\cH)\rightarrow\Pos(\cK)$ respecting addition and
 multiplication by positive scalars, thence to a $\Real$-linear map
 $\SA(\cH)\rightarrow\SA(\cK)$, and finally to a $\Cmplx$-linear map
 $\Blin(\cH)\rightarrow\Blin(\cK)$.
 The formulas are obvious; it is just a matter of verifying
 well-definedness.
 Start with $\Gamma \rho = \Tr\rho\cdot \Gamma\left(\tfrac{\rho}{\Tr\rho}\right)$,
 for $\rho\in\Pos(\cH)$.
Suppose $s\rho + t\sigma = s'\rho'+t\sigma'$ for
$\rho,\sigma,\rho',\sigma'\in\Pos_1(\cH)$ and $s,t,s',t'\in\Real^+$.
Then, $s+t = s'+t'$ is the trace $c$ of this operator.
$s\rho + t\sigma = c\left[c^{-1}{s}\rho + c^{-1}{t}\sigma\right]
= c\left[c^{-1}{s'}\rho' + c^{-1}{t'}\sigma'\right]$, so the extension
satisfies $\Gamma(s\rho+t\sigma) = s\Gamma(\rho)+t\Gamma(\sigma)$ for
all positive $\rho,\sigma$ and $0\le s,t$.
Now, define $\Gamma(-\rho) = -\Gamma(\rho)$.
With $0\le s,t,s',t'$, suppose that $s\rho-t\sigma = s'\rho'-t'\sigma'$.
Then,
$s\rho+t'\sigma' = s'\rho'+t'\sigma$, so
$s\Gamma(\rho)+t'\Gamma(\sigma') = s'\Gamma(\rho')+t\Gamma(\sigma)$
showing that the extension to $\SA(\cH)$ is $\Real$-linear.
The final step to complex scalars is simple.

\subsection{Convex sets and cones}\label{sec:cones}

$\Pos(\cH)$ is not a vector space, but a closed convex cone. This concept
plays a fundamental role in the sequel.

\begin{definition}[convex set]
  A subset $\cC$ of a \textit{real} vector space $\cV$ is \textit{convex}
  if whenever $x,y$ are in $\cC$, the entire line segment
  \mbox{$[x,y] = \{sx+(1-s)y\,:\, 0\le s\le 1\}$} is also in $\cC$.
\end{definition}
\begin{definition}[cones]
  $\cC\subseteq \cV$ is a \textit{cone} if whenever $x$ is in $\cC$, then so is
  the entire ray $\{s x\,:\, 0 < s\}$.
  $\cC$ is a convex cone if it is also convex.
  An alternative full characterization is:
  whenever $x,y$ are in $\cC$, then so are all \textit{positive}
  linear combinations $\{ax+by\,:\, 0 < a,b\}$.
  
  A \textit{proper cone} $\cC$ has the additional properties that
(i) it is topologically closed,
(ii) it has nonempty interior,
and (iii) $x,-x\in \cC$ implies $x=0$.
From (iii) and (i), it easily follows that $\cV = \cC - \cC$, i.e., every
$v\colon \cV$ is the difference of two vectors in $\cC$.
\end{definition}

\paragraph{Example: $\Pos(\cK)$ is a proper cone in $\SA(\cK)$.}
\label{sec:pos cone}\leavevmode

(i) $a,b\ge 0$ and $A,B\colon\Pos(\cK)$ immediately imply
$aA+bB \in \Pos(\cK)$, (ii) \mbox{$\inpr{v}{A_nv} \to \inpr{v}{Av}$} whenever
$A_n\to A$, (iii) any sufficiently small perturbation of $\Id_\cK$ is positive,
(iv) $\inpr{v}{Av} \ge 0$ and $-\inpr{v}{Av} \ge 0$ can hold for all $v$ only if $A=0$.

$\SA(\cH)$ is the real span,
and $\Blin(\cH)$ the complex span, 
of $\Pos({\cH})$.
This is so because any hermitian operator is, through its spectral representation,
canonically expressed as a difference of two positive operators.
We write $``A \ge 0$'' for ``$A$ is positive''. This provides a partial order
on $\SA(\cH)$: $A \le B$ means $B-A \ge 0$.

\subsection{Superoperators preserving hermiticity
  \paren{$\HP$} and positivity \paren{$\Mon$}}\label{sec:HP Mon}

At this point, we have concluded that a quantum operation corresponds to
an operator in $\Blin(\Blin(\cH),\Blin(\cK))$ which maps $\Pos(\cH)$ into
$\Pos(\cK)$. To discuss these things comfortably, we introduce notation
for subclasses of $\Blin(\Blin(\cH),\Blin(\cK))$ which preserve certain
properties:
\begin{equation}\label{eq:Mon-HP-Lin}
\Mon(\cH,\cK)
\subset  \HP(\cH,\cK)
\subset  \Blin(\Blin(\cH),\Blin(\cK)).
\end{equation}
The classes of \textit{hermiticity-preserving} and \textit{monotone} superoperators
are defined by
\begin{equation}
\begin{array}{lcl}
  \Lambda \in \HP(\cH,\cK) & \quad\equiv\quad
   & \quad \Lambda(\SA(\cH)) \subseteq \SA(\cK) \\
  \Lambda \in \Mon(\cH,\cK) & \quad\equiv\quad
   & \quad \Lambda(\Pos(\cH)) \subseteq \Pos(\cK).
\end{array}
\end{equation}
Note that hermiticity-preserving operators can alternatively be defined as
those which commute with hermitian conjugation ($\dag$).
$\HP(\cH,\cK)$ is a vector space over $\Real$, but not over $\Cmplx$.
In fact, 
$\Blin(\Blin(\cH),\Blin(\cK))$ is the direct sum
$\HP(\cH,\cK) \oplus_\Real i\HP(\cH,\cK)$ over $\Real$.
$\Mon(\cH,\cK)$, in contrast,
is not a vector space at all, but rather a pointed
closed convex cone. 

$\Mon$, by which we mean the full system over all pairs of Hilbert spaces,
is a category. This is simple: the composition of two operators which preserve
the property of positivity also does so.

\subsection{$\Mon$ may seem right, but it has a flaw}

At first sight, $\Mon$ looks like our sought-for category of quantum operations,
but there is a subtle problem.

\paragraph{Example of an inadmissible monotone map}\label{ex:Mon not enough}
\leavevmode

  Let $\Lambda$ act on density matrices of a single qubit by transposition
  in the computational basis; this clearly preserves positivity.
  Now consider the entangled two-qubit state $\Dket{10}+\Dket{01}$
  (normalization omitted for convenience), with corresponding density matrix
  \begin{equation}\nonumber
\rho =
\outpr{10}{10}
+ \outpr{01}{01}
+ \outpr{10}{01}
+ \outpr{01}{10}.
\end{equation}
Performing transposition on the first qubit without altering the second
delivers a non-positive density matrix:
\begin{equation}\nonumber
(\Lambda\otimes\Id)\rho
=  
 \outpr{10}{10}
+ \outpr{01}{01}
+ \outpr{00}{11}
+ \outpr{11}{00}.
\end{equation}
(For instance, take expectation value in the vector state $\Dket{00}-\Dket{11}$.)

\subsubsection{Operations should combine in parallel}

We should be able to combine quantum operations not only in series
(a requirement already made explicit), but also in \textit{parallel}.
The example shows that this is not automatic.
In quantum theory, parallel combination is achieved with tensor product,
so this says that whenever $\Lambda$ and $\Gamma$ are acceptable, so is
$\Lambda\otimes\Gamma$.
(The expression ``$\Lambda\otimes\Gamma$'' implies that $\Lambda$ and $\Gamma$
operate on disjoint sets of degrees of freedom, just as ``$\cH\otimes\cH$''
implies two \textit{distinct copies} of $\cH$.)

Therefore, we are looking for a subcategory of $\Mon$ which is closed under
tensor product. And, since we do not want to exclude anything without cause,
it should be maximal. Fortunately, there is only one such subcategory.
The problem is \textit{prima facie} collective, since whether a given operation
can be joined via $\otimes$ with all others depends upon what others there are.
In fact, there is a simple map-by-map test, as we shall see now.

\subsection{CP makes the world safe for tensor products}\label{sec:CP safe}

\subsubsection{The definition}\label{sec:CP def}

  $\Lambda \colon \Blin(\Blin(\cH),\Blin(\cK))$ is \textit{completly positive}
  (CP) if
\begin{equation}\label{eq:CP in symbols}
\forall N\ge 1,\;
\Lambda \otimes \Id_{\Blin(\Cmplx^N)} \in
\Mon(\cH\otimes\Cmplx^N,\cK\otimes\Cmplx^N)
\end{equation}
The class of CP operators in $\Blin(\Blin(\cH),\Blin(\cK))$ is denoted $\CP(\cH,\cK)$.
\subsubsection{$\CP$ is the unique subcategory of $\Mon$ closed under $\otimes$}
\label{prop:CP is OK}

  The case $N=1$ in the definition shows that $\CP\subset\Mon$.
  That it contains identities is obvious.
  If $\Lambda,\Gamma\in\CP$ are composable as linear maps, then
\mbox{$\Gamma\Lambda\otimes\Id_{\Blin(\cN)} = 
  (\Gamma\otimes\Id_{\Blin(\cN)}) (\Lambda\otimes\Id_{\Blin(\cN)})$}.
By assumption, this is a composition of monotone maps, which is itself monotone,
so $\Gamma\Lambda\in\CP$. This all means $\CP$ is a subcategory of $\Mon$.

Now for the crucial closure under $\otimes$.
If $\Lambda\in\Mon(\cH,\cK)$ and $\Gamma\in\Mon(\cN,\cM)$ are $\CP$,
composable as linear maps, then
\begin{equation}\nonumber
(\Lambda\otimes\Gamma)\otimes\Id_{\Blin(\cO)}
= (\Lambda\otimes\Id_{\Blin(\cM)})(\Id_{\Blin(\cH)}\otimes\Gamma)\otimes\Id_{\Blin(\cO)}
= (\Lambda\otimes\Id_{\Blin(\cM)}\otimes\Id_{\Blin(\cO)})
(\Id_{\Blin(\cH)}\otimes\Gamma\otimes\Id_{\Blin(\cO)}).
\end{equation}
Now, 
$\Lambda\otimes\Id_{\Blin(\cM)}\otimes\Id_{\Blin(\cO)}
= \Lambda\otimes\Id_{\Blin(\cM\otimes\cO)}$ is monotone, so if the
$\Id_{\Blin(\cH)}\otimes\Gamma\otimes\Id_{\Blin(\cO)}$ is also, the display shows
$\Lambda\otimes\Gamma\in\CP$ by the same argument as in the previous paragraph.
But this is simple since permutation of factors in a tensor product is an
isometric isomorphism: 
$\Id_{\Blin(\cH)}\otimes\Gamma\otimes\Id_{\Blin(\cO)}$ is monotone iff
$\Gamma\otimes\Id_{\Blin(\cH)}\otimes\Id_{\Blin(\cO)}$ is so, and that is the
case by hypothesis.
So, $\CP$ is a subcategory of $\Mon$ closed under tensor product. It is maximal
because the definition only requires that each morphism be compatible with
identities and those must be included.

\subsection{$\Hilb$ and $\CP$ have the same objects}

The reader has probably noticed a seeming quirk of our notation, namely
we write ``$\CP(\cH,\cK)$'' rather than ``$\CP(\Blin(\cH),\Blin(\cK))$''
for a subset of $\Blin(\Blin(\cH),\Blin(\cK))$,
and similarly ``$\Mon(\cH,\cK)$'' and ``$\HP(\cH,\cK)$''.
Part of the reason for this is simply that it lightens
the notation without creating ambiguity; ``$\Blin(\cH,\cK)$'', in contrast,
already means something. Slightly deeper is the idea that we want our two
categories to share the same set of \textit{objects} --- Hilbert spaces, or even
better the systems which those name. In $\Hilb$, the set of morphisms from
$\cH$ to $\cK$ is $\Blin(\cH,\cK)$, whereas in $\CP$ it is $\CP(\cH,\cK)$.
The notion of category does not require that the morphisms be literal functions
between the objects as sets. The objects really need do no more than serve as
labels to show which morphisms are composable.
Finally, writing $\CP(\cH,\cK)$ gives us room to maintain that $CP$ maps are not
\textit{literally} linear maps, i.e., the linear extension is just a convenience.
Nevertheless, the bulk of these notes will deal with \CP\ maps under their linear guise.

\subsection{\usebox{\boxt} is a monoidal functor}\label{sec:theta functor}

Now that the category $\CP$ is established, we can say that
$\btheta$ (\ref{eq:theta}) is a \textit{functor}, acting identically on objects.
This means that it maps identity morphisms to identity morphisms,
$\btheta(\Id_\cH) = \Id_{\Blin(\cH)}$, and compositions to corresponding
compositions, $\btheta(BA) = \btheta(B)\btheta(A)$. These are easily verified.

In $\Hilb$, $\otimes$ is an operation on both objects and morphisms and these two
operations are coherent in the sense that
whenever $A\colon\Blin(\cH,\cK)$ and $B\colon\Blin(\cN,\cM)$, then
$A\otimes B \in \Blin(\cH\otimes\cN,\cK\otimes\cM)$.
Also, the operation on objects has an ``almost identity'' $\Cmplx$, meaning
$\cH \cong \cH\otimes\Cmplx$. These properties make $\Hilb$ a \textit{monoidal category}.
$\CP$ has the exactly analogous properties (indeed, we insisted on them), so it too
is a monoidal category.
And, finally, $\btheta$ is a \textit{monoidal functor} because it 
respects these structures in the sense that
it (trivially) carries the $\otimes$-identity object in $\Hilb$ to that in $\CP$ and
\begin{equation}\nonumber 
\btheta(A\otimes B) = \btheta(A) \otimes \btheta(B).  
\end{equation}

\section{Generalized Jamio{\l}kowski isomorphism $\jam$: the power of swap}
\label{sec:Jamiolkowski}

This Section develops a version of the Jamio{\l}kowski isomorphism $\jam$
partially inspired by \cite{Grabowski+Kus+Marmo-07}.
Although the demonstration uses tensor products, the formulation is entirely in
the relevant function spaces, without the tensor product proxies used in
references \cite{Grabowski+Kus+Marmo-07} and \cite{Jamiolkowski-72}.
Not until \S\ref{sec:correspondence} will the usefulness of $\jam$ really
become clear.
There it will be turned into a powerful tool for uncovering the properties
of $\CP$ maps.
\subsection{Notational conventions}

Hilbert spaces are denoted by calligraphic letters $\cH$, $\cK$, etc.
Vectors are usually denoted by lower-case latin letters matching the Hilbert
space name, e.g., $h,h'\in\cH$; more-conventional
lower-case greek letters are sometimes used.
Upper-case italic letters $A$, $B$, etc. denote operators.
The space of linear operators from $\cH$ to $\cK$
is mostly denoted $\Blin(\cH,\cK)$ ($\Blin(\cH,\cH)$ is abbreviated to $\Blin(\cH)$),
but in this section we use $\LLin{\cH}{\cK}$. A right-to-left arrow is
somewhat unconventional, but corresponds with the order in which operators are
composed and can be helpful when there are many spaces to keep track of.
For instance, if $A\colon \cK\leftarrow\cH$ and $B\colon\cN\leftarrow\cK$,
then $BA\colon\cN\leftarrow\cH$.
Upper-case greek letters $\Lambda$, $\Gamma$, stand for operators on operators,
that is, elements of a space like \mbox{$\LLin{\LLin{\cM}{\cN}}{\LLin{\cH}{\cK}}$}.
These are sometimes called \textit{superoperators} in the literature.
We will use that terminology, too, but it will gradually give way to
unadorned ``operator''.

Simple juxtaposition indicates composition, as in $AB$,
or application, as in $A\psi$.
(Considering $\psi\in\cH$ as a linear map $\Cmplx\rightarrow\cH$ taking
$1$ to $\psi$, this is also composition.)
$\Lambda \cdot \psi\pam{\phi}$ is the same as $\Lambda (\psi\pam{\phi})$;
the is a form of punctuation, equivalent to an opening parenthesis with
an invisible closing parenthesis as far to the right as possible.
The composition symbol, ``$\circ$'', is often omitted unless it
is necessary for disambiguation.

Sometimes, the notation $A\colon\Blin(\cH,\cK)$ (for example) is used
where one might expect $A\in\Blin(\cH,\cK)$.
The distinction is that ``$A\colon\Blin(\cH,\cK)$'' is not a proposition or claim,
but merely a hypothesis (nothing to check).
Readers who find this disturbing are advised to mentally replace
``$\colon$'' with ``$\in$''.

\subsection{The Hilbert space $\HS(\cH,\cK)$}

\subsubsection{Hilbert-Schmidt inner product}

$\LLin{\cH}{\cK}$ has a {Hilbert} space structure 
given by the \textit{Hilbert-Schmidt} inner product
\begin{equation}
  \label{eq:HS inner prod}
\inpr{A}{B} \defeq \Tr A^\dag B,  
\end{equation}
where the hermitian adjoint of $A\colon{\cK}\leftarrow{\cH}$ is the operator 
$A^\dag\in\LLin{\cK}{\cH}$ defined by
$\inpr{A^\dag k}{h}_\cH  = \inpr{k}{A h}_\cK$.
Subscripting of inner products like this is unnecessary, hence will not be used.
For instance, from the type of $A$ in the above, one deduces $h\in\cH$
and $k\in\cK$.

\subsubsection{Useful identities}

\paragraph{Composition with a fixed operator as a superoperator}\label{sec:HS multiplication}\leavevmode

for $A\colon{\cK}\leftarrow{\cH}$, $B\colon{\cK}\leftarrow{\cN}$,
the Hilbert-Schmidt inner product satisfies the identities
\begin{equation}\nonumber
 \inpr{B^\dag A}{C}  
 =  \inpr{A}{BC}
 = \inpr{AC^\dag}{B}
\end{equation}
In general, each of
the three inner products in the above display are in different spaces.
The first equality says that left-multiplication by $B^\dag$ is the adjoint
of left-multiplication by $B$, and the second equality says something similar
about right-multiplication by $C$ or $C^\dag$.

\paragraph{Matrix elements and the Hilbert-Schmidt inner product}\leavevmode

Matrix elements of an operator $A\colon\cH\rightarrow\cK$
can be expressed in terms of the associated Hilbert-Schmidt inner product as
\begin{equation}\label{eq:swap 1}
\inpr{k}{Ah} = \inpr{k\pam{h}}{A}.  
\end{equation}
Note that the two inner products here are (LHS) in $\cK$ and (RHS) in $\LLin{\cH}{\cK}$.
The shifting maneuver in the above display will be useful on many occasions.

\subsubsection{Iterating to higher levels}

$\Blin(\cH,\cK)$ and $\Blin(\cM,\cN)$ are thus themselves Hilbert spaces.
Hence, in exactly the same way, we obtain a Hilbert space structure on
$\Blin(\Blin(\cH,\cK),\Blin(\cM,\cN))$. We will actually use the
Hilbert-Schmidt structure at this level. 
One can go on \textit{ad infinitum}.

\subsubsection{Operational significance of the inner product}
\label{sec:inner prod operational}

We make a short aside here on the physical/operational significance of
the inner products we are working with.
Here and throughout the paper, we will be using the following notation.

In quantum mechanics, the inner product on a state-vector-space
$\cH$ is fundamental, and the Born interpretation explains its
physical significance: $|\inpr{\phi}{\psi}|^2$ is a transition probability.
If we prepare a projective measurement to test whether the state is
$\bpi(\phi)$, the probability of success, assuming the state is actually
$\bpi(\psi)$ is $|\inpr{\phi}{\psi}|$. But, this is just
$\inpr{\bpi(\phi)}{\bpi{\psi}}$ (no modulus squared).
It looks as if the Hilbert-Schmidt inner product on $\Blin(\cH)$
--- or $\Pos(\cH)$, anyway --- might be at least as well-founded as that
on $\cH$. But what of the extension to general (i.e., mixed) states?
Consider such a normalized state $\rho = \sum c_i \bpi(\psi_i)$,
with normalized $\psi$. Preparing this state is generally accepted as
equivalent to preparing $\bpi(\psi_i)$ with probability $c_i$.
In that case, $\inpr{\bpi(\phi)}{\rho}$ still makes sense as
a success probability. To replace $\phi$ with a mixed state requires
a little asymmetry. We should interpret $\sigma = \sum d_i \bpi(\phi_i)$
as a randomized measurment. Then, $\inpr{\sigma}{\rho}$ is again the
corresponding success probability.

The operational credentials of the induced inner product on
$\Blin(\Blin(\cH),\Blin(\cK))$ are much murkier.
The best we can do is to observe that
$\inpr{A\pam{B}}{\Lambda} = \inpr{A}{\Lambda B}$, where the
inner product on the right is already interpreted.
Perhaps this failure is to be expected, however.
\subsubsection{A fundamental isomorphism: $\HS(\cH,\cK) \cong \cK\otimes \cH^*$}
\label{sec:HS(H,K)=KxH*}

There is a natural Hilbert isomorphism, i.e., unitary mapping
\begin{equation}
\label{eq:function-tensor iso}
\begin{array}{ccc}
\HS(\cH,\cK) & \cong & \cK \otimes \cH^* \\
 k \pam{h}     & \leftrightarrow & k\otimes\pam{h}.
\end{array}
\end{equation}
The indicated correspondence between rank-one operators $k\pam{h}$ in
$\Blin(\cH,\cK)$ and decomposable tensors $k\otimes\pam{h}$ in $\cK\otimes \cH^*$
extends to sums of such things linearly. Coherence of the definition is guaranteed
by the identical distributivity and scalar multiplication properties:
$k\otimes\pam{h} + k'\otimes\pam{h} = (k+k')\otimes\pam{h}$ and
$c(k\otimes \pam{h}) = (ck)\otimes\pam{h} = k\otimes \pam{\overline{c}h}$ are
still correct equations if `$\otimes$' is replaced by `$\circ$'.
Thus, (\ref{eq:function-tensor iso}) defines a linear isomorphism.
Because it preserves inner products,
it is even a Hilbert isomorphism (unitary mapping).
To see this, it suffices to verify inner products on dyadics.
\begin{equation}
  \label{eq:L(H,K) inner prod}
  \begin{array}{rcll}
  \inpr{k\pam{h}}{k'\pam{h'}}
& = & \Tr ( (k\pam{h})^\dag\cdot k'\pam{h'})   
& \quad { \bm\{} \text{Hilbert-Schmidt inner product} { \bm\} } \\
& = & \Tr (h\pam{k}\cdot k'\pam{h'})
& \quad {\bm\{} (k\pam{h})^\dag = h\pam{k} {\bm\}} \\
&  = & \inpr{k}{k'}\Tr h\pam{h'} & \\
& = & \inpr{k}{k'}\inpr{h'}{h} & \\
& = & \inpr{k}{k'}\inpr{\pam{h}}{\pam{h'}} 
& \quad {\bm\{} (\ref{eq:dual inner product}) {\bm\}} \\
& = & \inpr{k\otimes\pam{h}}{k'\otimes\pam{h'}}  
& \quad {\bm\{} (\text{see below}) {\bm\}} 
  \end{array}
\end{equation}
Recall that the inner product on the Hilbert tensor product $\cM\otimes\cN$
is defined by $\inpr{m\otimes n}{m' \otimes n'} = \inpr{m}{m'} \inpr{n}{n'}$.

\paragraph{Caution.}
It may be tempting to consider $k\pam{h}$ and $k\otimes\pam{h}$ as literally
identical. That is, however, a serious semantic mistake. `$k\pam{h}$' represents
a combination \textit{in series}, whereas `$k\otimes\pam{h}$' represents a
combination \textit{in parallel}.

\subsection{Jamio{\l}kowski transformation}

\subsubsection{Definition}\label{sec:jam defined}

Swapping the order of factors in a tensor product is trivially a Hilbert
space isomorphism.
The Jamio{\l}kowski isomorphism 
\begin{equation}\label{eq:jam}
  { \LLin{\LLin{\cH}{\cK}}{\LLin{\cM}{\cN}} }
  \;\bijam\;
  { \LLin{\LLin{\cH}{\cM}}{\LLin{\cK}{\cN}} }
\end{equation}
is obtained by this simple maneuver, together with the isomorphism
(\ref{eq:function-tensor iso}).
The following display shows in detail how $\jam$ is assembled from
elementary isomorphisms, tracking a generic
separable element in the middle column, with type shown on the right.
\begin{equation}
\begin{array}{lllr}
&  & n\pam{m}\circ \pam{k\pam{h}}
  &\qquad
    (\cN\leftarrow\cM)\leftarrow(\cK\leftarrow\cH)
  \\
  {\bm\{}\ref{eq:function-tensor iso}{\bm\}}
& \leftrightarrow
& (n\pam{m})\otimes(\pam{k}h) 
  &\qquad
    (\cN\leftarrow\cM)\otimes(\cK\leftarrow\cH)^*
  \\
  {\bm\{}\ref{eq:function-tensor iso}{\bm\}}
&\leftrightarrow
& (n\otimes\pam{m})\otimes(\pam{k}\otimes h) 
  &\qquad
    (\cN\otimes\cM^*)\otimes(\cK^* \otimes\cH)
  \\
{\bm\{}\text{associativity}{\bm\}}
&=
& n\otimes\pam{m}\otimes\pam{k}\otimes h 
  &\qquad
    \cN\otimes\cM^*\otimes\cK^* \otimes\cH
  \\
  {\bm\{}\text{swap}{\bm\}}
& \leftrightarrow
& n\otimes\pam{k}\otimes\pam{m}\otimes h 
  &\qquad
 \cN\otimes\cK^*\otimes\cM^* \otimes\cH \\
  {\bm\{}\text{reverse}{\bm\}}
& \leftrightarrow
& n\pam{k} \circ \pam{m\pam{h}} 
  &\qquad
 (\cN\leftarrow\cK)\leftarrow(\cM \leftarrow \cH)
\end{array}
\end{equation}
From the penultimate to final line, the first three steps are simply reversed,
and the isomorphism is extended to the entire spaces by linearity.
\subsubsection{Description in terms of ground matrix elements}
\label{sec:J in terms of GMEs}

A (super)operator $\Lambda$ in $\Blin(\cH,\cK)$ is completely described by its
\textit{matrix elements} $\inpr{A}{\Lambda B}$ for $A\colon\Blin(\cK)$ and $B\colon\Blin(\cH)$.
However, this corresponds to a linear combination of \textit{ground matrix elements} of the form
$\inpr{k'\pam{k}}{\Lambda\cdot h'\pam{h}}$ with $k,k'\in\cK$, $h,h'\in\cH$. 
In terms of such ground matrix elements, $\jam$ is expressed as
\begin{equation}\label{eq:J in GME form}
\inpr{k'\pam{h'}}{\jam\Lambda\cdot k\pam{h}} = \inpr{k'\pam{k}}{\Lambda\cdot h'\pam{h}}.
\end{equation}
 
\subsection{Key properties of $\jam$}

\begin{tcolorbox}[standard jigsaw,opacityback=0]
  \begin{align}
  \jam \text{ is its own inverse}
  \label{eq:own inverse} \\
  A\pam{B}  \bijam  A \blank B^\dag
  \label{eq:jam rank 1} \\
 \jam(\Lambda\otimes\Gamma) = \jam \Lambda \otimes \jam \Gamma
 \label{eq:jam distributes over tensor}
  \end{align}
\end{tcolorbox}


Since any superoperator of type $\Blin(\Blin(\cH,\cK),\Blin(\cM,\cN))$ can
be written as a sum of superoperators $A\pam{B}$ for $A\colon\Blin(\cM,\cN)$
and $B\colon\Blin(\cH,\cK)$, (\ref{eq:jam rank 1}) is equivalent to the
definition of $\jam$. Usually, it will be more convenient to use
(\ref{eq:jam rank 1}) rather than direct appeal to the definition.

\subsubsection{Demonstrations}

\begin{proof}
[Proof that $\jam$ is its own inverse]
  Immediate from the definition.
\end{proof}

\begin{proof}[Proof of $A\pam{B}  \bijam  A \blank B^\dag$]
Consider the special case  
\begin{equation}\label{eq:special case}
  A = n\pam{k}, \quad 
  B = m\pam{h}, \quad 
  X = k'\pam{h'}.
\end{equation}
Then,
$\jam (A\pam{B})
= \jam ( n\pam{k}\, (\pam{m\pam{h}}) )
= n\pam{m}\cdot \pam{k\pam{h}}$,
so
\begin{equation}\nonumber 
\begin{split}
\jam (A\pam{B})X
  = \, &
\Big( n\pam{m}\, (\pam{k\pam{h}}) \Big) k'\pam{h'}
\\
= \, & n\pam{m} \inpr{k\pam{h}}{k'\pam{h'}}.
\end{split}
\end{equation}
On the other hand, 
\begin{equation}\nonumber 
  \begin{split}
  n\pam{k} (k'\pam{h'}) {h\pam{m}}
&  = 
  n\inpr{k}{k'} \inpr{h'}{h} \pam{m} \\
&  = 
  n\pam{m} \inpr{k\pam{h}}{k'\pam{h'}}.
  \end{split}
\end{equation}
Hence, since $(m\pam{h})^\dag = h\pam{m}$,
the special case (\ref{eq:special case}) of $\jam(A\pam{B})X = AXB^\dag$ is proven.
But both sides of this special case equation are linear in $A$ and $X$ and
conjugate linear in $B$, so this is enough.
\end{proof}

\begin{proof}[Proof of 
$\jam(\Lambda\otimes\Gamma) = \jam \Lambda \otimes \jam \Gamma$]
  
By (\ref{eq:jam rank 1}),
\begin{equation}\nonumber
  A\pam{B} \otimes S\pam{T}
  =  (A\otimes S) (\pam{B\otimes T})
  \bijam
  (A\otimes S) \blank (B^\dag \otimes T^\dag).
\end{equation}
On the other hand,
\begin{equation}\nonumber
\begin{split}
(A \otimes S)(X\otimes Y)(B^\dag\otimes T^\dag)
=\, & 
    A X B^\dag \otimes S YT^\dag
    \\
=\, & 
    \jam(A\pam{B})X \otimes \jam(S\pam{T})Y
   \\
=\, & 
    (\jam(A\pam{B}) \otimes \jam(S\pam{T})) (X \otimes Y).
\end{split}
\end{equation}
Thus, $\jam(A\pam{B}\otimes S\pam{T}) = 
\jam(A\pam{B})\otimes \jam(S\pam{T})$.
This suffices by linearity ($\Lambda\otimes\Gamma$ is a sum of terms like those).
\end{proof}

\paragraph{Example: orthonormal bases.}\label{sec:ONB}\leavevmode

$\LLin{\cH}{\cK}$ is spanned by rank-one operators, i.e., those of the form
$k\pam{h}$. From these, according to (\ref{eq:L(H,K) inner prod}),
we can extract an orthonormal basis (ONB) $\{k_i\pam{h_j}\}$, where $\{k_i\}$ and
$\{h_i\}$ are orthonormal bases of $\cK$ and $\cH$, respectively.

Now, specialize this to the case of spaces of form
$\Blin(\cH,\cK)$ and $\Blin(\cM,\cN)$.
If $\{A_i\}$ is an ONB of $\Blin(\cH,\cK)$, and
$\{S_\alpha\}$ of $\Blin(\cM,\cN)$, 
then $\{A_i\pam{S_\alpha}\}$ is an ONB of 
$\Blin(\Blin(\cH,\cK),\Blin(\cM,\cN))$.
Because $\jam$ is a Hilbert isomorphism,
$\{A_i\blank S_\alpha^\dag\}$, is \textit{also} an ONB.

\section{The Jamio{\l}kowski correspondence
  $\CP(\cH,\cK)\bijam\Pos(\Blin(\cH,\cK))$}
\label{sec:correspondence}

In developing the generalized Jamio{\l}kowski isomorphism, we used
four different underlying Hilbert spaces ($\cH$, $\cK$, $\cM$, $\cN$)
because that helped to keep the distinct r\^{o}les of the spaces
clearer, and cost nearly nothing. However, we are not really interested
in such generality. This section specializes to the cases appropriate
for discussing $\CP$ maps, namely
\begin{equation}\label{eq:pi-theta}
  \begin{split}
  \Blin(\Blin(\cH),\Blin(\cK))
& \bijam 
  \Blin(\Blin(\cH,\cK)),  \\
  \btheta(A)  & \bijam   \bpi(A) 
  \end{split}
\end{equation}
The correspondence $\btheta(A)\bijam\bpi(A)$
is an obvious specialization of
\mbox{$A\pam{B}  \bijam  A \blank B^\dag$} (\ref{eq:jam rank 1}).
That it suffices to specify $\jam$ in this case follows from the
fact that one-dimensional projectors span $\Blin(\cH)$ for any
Hilbert space $\cH$, in particular $\Blin(\cH,\cK)$.
The point is made more concretely by the polarization identity
\mbox{$A\pam{B} = \tfrac{1}{4}\sum i^{-n} \bpi(A+i^n B)$}.  

The main goal of this section is to show that $\jam$ gives a
canonical isometry between the class of CP operators in
$\Blin(\Blin(\cH),\Blin(\cK))$ and the positive operators in
$\Blin(\Blin(\cH,\cK))$.
This actually involves a couple layers of nesting.

\subsection{The central diagram for $\jam$}

The remainder of this section will be spent justifying the 
following ``central'' commutative diagram:
\begin{equation}\label{eq:central diagram}
  \begin{tikzcd}
&  \CP(\cH,\cK)
   \arrow[dd,leftrightarrow,"\jam"']
   \arrow[r,hookrightarrow,"\subset"]
&  \HP(\cH,\cK)
   \arrow[dd,leftrightarrow,"\jam"']
   \arrow[r,hookrightarrow,"\subset"]
&  \Blin(\Blin(\cH),\Blin(\cK))
   \arrow[dd,leftrightarrow,"\jam"']
   \\
\Blin(\cH,\cK)   
\arrow[ur,rightarrow,"\btheta"]
\arrow[dr,rightarrow,"\bpi"]
   \\
&   \Pos(\Blin(\cH,\cK))
    \arrow[r,hookrightarrow,"\subset"] 
&   \SA(\Blin(\cH,\cK))
    \arrow[r,hookrightarrow,"\subset"] 
&   \Blin(\Blin(\cH,\cK)) 
  \end{tikzcd}
\end{equation}
All geometrical properties of $\CP(\cH,\cK)$ can be studied through
$\Pos(\Blin(\cH,\cK))$. Following sections will amply demonstrate
the power of this approach.
The middle items in the rows here are real Hilbert spaces (i.e., over the
field $\Real$ of scalars). The items on the right are their complexifications.
This is familiar for the bottom row through the splitting of an arbitrary
operator into hermitian and anti-hermitian parts and then follows for
the top row through $\jam$, as soon as the middle $\jam$ arrow is
established.

\subsection{Differences from conventional terminology}

The literature
Traditionally, the literature uses the term ``positive''
rather than ``monotone''.
This usage is adopted from a context ($C^*$ algebras) where it is unambiguous,
but for us it can cause confusion because \mbox{$\Lambda\in \Blin(\Blin(\cH))$}
might be monotone, and it might be positive
(as an operator on the Hilbert space $\Blin(\cH)$).
These are very different things.
Consistency would recommend ``completely monotone'' over ``completely positive'',
but the latter is far too established for that.
We escape the dilemma by always using the acronym ``CP'' rather than the full phrase.
Similarly, we will below use ``$N$-monotone'' instead of the usual ``$N$-positive''.
\subsection{Demonstration of the central diagram (\ref{eq:central diagram})}

In demonstrating the central diagram, we will use some intermediate
correspondences, as indicated in this table:
\renewcommand{\arraystretch}{1.3}
\begin{center}
\begin{tabular}{ll}
    \hline 
 ${\Blin(\cH)}\rightarrow{\Blin(\cK)}$ & $\Blin({\cH}\rightarrow{\cK})$
     \\
    \hline
   hermiticity-preserving & hermitian \\
   monotone & rank-$1$-positive \\
   $N$-monotone & rank-$N$-positive \\
   CP  & positive \\
    \hline
\end{tabular}
\end{center}

\subsubsection{Definition: $N$-monotone, rank-$N$ positive}\label{sec:N-pos def}

\noindent $\Lambda\in\Blin(\Blin(\cH),\Blin(\cK))$ is \textit{$N$-monotone}
if $\Lambda\otimes\Id_{\Blin(\Cmplx^N)}$ is monotone.
($1$-monotone is the same as plain monotone.)

\noindent \mbox{$\Lambda\in\Blin(\Blin(\cH,\cK))$} is \textit{rank-$N$-positive}
if \mbox{$\inpr{T}{\Lambda T}\ge 0$} for all $T$ of rank not exceeding $N$.

We will not have much use for these intermediate notions after this section.

\subsubsection{$\SA(\Blin(\cH,\cK)) \bijam \HP(\cH,\cK)$}
  
$\SA(\Blin(\cH,\cK))$ is the real span of $\setof{\bpi(A)}{A\in\Blin(\cH,\cK)}$,
and if $X\in\SA(\cH)$, then $AXA^\dag\in\SA(\cK)$,
i.e., $\btheta(A)\in\HP(\cH,\cK)$.
So, $\jam$ carries $\SA(\Blin(\cH,\cK))$ \textit{into} $\HP(\cH,\cK)$
But, $\Blin(\Blin(\cH,\cK))$ is the complexification of $\SA(\Blin(\cH,\cK))$,
so $\Blin(\Blin(\cH),\Blin(\cK))$ is the complexification of
$\jam\,[ \SA(\Blin(\cH,\cK))]$. Since no nonreal multiple of a nonzero
hermiticity-preserving superoperator can be hermiticity-preserving,
$\jam\,[ \SA(\Blin(\cH,\cK))] = \HP(\cH,\cK)$.

\subsubsection{Rank-$1$-positive $\bijam$ Monotone}\label{sec:rk-1-pos <-> mon}

From $\bpi(k)A\bpi(h) = k\inpr{k}{Ah}\pam{h}
= k\pam{h}\inpr{k\pam{h}}{A}$,
\begin{equation}
  \bpi(k)\blank\bpi(h)
 = \bpi(k\pam{h}) 
 \bijam
  \btheta(k\pam{h}) 
=  \bpi(k)\pam{\bpi(h)}. 
\label{eq:jam pis} 
\end{equation}
Therefore, 
\begin{equation}\nonumber
\begin{split}
\inpr{k}{(\Lambda \cdot \bpi(h))k}      
& = \inpr{\bpi(k)}{\Lambda \cdot \bpi(h)} \\     
& = \ilinpr{\bpi(k)\pam{\bpi(h)}}{\Lambda} \\
& = \inpr{\bpi(k\pam{h}) }{\Lambda} \\
& = \inpr{k\pam{h}}{\Lambda\cdot k\pam{h}} 
  \end{split}
\end{equation}
$\Lambda$ is monotone if and only if the LHS is non-negative for all $h$ and $k$.
On the other hand, $k\pam{h}$ ranges over the
rank-1 operators in $\LLin{\cH}{\cK}$ as $h\colon\cH$ and $k\colon\cK$ are
varied.

\subsubsection{Purification}\label{sec:purification}

\paragraph{Definition.}\leavevmode

  A \textit{rank-1} operator $\widetilde{A}\in\Blin(\cH\otimes\cK)$
  such that $A = \Tr_{\cK} \widetilde{A}$ is a \textit{$\cK$-purification} of $A$.
  (Or, just \textit{purification} if the specificity is not needed.)

\paragraph{Lemma.}\label{sec:purification lemma}\leavevmode

\begin{enumerate}
\item 
    $\cK$ can purify any operator of rank up to $\dim\cK$.
\item If $\widetilde{A}$ is a $\cK$-purification of $A$, then
      $\inpr{B}{A} = \langle{B\otimes\Id_{\cK}}|{\widetilde{A}}\rangle$.
    \item If $\widetilde{A}$ and $\widetilde{B}$ are $\cK$-purifications of $A$ and $B$,
      then
\mbox{$\inpr{B}{\Lambda\cdot A} 
  =  \inpr{ \widetilde{B} } { \Big(\Lambda \otimes\bpi(\Id_{\cK})\Big) \widetilde{A}}$}
\end{enumerate}
\begin{remark}
  Item 2 can be rephrased as:
  \mbox{$\text{ -- }\otimes \Id_\cK \colon \Blin(\cH,\cN)\rightarrow
    \Blin(\cH\otimes\cK,\cN\otimes\cK)$} and \\
  \mbox{$\Tr_\cK \colon
    \Blin(\cH,\cN)\leftarrow \Blin(\cH\otimes\cK,\cN\otimes\cK)$} are adjoint.
\end{remark}
\begin{proof}\leavevmode
  
\noindent 1.
If $e_1,\ldots,e_N$ is an orthonormal basis of $\cK$, then  
\begin{equation}\nonumber
\Tr_{\cK} \left\{ \left(\sum_{i=1}^N k_i\otimes e_i\right)
  \left(\pam{\sum_{j=1}^N h_j\otimes e_j}\right) \right\}
=
\sum_{i,j=1}^N k_i\pam{h_j} \Tr_{\cK} (e_i\pam{e_j})
=
\sum_{i,j=1}^N k_i\pam{h_j} \delta_{ij}
=
\sum_{i=1}^N k_i \pam{h_i}.
\end{equation}
Operators of rank not exceeding $N$ can be written as in the final expression
above.

\noindent 2. If $A,B\colon\Blin(\cH,\cN)$,
\begin{equation}\nonumber
    \langle{B\otimes\Id_{\cK}}|{\widetilde{A}}\rangle
    = \Tr_{\cH\otimes\cK} (B^\dag\otimes\Id_{\cK})\widetilde{A} 
    = \Tr_{\cH} \Big( B^\dag \Tr_\cK \widetilde{A} \Big) 
    = \Tr_{\cH} B^\dag A
    =  \inpr{B}{A}.
\end{equation}

\noindent 3.
It suffices to prove this for $\Lambda$ of the form  $C\pam{D}$. But,
\begin{equation}\nonumber
  \inpr{B}{(C\pam{D}) A} 
=  \inpr{B}{C} \inpr{D}{A} 
=  \langle{\widetilde{B}}|{C\otimes\Id_{\cK}}\rangle
  \langle{D\otimes\Id_{\cK}}|{\widetilde{A}}\rangle
  =  \inpr{ \widetilde{B} } { \Big(C\pam{D}\otimes\bpi(\Id_{\cK})\Big) \widetilde{A}},
\end{equation}
the second equality by a double application of part 2 above.
\end{proof}

\subsubsection{Rank-$N$-positive $\bijam$ $N$-Monotone}
\label{sec:rk-N-pos <-> N-mon}

%
\begin{lem*}
The following are equivalent.

\begin{tabular}{lll}
a. & $\Lambda$ & is rank-$N$-positive \\
b. & $\Lambda \otimes \bpi(\Id_{\Cmplx^N})$ & is rank-1-positive \\
c. & $\jam\Lambda$ & is $N$-monotone
\end{tabular}
\end{lem*}
\begin{proof}
If $\rk A \le N$, and $\widetilde{A}$ is a $\Cmplx^N$-dilation of $A$,
Lemma \ref{sec:purification lemma}(3) gives
$\inpr{A}{\Lambda\cdot A} 
  =  \ilinpr{ \widetilde{A} } { (\Lambda \otimes\bpi(\Id_{\Cmplx^N})) \widetilde{A}}$.
Positivity of the LHS and RHS as $A$ varies is the definition of
``$\Lambda$ is rank-$N$-positive''
and ``$\Lambda \otimes\bpi(\Id_{\Cmplx^N})$ is rank-$1$-positive'', respectively.
This is a $\Leftrightarrow$ b.
For b $\Leftrightarrow$ c, note that
$\jam$ distributes over $\otimes$, according to
(\ref{eq:jam distributes over tensor}), so
\begin{equation}\nonumber
  \Lambda\otimes\bpi(\Id_{\Cmplx^N})
  \bijam
  \jam\Lambda\otimes\btheta(\Id_{\Cmplx^N})
  = \jam\Lambda\otimes \Id_{\Blin(\Cmplx^N)}.
\end{equation}
Thus, $\Lambda\otimes\bpi(\Id_{\Cmplx^N})$ is rank-$1$-positive iff 
$\jam\Lambda\otimes \Id_{\Blin(\Cmplx^N)}$ is monotone,
by \S~\ref{sec:rk-1-pos <-> mon}.
But this latter condition is the definition of ``$\jam\Lambda$ is $N$-monotone''.
\end{proof}

\subsubsection{$\Pos(\Blin(\cH,\cK)) \bijam \CP(\cH,\cK)$}

This follows immediately from the previous step because
the definition of $\CP$ can be slightly rephrased as ``$\forall N\ge 1$, $N$-monotone'',
and that of positive as ``$\forall N\ge 1$, rank-$N$-positive''.

Looking more closely we see that $\Lambda\in\Blin(\Blin(\cH),\Blin(\cK))$ is
$\CP$ as soon as it is $\dim\cH\cdot\dim\cK$-monotone because the maximum rank
of an operator in $\Blin(\Blin(\cH,\cK))$ is $\dim\cH\cdot\dim\cK$.

%
%
\section{Kraus decomposition via Jamio{\l}kowski transform}\label{sec:Kraus}

\subsection{Kraus decomposition is extremal decomposition}\label{sec:Kraus and extreme}

$\Pos(\Blin(\cH,\cK))$ is a proper cone in $\SA(\cH,\cK)$ [Example \ref{sec:pos cone}],
hence $\CP(\cH,\cK)$ is a proper cone in $\HP(\cH,\cK)$. There is more to be said,
however, especially in connection with extremality.

\subsubsection{Definitions (extreme/pure CP map,  extremal decomposition, $\extr$)}
\label{sec:extremality defs}

A point of a convex set $\cC$ is \textit{extreme} if it cannot be expressed
as a convex combination \mbox{$\lambda y + (1-\lambda) z$, $0<\lambda<1$}
$y$ and $z$ \textit{distinct} points in $\cC$.

For a convex cone $\cC$, a slight variation on this concept is more useful,
because at most the origin is an extreme point, in the preceding sense.
Vector $x$ in $C$ is called \textit{extreme vector}, and the ray
it is on is an \textit{extreme ray}, if it cannot be expressed as a sum $y+z$
of linearly independent nonzero $y,z$ in $\cC$.

\textit{Extremal decomposition} refers to either 
expression of a vector in a convex cone as a sum of extreme vectors,
or expression of a point in a convex set as a convex combination of
extreme points. Context should disambiguate.

Since the term ``vector'' can be confusing in our context, we will refer
to elements of an extreme ray of $\CP(\cH,\cK)$ 
as \textit{extreme CP maps} or \textit{pure CP maps}.
The motivation for the latter terminology will become clear in
\S~\ref{sec:Kraus and extreme}. The set of extreme maps in
$\CP(\cH,\cK)$ is denoted $\extr\CP(\cH,\cK)$.


\subsubsection{The extreme positive operators are those of rank 1}
\label{sec:extreme pos are rank 1}

Every operator in $\Pos(\cH)$ is decomposable into a sum of extreme (or pure)
positive operators;
the latter are the rank-1 elements, the image of $\bpi$.
It is a familiar matter to write a positive operator as a sum $\sum \bpi(h_i)$
where the $h_i$ are eigenvectors. Another way to make such a decomposition is
described in \S~\ref{sec:Pos red} below.

It is perhaps less familiar that $\bpi(h)$ is extreme, i.e., in any decomposition as
a sum of positive operators all terms are proportional to $\bpi(h)$. Therefore,
we provide a proof.

\begin{lem}\label{lem:range-span}
If $S,T,W\colon \Pos(\cH)$ satisfy $S+T=W$, then
$\Ran S + \Ran T = \Ran W$.
\end{lem}
\begin{proof}
  We rely on the basic fact that for any positive operator $M$,
  \begin{equation}
  \ker A = \setof{v\colon \cH}{\inpr{v}{Av} = 0}.
  \end{equation}
$\ker W = \ker S \cap \ker T$ follows immediately.
Now use the $(\Ran A)^\perp = \ker A^\dag$, valid for any operator, together with
\mbox{$(\Ran S)^\perp \cap (\Ran T)^\perp = (\Ran S + \Ran T)^\perp$}.
\end{proof}

\paragraph{Comment: more-geometrical methods.}
Our proof of the existence of extremal decomposition relied on the existence of
a complete set of eigenvectors.
This is certainly a good argument, but one more in keeping with the geometric flavor
can be made by appeal to the (finite-dimensional) Krein-Milman theorem which says that
extremal decompositions always exist for a compact (i.e., closed and bounded) convex set.
Every ray in the cone $\Pos(\cH)$ intersects the affine hyperplane of trace-1 operators,
and that intersection is compact.
Thus, for a normalized multiple of any positive operator
a decomposition into a convex combination of extreme trace-1 positive operators
exists by Krein-Milman, and this can be used to obtain a decomposition of the
original operator.

\subsubsection{Extreme and boundary are not the same thing}

\begin{equation}\nonumber
  \begin{tikzpicture}[scale=0.6,thick]
\coordinate (apex) at (0,0); 
\coordinate (corner1) at (5,3);
\coordinate (corner2) at (6,1.0);
\coordinate (corner3) at (5.5,4);
\coordinate (corner4) at (6,2.6);
\coordinate (mid) at ($ (apex)!50!(corner2) $);
\draw (apex) -- (corner1);
\draw (apex) -- (corner2);
\draw (apex) -- (corner3);
\draw[dashed] (apex) -- (5.4,2.34);
\draw (5.4,2.34) -- (corner4);
\draw[dashed,white!60!black] (corner2) -- (corner1) -- (corner3) -- (corner4) -- cycle;
\draw[fill] (apex) circle (2pt);
\node at (-0.2,-0.3) {$0$};
\draw[-Stealth,red,very thick] (apex) -- (5.4,2.1);
\node at (5.8,2.1) {\color{red}{$\rho$}};
\draw[-Stealth,blue,very thick] (apex) -- (2.8,0.46);
\node at (3.0,-0.1) {\color{blue}{$\bpi(A_1)$}};
\draw[-Stealth,blue,very thick] (apex) -- (3.2,1.9);
\node at (2.6,2.5) {\color{blue}{$\bpi(A_2)$}};
\end{tikzpicture}
\end{equation}

A point $x$ is a \textit{boundary point} of a set $S$ precisely when
every neighborhood of $x$ intersects both $S$ and its complement.
In particular, $x$ need not belong to $S$. This is a topological concept.
The concept of extreme point is different. In $\Pos(\cH)$, the extreme
points are the operators of rank 1. Interior points have full rank
($\dim\cH$). A positive operator with rank strictly between these extremes is
a boundary point of $\Pos(\cH)$, but not extreme.

\subsubsection{Kraus decomposition from extremal decomposition of
  positive operators}

We saw in \S\ref{sec:extreme pos are rank 1} that the extreme elements of
$\Pos(\Blin(\cH))$ are those of the form $\pi(A)$.
Since $\jam$ is an isometry between $\Pos(\Blin(\cH))$ and $\CP(\cH)$,
we immediately deduce the following
\begin{itemize}
\item 
the extreme elements of the cone $\CP(\cH,\cK)$
are precisely those of form $\btheta(A)$,
\item any $\Lambda\colon\CP(\cH,\cK)$ can be expressed as
$\sum \btheta(A_i)$, for some collection of operators $A_i\colon\Blin(\cH,\cK)$,
\item the $A_i$ may even be chosen mutually orthogonal.
\end{itemize}

A decomposition of a $\CP$ map $\Lambda$ of the form
\begin{equation}\nonumber
\Lambda = \sum \btheta(A_i)
\end{equation}
is known as a \textit{Kraus decomposition}.
We have just seen that Kraus decompositions exist and 
that they are decompositions into pure $\CP$ maps.
The use of the term ``pure'' is doubly justified: via $\jam$ and via $\btheta$.

\subsubsection{The set of extreme (pure) CP maps is closed}\label{sec:extr CP closed}

The \textit{Kraus rank} of a $\CP$ map $\Lambda$ is the
minimal number of $\btheta$-maps in a Kraus decomposition.
It is equal to the ordinary rank of $\jam\Lambda$.
\begin{lem*}
  {Rank}, {denoted $\rk$}, is a lower semicontinuous function
  on $\Blin(\cH,\cK)$.
  That is, \\
  $\setof{A\in\Blin(\cH,\cK)}{\rk A \le m}$ is closed.
\end{lem*}
\begin{proof}
  $\rk A > m$ if and only if there are linearly independent
  $k_1,\ldots,k_{m+1}$ such that $A^\dag k_i \neq 0$ for every $i$.
  This condition is clearly stable under sufficiently small
  perturbation of $A$, i.e., $\setof{A}{\rk A > m}$ is open.
\end{proof}
In particular, the positive maps with rank not exceeding one, which is
exactly the extreme positive maps, is closed. Since $\jam$ is an
isometry, the same is true of the set of extreme $\CP$ maps.

\subsection{A quasi-Cauchy-Schwarz inequality for \CP\ maps}\label{sec:quasi-CS ineq}

Any operator $\Lambda$ in $\Blin(\Blin(\cH),\Blin(\cK))$ is
determined by its ground matrix elements (\S~\ref{sec:J in terms of GMEs}).
In turn, all of those matrix elements can be expressed as linear
combinations of matrix elements between pure states, according to
the polarization identity
\begin{equation}
  \inpr{k'\pam{k}}{\Lambda\cdot h'\pam{h}}
  =
  \tfrac{1}{16}\sum_{n,m=0}^3 i^{n-m} 
  \inpr{\bpi(k'+i^mk)}{\Lambda\cdot \bpi(h'+i^nh)},
\end{equation}
which is verified straightforwardly by expansion of the RHS.

For $\CP$ maps, we can go further.
$\Lambda\colon\CP(\cH,\cK)$ satisfies the quasi-Cauchy-Schwarz inequality
\begin{tcolorbox}[standard jigsaw,opacityback=0]
\begin{equation}
  \label{eq:quasi-CS}
|\inpr{k'\pam{k}}{\Lambda\cdot h'\pam{h}}|^2  
\le
\inpr{\bpi(k')}{\Lambda\cdot \bpi(h')}
\inpr{\bpi(k)}{\Lambda\cdot \bpi(h)}
\end{equation}
\end{tcolorbox}
\begin{proof}
  The Cauchy-Schwarz inequality applies to any sesquilinear form, e.g.,
  for a positive operator $T$, $|\inpr{x}{Ty}|^2 \le |\inpr{x}{Tx}\inpr{y}{Ty}|$,
  and $T$ here could be $\jam\Lambda$. Now calculate
\begin{equation}\nonumber
  \begin{split}
|\inpr{k'\pam{k}}{\Lambda\cdot h'\pam{h}}|^2  
& =
| \inpr{k'\pam{h'}}{\jam\Lambda\cdot k\pam{h}}|^2 \\
& \le
|\inpr{k\pam{h}}{\jam\Lambda\cdot k\pam{h}}
\inpr{k'\pam{h'}}{\jam\Lambda\cdot k'\pam{h'}}| \\
& =
|\inpr{k'\pam{k'}}{\Lambda\cdot h'\pam{h'}}
\inpr{k\pam{k}}{\Lambda\cdot h\pam{h}}|
\end{split}
\end{equation}
\end{proof}
\subsection{A construction of Kraus decomposition}\label{sec:Kraus construction}

This subsection exposes an algorithm for constructing Kraus decomposition which
does not require explicit Jamio{\l}kowski transformation, much less finding
eigenvectors. We will use it to prove the existence of Kraus decomposition
in the separable setting, in \S\ref{sec:Kraus construction 2}.

\subsubsection{Reducing a positive operator}\label{sec:Pos red}

\begin{lem}\label{lem:Pos red}
  The (nonlinear) map
  \mbox{$\reduce\colon\cH\times\Pos(\cH)\rightarrow\cH\times\Pos(\cH)$}
  defined by:
  $\reduce(h,T) = (h',T')$, where
  \begin{equation}\label{eq:red}
    h' =
    \begin{cases}
      0, & h\in\ker T \\
\frac{T{h}}{\sqrt{\inpr{h}{Th}}}, & \text{otherwise},
    \end{cases}
  \end{equation}
  and $T = \bpi(h') + T'$, satisfies
  \begin{equation}\label{lem:Pos reduction}
\ker T' \supseteq \ker T + \Cmplx h.
  \end{equation}
\end{lem}
\begin{proof}
\begin{equation}\label{eq:lem red 1}
 \inpr{h}{Th} \inpr{k}{\bpi(h')k}
 = |\ilinpr{k}{T h}|^2
 \le \inpr{h}{T h} \inpr{k}{T k}.
\end{equation}
The inequality here is Cauchy-Schwarz (see \S~\ref{sec:quasi-CS ineq}).
This shows $\bpi(h') \le T$, so that $0 \le T' \le T$, i.e.,
$T'$ is positive.
In case $k=h$, `$\le$' in (\ref{eq:lem red 1}) becomes `$=$', so
$\inpr{h}{T' h} = 0$, and therefore $T' h = 0$.
\end{proof}

The positive operator $T'$ is a \textit{reduction} of $T$ with respect to $h$,
meaning 
\subsubsection{Recursive application}\label{sec:Pos recursive red}

Now, given $T_0\in\Pos(\cH)$, and an orthogonal basis $h_1,\ldots,h_{\dim\cH}$ of $\cH$,
define recursively
\begin{equation}\label{eq:Pos recursive}
  (h'_n,T_n) \defeq \reduce(h_n,T_{n-1}), \quad n=1,\ldots,\dim{\cH}.
\end{equation}
Then,
\begin{equation}
\ker T_n \supseteq \cH_n \defeq \Span \{h_1,\ldots,h_n\}.
\end{equation}
Between vectors in $\cH_n$, the operators $\sum_{i=1}^n \bpi(h'_i)$ and $T_0$ have
the same matrix elements, while those of $T_n$ are zero.
It follows in particular that $T_{n+1}=0$, or 
\begin{equation}
T_0 = \sum_{i=1}^{\dim\cH} \bpi(h'_i).
\end{equation}

\subsubsection{Reducing a $\CP$ map}\label{sec:CP red}


To turn this into an algorithm for construction of Kraus decomposition, specialize
to the case $\cH=\Blin(\cK,\cN)$, use Jamio{\l}kowski transform, and work in terms
of ground matrix elements, using ({\ref{eq:J in GME form}})
from \S~\ref{sec:J in terms of GMEs}, namely
\begin{equation}\tag{*}
\inpr{n'\pam{n}}{\Lambda\cdot k'\pam{k}}
=
\inpr{n'\pam{k'}}{\jam\Lambda\cdot n\pam{k}}.
\end{equation}
After the formulas are established, $\jam$ will
no longer be needed.

Define (caution: we are overloading the name `$\reduce$')
\begin{equation}
    \reduce(n,k,\Lambda) \defeq (A,\Lambda'),
    \quad \text{where }
    (A,\jam\Lambda') = \reduce(n\pam{k},\jam\Lambda) 
\end{equation}
Then, taking $T=\jam\Lambda$ and $h=n\pam{k}$ in Lemma \ref{lem:Pos red},
$h'$ there becomes $A$, written in terms of matrix elements,
with the aid of (*) above, as
\begin{equation}
  \label{eq:CP reduction}
  \inpr{n'}{A k'} =
  \left\{
  \begin{array}{ll}
0, & c=0 \\    
c^{-1/2} \inpr{n'\pam{n}}{\Lambda\cdot k'\pam{k}}, & \text{otherwise}
  \end{array}
  \right\}
\quad \text{where }
c = {\inpr{\bpi(n)}{\Lambda\cdot \bpi(k)}}.
\end{equation}
Similarly, $T'$ in the Lemma becomes $\jam\Lambda'$, so that
\begin{equation}
\Lambda = \btheta(A) + \Lambda'.
\end{equation}

An immediate consequence of (*) and the quasi-Cauchy-Shwarz inequality is,
\begin{equation}\label{eq:CP kernel}
\jam\Lambda'\cdot k\pam{h} = 0
\quad\Leftrightarrow\quad
\inpr{\blank\pam{k}}{\Lambda'\cdot \blank\pam{h}} = 0
\quad\Leftrightarrow\quad
\inpr{\bpi(k)}{\Lambda'\cdot \bpi(h)}=0,
\end{equation}
the middle expression being a function on $\cK\times\cH$ with $\blank$'s
for the arguments.
Therefore,
\begin{equation}\label{eq:?}
  \inpr{\bpi(n')}{\Lambda' \cdot \bpi(k')} = 0
  \;\Leftarrow\;
  \left\{
  \begin{array}{c}
  (n',k')=(n,k) \\
  \text{ or } \\
  \inpr{\bpi(n')}{\Lambda \cdot \bpi(k')} = 0
  \end{array}
\right.{.}
\end{equation}

\subsubsection{Sanity check: {\usebox\boxt}-map}

It is instructive to see how this works in case $\Lambda = \btheta(B)$ is
itself a $\btheta$-map.
Then, in \ref{eq:CP reduction},
\mbox{$\inpr{n'\pam{n}}{\Lambda\cdot k'\pam{k}} = \inpr{n'}{B k'} \inpr{B k}{n}$}
and $c = |\langle n | Bk \rangle|^2$. Hence, if $c\neq 0$, 
$A$ is equal to $B$ times a phase, so that $\btheta(A) = \btheta(B)$,
which is reassuring.

\subsubsection{Iterative decomposition algorithm:  reducing one dimension at a time}
\label{sec:Kraus algorithm}

A $\CP$ map can be reduced recursively, just as was done for
a positive operator in \S~\ref{sec:Pos recursive red}.
Given orthogonal bases $n_1,\ldots,n_{\dim\cN}$ (of $\cN$) and $k_1,\ldots,k_{\dim\cK}$
(of $\cK$),
enumerate the corresponding basis of $\Blin(\cK,\cN)$
as $B_k \defeq n_{l(k)}\pam{k_{r(k)}}$, $k=1,\ldots,\dim\cN\cdot\dim\cK$.
Successively reduce by $B_1,B_2,\ldots$:
\begin{equation}
  (A_m,\Lambda_m) \defeq
  \reduce\Big(B_m,\Lambda_{m-1} \Big),
  \quad m=1,\ldots,\dim{\cK}\cdot\dim\cN.
\end{equation}
Then we have
\begin{equation}\label{eq:reduction n}
\Lambda_0 = \sum_{i=1}^{n} \btheta(A_i) + \Lambda_n,
\end{equation}
and
\begin{equation}
\Lambda_{1+\dim\cK\cdot\dim\cN} = 0.
\end{equation}

\paragraph{Example reduction schedule.}\label{sec:red schedule}
\leavevmode

The following figure shows an example of a useful basis enumeration.
\begin{center}
\begin{tikzpicture}[scale=0.7]
  \foreach \x in {0,1,2,3,4,5}{
    \foreach \y in {0,1,2,3,4,5}{
      \draw[fill] (\x,\y) circle (2pt);
    }
  };
  \coordinate (shift) at (0.25,0.25);
  \draw (0,0)+(shift) node{1};
  \draw (1,0)+(shift) node{2};
  \draw (0,1)+(shift) node{3};
  \draw (2,0)+(shift) node{4};
  \draw (1,1)+(shift) node{5};
  \draw (0,2)+(shift) node{6};
  \draw (3,0)+(shift) node{7};
  \draw (2,1)+(shift) node{8};
  \draw (1,2)+(shift) node{9};
  \draw (0,3)+(shift) node{10};
  \draw (4,0)+(shift) node{11};
  \draw (3,1)+(shift) node{12};
  \draw (2,2)+(shift) node{13};
  \draw (1,3)+(shift) node{14};
  \draw (0,4)+(shift) node{15};
  \draw (5,0)+(shift) node{16};
  \draw (4,1)+(shift) node{17};
  \draw (3,2)+(shift) node{18};
  \draw (2,3)+(shift) node{19};
  \draw (1,4)+(shift) node{20};
  \draw (0,5)+(shift) node{21};
  \draw[dashed] (-0.25,-0.25) rectangle (2.6,2.6);
  \coordinate (l) at (-0.6,0.1);
  \draw (0,0) + (l) node{$k_1$};
  \draw (0,1) + (l) node{$k_2$};
  \draw (0,2) + (l) node{$k_3$};
  \draw (0,3) + (l) node{$k_4$};
  \draw (0,4) + (l) node{$k_5$};
  \draw (0,5) + (l) node{$k_6$};
  \coordinate (b) at (0.1,-0.6);
  \draw (0,0) + (b) node{$h_1$};
  \draw (1,0) + (b) node{$h_2$};
  \draw (2,0) + (b) node{$h_3$};
  \draw (3,0) + (b) node{$h_4$};
  \draw (4,0) + (b) node{$h_5$};
  \draw (5,0) + (b) node{$h_6$};
  \draw[dashed] (-0.25,6) -- ++(0,-6.25) -- ++(6,0);
\end{tikzpicture}
\end{center}
The dot in the $j^{\text{th}}$ row from the bottom and $i^{\text{th}}$
column from the left corresponds to $k_j\pam{h_i}$. If we enumerate in
the order shown, then after 13 steps, we will have reduced out
$\Span\{k_1,k_2,k_3\}$ and $\Span\{h_1,h_2,h_3\}$. That is,
for $k,k'$ and $h,h'$ in those subspaces,
$\inpr{k'\pam{k}}{\Lambda_{13}\cdot h'\pam{h}} = 0$.

\section{Miscellany}\label{sec:miscellany}

This section further develops some of the ideas we have been
working with and demonstrates more uses for $\jam$.
This material is not called on in later sections.

\subsection{More on representations of $\CP$ maps}

\subsubsection{Freedom in Kraus decomposition}

It is obvious that the particular operators $A_i$ appearing in
a Kraus decomposition $\Lambda = \sum \btheta(A_i)$ are not uniquely determined.
For a somewhat silly example,
$\btheta(A) = \btheta(\tfrac{A}{\sqrt{2}})+\btheta(\tfrac{A}{\sqrt{2}})$.
In fact, we have the following.
\begin{prop*}
  $\sum \btheta(A_i) = \sum \btheta(B_j)$ if and only if
  $A_i = \sum_j U_{ij} B_j$ for some unitary matrix $U$,
  where the smaller of the sets $\{A_i\}$ and $\{B_j\}$
  is padded out with zeros so that $U$ can be square.
\end{prop*}

Judging by other sources, this seems to be considered of some importance.
Nielsen \& Chuang (p. 372, \cite{Nielsen+Chuang}) and
Breuer \& Petruccione (p. 91, \cite{Breuer+Petruccione}) both expend
nontrivial effort reducing it to an analogous statement about density operators
(to be given momentarily). Vacchini (pp.~116--119) proves it directly.
We will have no use for this Proposition. The reason for discussing it is
to point out that, with our machinery, the reduction to the analogous
statement for density operators is immediate via $\jam$.
Indeed, $\sum \btheta(A_i) = \sum \btheta(B_j)$
if and only if $\sum \bpi(A_i) = \sum \bpi(B_j)$.

\paragraph{}

Having come this far, though, we might as well address the freedom in
extremal decomposition of a positive operator.
To make it more interesting, we will deduce it from a statement,
not about how we can represent things, but about the things themselves.
Consider a bipartite system $\cH\otimes\cK$.
How much can we learn about a pure state of this system through
measurements only on $\cH$?
Prop. \ref{sec:bipartite identifiability} just below says that
we can narrow it down only to an orbit of $\Id_{\cH}\otimes U(\cK)$,
where $U(\cK)$ is the unitary group on $\cK$.
\S\ref{sec:freedom in extremal decomp} explains how this solves
the posed question about extremal decomposition and makes contact
with the idea of purification (\S\ref{sec:purification}) again.

\paragraph{Proposition.}\label{sec:bipartite identifiability}\leavevmode

If
\begin{equation}
\forall A\colon\Blin(\cH),\quad
\inpr{X}{(A\otimes\Id_\cK) X}  
= \inpr{Y}{(A\otimes\Id_\cK) Y},
\end{equation}
then $X = (\Id_\cH\otimes U)Y$ for some $U$ in $\Uni(\cK)$,
the unitary group over $\cK$.
\begin{proof}
With $(h_i)$ an ONB of $\cH$,
ecompose the vectors as
  $X=\sum h_i\otimes x_i$, $Y=\sum h_i\otimes y_i$.
  Then
  \begin{equation}\nonumber
  \inpr{X}{(A\otimes\Id_\cK) X}  = \sum_{ij} A_{ij}\inpr{x_i}{x_j},
\end{equation}
and similarly for $Y$. But, the matrix elements $A_{ij}$ can be anything,
so $\inpr{x_i}{x_j} = \inpr{y_i}{y_j}$. The conclusion follows by appeal
to the following lemma.
\end{proof}
\begin{lem*}
  If
  $\forall i,j=1,\ldots,N$,
  \mbox{$\inpr{\varphi_i}{\varphi_j} = \inpr{\psi_i}{\psi_j}$},
  then $\varphi_i \mapsto \psi_i$ is a well-defined unitary transformation
  from $\Span\{\varphi_i\}$ to $\Span\{\psi_i\}$.
\end{lem*}
\begin{proof}
  It is clear that the suggested mapping is unitary if it is
  a well-defined linear map. To show that, we must show
  that it respects linear dependence.
  Thus, suppose $c_i\varphi_i = 0$. This is equivalent to
  $[\inpr{\varphi}{\varphi}][c]=0$ in a schematic matrix notation.
  But, this in turn is equivalent to
  $[\inpr{\psi}{\psi}][c]=0$ and to $c_i\psi_i = 0$. 
\end{proof}
\paragraph{Application to freedom in extremal decomposition.}
\label{sec:freedom in extremal decomp}\leavevmode

$\rho=\sum \bpi(x_i)$ has no more than $\dim\cK$ terms, it
has a $\cK$-purification $\bpi(X)$ with $X=\sum x_i\otimes k_i$
and $k_1,\ldots,k_{\dim\cK}$ an ONB of $\cK$. For convenience, we
refer to $X$ itself also as a purification of $\rho$.
Now, if $\rho=\sum\bpi(y_i)$ is a second extremal decomposition,
we get a second $\cK$-purification $Y=\sum y_j\otimes k_j$
(if $\dim\cK$ is large enough).

That $X$ and $Y$ are both purifications of $\rho$ is equivalent to
$\inpr{X}{(A\otimes\Id_\cK)X}=\inpr{Y}{(A\otimes\Id_\cK)Y}$ for all
$A\colon\Blin(\cH)$.
Applying Prop. \ref{sec:bipartite identifiability}, $X=(\Id_\cH\otimes U)Y$,
and
\begin{equation}\nonumber
  (\Id_\cH\otimes U)Y
  = \sum_{i} y_i \otimes \Big(\sum_j U_{ji}k_j\Big)
  = \sum_{j} \Big(\sum_i U_{ji} y_i\Big) \otimes k_j
\end{equation}
yields $x_j = \sum_i U_{ji} y_i$.

\subsubsection{Representating a $\CP$ map as a sum of rank-1 operators}
\label{sec:sum of rank-1}

Consider the Jamio{\l}kowski correspondence
\begin{equation}\nonumber
  \sum B_i \pam{A_i} \bijam \sum B_i\blank A_i^\dag.
\end{equation}
If $A_i , B_j\in\Blin(\cH,\cK)$, then the LHS may be positive
or the RHS $\CP$, and either one of those possibilities implies the other.
This was the situation which largely concerned us in \S\ref{sec:Kraus}.
Now we want to consider
$B_j\colon\Blin(\cK)$, $A_i\colon\Blin(\cH)$.
Matters are then reversed,
in that the LHS is in $\Blin(\Blin(\cH),\Blin(\cK))$, hence is conceivably $\CP$.

Representing an operator as a sum of rank-1 operators in this way is often
very practical, so we are naturally interested in rewriting positivity of
the RHS is a more digestible form.
Breuer\cite{Breuer-07} develops the criteria for such a sum of rank-1 operators
to be a trace-preserving $\CP$ projection, where by ``projection'' we mean
merely an idempotent operator ($\Lambda\circ\Lambda = \Lambda$).
This theorem is also discussed on pp. 283--286 of Vacchini\cite{Vacchini}.
We reproduce it here in our idiom, illustrating the use of $\jam$
for that purpose. 

\paragraph{$\sum B_i\pam{A_i}$ is $\CP$ $\Leftrightarrow$
  $\sum B_i \otimes (A_i^\dag)^T \ge 0$.}
\leavevmode

If $A$ is in $\Blin(\cH,\cK)$, then its transpose $A^T\in\Blin(\cK^*,\cH^*)$
is defined via
\begin{equation}
  (A^T\pam{k}) h
  = \pam{k} A h.
\end{equation}
We can even write this as $A^T = \blank A$.
No basis is needed to make this definition.
However, note that,
if $[A]$ is the matrix of $A$ with respect to ONBs
$(k_i)$ and $(h_i)$, then $[A]^T$ --- the transpose of that matrix ---
is the matrix of the operator $A^T$ with respect to the dual
ONBs $(\pam{h_i})$ and $(\pam{k_i})$.

Recall from \S\ref{sec:HS(H,K)=KxH*} the Hilbert isomorphism
$\HS(\cH,\cK)\cong \cK\otimes\cH^*$ under which
$k\pam{h} \leftrightarrow k\otimes \pam{h}$. This induces a corresponding
Hilbert isomorphism 
\begin{equation}
B\blank A^\dag \colon\Blin(\Blin(\cH,\cK))
\leftrightarrow B\otimes (\blank A^\dag) \colon\Blin(\cK\otimes\cH^*).
\end{equation}
Hence, positivity of $\sum B_i\blank{A_i^\dag}$ is 
equivalent to positivity of $\sum B_i\otimes (\blank{A_i^\dag})
= \sum B_i\otimes (A_i^\dag)^T$.
Of course, we may cash this out in the form of a matrix.

%
%
%
%
\paragraph{$\sum B_i\pam{A_i}$ is trace-preserving $\Leftrightarrow$
  $\sum (\Tr B_i)^* A_i = \Id_\cH$.}
\leavevmode

This follows simply from
$\Tr((B\pam{A})\rho)
= \inpr{\Id_\cK}{(B\pam{A})\rho}
= \inpr{(A\pam{B})\Id_\cK}{\rho}
= \inpr{B}{\Id_\cK} \inpr{A}{\rho}
= (\Tr B)^* \inpr{A}{\rho}$.

\paragraph{$\sum B_i\pam{A_i}$ is idempotent $\Leftarrow$ $\inpr{A_j}{B_i} = \delta_{ji}$.}
\leavevmode

We have
\begin{equation}\nonumber
  \left( \sum_i B_i\pam{A_i}\right)^2
  = \sum_{i,j} \inpr{A_j}{B_i} B_j\pam{A_i}. 
\end{equation}
Hence, $\sum B_i\pam{A_i}$ is idempotent if the condition $\inpr{A_j}{B_i}=\delta_{ji}$
holds. That is not \textit{necessary}, as two of the $(B_i,A_i)$ pairs might well
be identical. However, if we assume that either $(A_i)$ or 
$(B_j)$ is linearly independent, the reverse implication holds.
Actually, it is possible to do better than that.

\paragraph{The $A$'s may be chosen mutually orthogonal,  and the $B$'s as well.}
\leavevmode

This is just a special case of Schmidt decomposition.

\paragraph{$\sum B_i\pam{A_i}\in\HP(\cH,\cK) \Leftrightarrow$ 
  $A$'s and $B$'s may be chosen hermitian.}
\leavevmode

Essentially by definition
\begin{equation}\nonumber
\HP(\cH,\cK) = \Blin_\Real(\SA(\cH),\SA(\cK)).
\end{equation}
Everything in sight here is an $\Real$-Hilbert space (the subscript
emphasizes this),
and $\Blin_\Real(\SA(\cH),\SA(\cK))$ is embedded in
$\Blin(\Blin(\cH),\Blin(\cK))$ simply by declaring that its operators
commute with multiplication by $i$.
With this observation, the conclusion ``$\Rightarrow$'' is immediate.
We can even carry
out a Schmidt decomposition so that the $A$'s are not only hermitian,
but mutually orthogonal, and similarly the $B$'s.

The ``$\Leftarrow$'' direction is obvious, since, again, the inner product
of hermitian operators is real.


%
%
%

\subsection{More about the category $\CP$}

\subsubsection{States as morphisms and \usebox{\boxp} as a special case of \usebox{\boxt}}\label{sec:states as morphisms}
  \newcommand{\ssto}{{\scriptscriptstyle{\to}}}
  
  There is a natural isomorphism $\cH \rightarrow \Blin(\Cmplx,\cH)$,
  given by $h\mapsto h^\ssto$, where $h^\ssto c = ch$.
  Although this isomorphism is natural, we are wary of 
  outright identification as confusion may result.
  Then, $c h_{\scriptscriptstyle{\to}}^\dag h' = \inpr{(h^\ssto)^\dag h'}{c}
  = \inpr{h'}{h^\ssto c}
  = c\inpr{h'}{h}$ implies that $(h^\ssto)^\dag = \pam{h}$;
we \textit{do} identify $\cH^*$ with $\Blin(\cH,\Cmplx)$ here,
as we have done all along.
Now, indentifying $c\colon\Cmplx$
with the multiplication operator by $c$ in $\Blin(\Cmplx)$,
  $\btheta(h_{\scriptscriptstyle{\to}})\in\CP(\Cmplx,\cH)$ is 
\begin{equation}\nonumber
  \btheta(h^\ssto)
  = h^\ssto\blank (h^\ssto)^\dag
  = h^\ssto \blank \pam{h}
  = (c\colon\Cmplx) \mapsto c \bpi(h).
\end{equation}
And, running our isomorphism backward, this is ``just'' $\bpi(h)$.
In this way, $A\colon\Pos(\cH)$ is
essentially the same as the $\CP$ map $c \mapsto c A$,
and
\begin{equation}
\Pos(\cH) \cong \CP(\Cmplx,\cH). 
\end{equation}
As for $h$ above, when reading $A\ge 0$ as a $\CP$ map, we write it as
$A^\ssto$.
Identification of $h$ with $h^\ssto$ turns the relation
$\btheta(A)\bpi(h) = \bpi(Ah)$,
which we have used several times, into
an expression of functoriality: $\btheta(A)\btheta(h)=\btheta(Ah)$.
  
Similarly,
$\btheta(\pam{h}) = \pam{h} \blank \pam{h}^\dag
= \pam{h} \blank {h^\ssto}$,
which takes $A\colon\Blin(\cH)$ to the map
$c \mapsto c\inpr{h}{Ah} =
c\inpr{\bpi(h)}{A} = c\pam{\bpi(h)}(A)$, thereby identifying
$\btheta(\pam{h})$ with $\pam{\bpi(h)}$.

\begin{equation}
A\in\Pos(\cH) \Leftrightarrow \pam{A}\in\CP(\cH,\Cmplx)  
\end{equation}
%

\subsubsection{
Some elementary  $\CP$ operations, with partial trace as a special case}

$\text{ -- }\otimes\rho$ is shorthand for the map
$(A\colon\Blin(\cH)) \mapsto A\otimes\rho$, for each $\cH$.
With $\rho\in\Pos(\cK)$ read as an element of $\CP(\Cmplx,\cK)$,
as in \S~\ref{sec:states as morphisms}, then
$\text{--}\otimes\rho$ is the same thing as
$\Id_{\Blin(\cH)}\otimes\rho^\ssto
  \in \CP(\cH)\otimes\CP(\Cmplx,\cK)$,
  hence is $\CP$.

  Then, 
  $(\Id_{\Blin(\cH)}\otimes\rho^\ssto)^\dag
= \Id_{\Blin(\cH)}\otimes(\rho^\ssto)^\dag
= \Id_{\Blin(\cH)}\otimes\pam{\rho}$
is $\CP$ by paragraph \ref{sec:CP creation} below.

\paragraph{$\Lambda\otimes\Gamma\in\CP$ implies that
$\Lambda$ and $\Gamma$ are $\CP$ up to a phase.}\label{sec:CP creation}

We know that tensor products of monotone operators are not necessarily monotone.
Suppose, conversely, $\Lambda\otimes\Gamma$ is monotone. Then,
\begin{equation}
  0 \le
  \inpr
  {\bpi(l\otimes m)}
  {(\Lambda\otimes\Gamma)\cdot \bpi(h\otimes k)}
  =
  \inpr{\bpi(l)}{\Lambda\cdot \bpi(h)}
  \inpr{\bpi(m)}{\Gamma\cdot \bpi(k)}.
\end{equation}
The complex phases of the factors on the RHS must therefore be opposite
for any $h,l,k,m$. But that means that the phase of
$\inpr{\bpi(l)}{\Lambda\cdot \bpi(h)}$ must be independent of $h,l$,
and similarly for the other factor. In other words, for some $\varphi$,
$e^{i\varphi}\Lambda$
and $e^{-i\varphi}\Gamma$ are $\Mon$.

Next, suppose that $\Lambda\otimes\Gamma$ is $\CP$.
Then, by definition of $\CP$,
$\Lambda\otimes\Id_{\Blin(\Cmplx^N)}\otimes\Gamma\otimes\Id_{\Blin(\Cmplx^M)}$ is
monotone for any $N,M$. By the previous paragraph,
$e^{i\varphi}\Lambda$ and $e^{-i\varphi}\Gamma$ are monotone (the phase must clearly
be independent of $M$ and $N$). Therefore,
$e^{i\varphi}\Lambda$ and $e^{-i\varphi}\Gamma$ are $\CP$.

\subsubsection{Isomorphisms in CP 
  are the \usebox{\boxt}-image of  isomorphisms in $\Hilb$}
\label{sec:Iso(CP) example}

[Cf. Thm.~3.4.1 of \cite{Rivas+Huelga}]

An isomorphism is a morphism with both a left and right inverse,
in which case those two are equal.
It is easy to see that if $A:\Blin(\cH,\cK)$ is invertible, then
$\btheta(A)^{-1} = \btheta(A^{-1})$.
This is a general property of functors, in fact.

For the converse, suppose that
\mbox{$\Lambda = \sum_i \btheta(A_i)\colon\CP(\cH,\cK)$}
is a two-sided inverse of
\mbox{$\Gamma = \sum_j \btheta(B_j)\colon \CP(\cK,\cH)$}.
Then,
\begin{equation}\nonumber
\sum_{ij} \btheta(B_jA_i) = \Id_{\Blin(\cH)}, \qquad
\sum_{ij} \btheta(A_iB_j) = \Id_{\Blin(\cK)}.
\end{equation}
But $\Id_{\Blin(\cH)} = \btheta(\Id_\cH)$
is an extreme map in $\CP(\cH)$,
and similarly $\Id_{\Blin(\cH)}$ in $\CP(\cK)$.
Therefore, \mbox{$A_iB_j \propto \Id_{\cK}$} and
\mbox{$B_jA_i \propto \Id_{\cH}$} for all $i,j$.
This easily implies the existence of $A$ such that $\forall i, A_i\propto A$ and
$\forall j, B_j\propto A^{-1}$.

\subsection{Trace-related properties}\label{sec:trace}

If a CP map is to preserve the normalization of density operators, it
must be trace-preserving in the sense of the following

\paragraph{Definitions.}
$\Lambda\colon\HP(\cH,\cK)$ is \textit{trace-non-increasing}, respectively
\textit{trace-preserving} when
$\Tr \Lambda\rho \le \Tr\rho$,
respectively $\Tr \Lambda\rho = \Tr\rho$,
for every $\rho\colon\Pos(\cH)$.
Some authors call trace-non-increasing CP maps \textit{quantum operations} and
trace-preserving ones \textit{quantum channels}.

It may seem at first glance that only strictly trace-preserving operations are
of interest in quantum physics, that is not so\cite{Kraus-83,Alicki+Lendi,Nielsen+Chuang}.

\paragraph{Adjoint formulation of trace-preserving.}

$\Tr A = \inpr{\Id_\cH}{A}$ for $A\colon\Blin(\cH)$. Therefore
a superoperator $\Lambda\colon\HP(\cH,\cK)$ is trace-non-increasing precisely when
\begin{equation}
\Lambda^\dag \Id_\cK \le \Id_\cH,  
\end{equation}
and trace-preserving when equality holds there.
Small enough multiples of any CP (or even monotone) $\Lambda$
are trace-non-increasing. In fact, $\| \Lambda \| \le 1$ (operator norm) suffices.
On the other hand, there may be \textit{no} multiple which is trace-preserving.
For instance, if $\Lambda = \btheta(A)$, there will be one only if
$A$ is proportional to a unitary.

\section{The separable setting: things fall apart}
\label{sec:things fall apart}

Starting in this section, we pivot toward the infinite-dimensional setting,
more accurately, the separable setting.
A separable Hilbert space is one which has a countable complete orthonormal
system. More generally --- and this applies equally to spaces like
$\Blin(\cH)$ --- a separable space is one which contains a countable dense
subset.

The most important points touched on in \S\ref{sec:things fall apart}
are the duality between the
space $\TC(\cH)$ of trace-class operators and the space $\Blin(\cH)$ of
bounded operators, and also
$\Blin(\TC(\cH),\TC(\cK))$
which plays the r\^{o}le that $\Blin(\Blin(\cH),\Blin(\cK))$
did in the finite-dimensional setting.

Norm topologies are important, but not enough for all our needs.
\S\ref{sec:SOT} discusses the strong-operator topology, partially
as a prototype for the GMET we will meet in \S\ref{sec:GMET}.
The very notion of topology is also discussed there.

\paragraph*{Textbook recommendations}

For more systematic developments of this material,
see Chapter 3 of Ref. \cite{BLPY}, \S 3.4 of Ref. \cite{Pedersen-Analysis},
or Chapter 7 exercises of Ref. \cite{Lang-Analysis}.

\subsection{Classes of bounded operators}

Since Hilbert space structures on all our spaces played such a
central r\^{o}le in the finite-dimensional theory, e.g., the
Jamio{\l}kowski transform was a unitary map,
we begin our exploration of the separable setting by seeing how that falls apart.
There is a space of linear operators on $\cH$ --- the Hilbert-Schmidt
operators --- for which a Hilbert space structure can be defined just
as before. But the associated topology is at the same time too strong
and too weak. We wind up with a rigged Hilbert space
\begin{equation}\nonumber
\TC(\cH) \subset \HS(\cH) \subset \Blin(\cH).
\end{equation}
All three of the spaces here are Banach spaces, and 
$\HS(\cH)$ is a Hilbert space. We discuss each in turn.

\subsection{The spaces and their norms}

\subsubsection{Bounded operators, $\Blin(\cH)$}

$\Blin(\cH)$ is the space of bounded operators on $\cH$,
with operator norm $\|A\| = \sup\{\|Ah\|\,:\, \|h\| = 1\}$.
It is a Banach space, which, we recall merely means that it
is sequentially complete: if $(A_n)$ is a Cauchy sequence
($\|A_n-A_m\| \to 0$ as $n,m\to\infty$), then there is some
operator $A$ to which the sequence converges. We might say that,
as far as the norm is concerned, there is nothing missing.

\subsubsection{Trace-class operators, $\TC(\cH)$}\label{sec:TC}

A {positive} operator $A$ has a trace in $[0,\infty]$
unambiguously defined by
\begin{equation}\label{eq:trace}
  \Tr A = \sum_i \inpr{h_i}{Ah_i},
\end{equation}
where $\{h_i\}$ is any complete orthonormal system in $\cH$.
For such an operator, the trace quasi-norm $\|A\|_1$ is equal to its trace.
By use of polar decomposition, we extend $\|\cdot\|_1$ to a all bounded operators:
if $B=MU$ with $M\ge 0$ and $U$ unitary, then
$\|B\|_1 \defeq \|M\|_1$. If $\|A\|_1 < \infty$, then $A$ is said to
be \textit{trace-class}. In that case, its trace is unambiguously
given by (\ref{eq:trace}).
The space of all trace-class operators equipped with this norm
is a Banach space denoted $\TC(\cH)$.

\subsubsection{Hilbert-Schmidt space, $\HS(\cH)$}

$A$ is Hilbert-Schmidt if $A^\dag A$ is trace class.
On the Hilbert-Schmidt operators we have the by-now familiar inner
product $\inpr{A}{B} = \Tr A^\dag B$, and norm $\|A\|_2 = \sqrt{\inpr{A}{A}}$.

\subsubsection{$\Blin(\TC(\cH),\TC(\cK))$}\label{sec:B(B1(H),B1(K))}

The superoperators which interest us are natively in $\Blin(\TC(\cH),\TC(\cK))$,
the space of all bounded linear operators from $\TC(\cH)$ to $\TC(\cK)$.
As always, ``bounded'' means that the operator quasi-norm
\begin{equation}
\|\Lambda\|_{1,1} = \sup\setof{\|\Lambda\rho\|_1}{\|\rho\|_1\le 1}
\end{equation}
is finite.
Instead of the generic ``$\|\cdot\|$'', we choose to decorate the
norm with subscripts corresponding to the two norms in the definiens.

\subsection{Trace-class/bounded duality}

The Hilbert-Schmidt inner product $\inpr{A}{B}$ does not extend to
$\Blin(\cH)$, but we can allow $A$ here to be any bounded operator if we
simultaneously restrict $B$ to be trace-class.
This pairing identifies $\Blin(\cH)$ with the Banach space
dual of $\TC(\cH)$ as $B \mapsto \inpr{B}{\;\cdot\;}$.
However, $\TC(\cH)$ is not the dual space of $\Blin(\cH)$,
i.e., $\TC(\cH)$ is not reflexive. 
That $\Blin(\cH)$ is the dual space of $\TC(\cH)$ means that the
operator norm satisfies
\begin{equation}\nonumber
\|B\| = \sup \setof{ |\inpr{B}{\rho}|}{\|\rho\|_1 \le 1}.
\end{equation}
Conversely,
\begin{equation}\nonumber
\|\rho\|_1 = \sup \setof{ |\inpr{B}{\rho}|}{\|B\| \le 1}.
\end{equation}
Actually, in this last formula, it is enough to consider finite-rank
operators $B$.
Here is another fact about the interplay of these spaces which will be useful
in the following. 
Since $BA$ runs over some subset of the ball of radius $\|A\|$ as $B$ runs over
the unit ball, \mbox{$\langle B,A\rho\rangle = \langle BA,\rho\rangle$}
implies that \mbox{$\|A\rho\|_1 \le \|A\| \|\rho\|_1$}.

\subsubsection{Bilinear versus sesquilinear pairing.}

\mbox{$\Lambda\colon\Blin(\TC(\cH),\TC(\cK))$}
has adjoint \mbox{$\Lambda^\dag\in\Blin(\Blin(\cK),\Blin(\cH))$}
defined by
\begin{equation}\nonumber
\inpr{\Lambda^\dag B}{\rho} = \inpr{B}{\Lambda \rho}.
\end{equation}

Conventionally, one uses a pairing $\Blin(\cH)\times\TC(\cH)$ which is
  linear in both variables: $(B,\rho) \mapsto \langle B, \rho\rangle
  =\Tr B\rho$. Then, one defines the dual of a superoperator as
  $\langle \Lambda^*,\rho\rangle = \langle B,\Lambda \rho\rangle$.
  For a general Banach space in place of $\TC(\cH)$, this would be
  the only choice.
  We are using instead a sesquilinear pairing in keeping with the
  rigged Hilbert space idea and for maximum continuity with the
  finite-dimensional theory.

\subsection{Meanings of old notations in the separable setting}

Much of the notation used in the finite-dimensional setting transfers
smoothly. The source of most of the clarifications needed arise from the
fact that many of the occurences of ``$\Blin(\cH)$'' should be replaced by
``$\TC(\cH)$''. In particular, the superoperators which interest us are
natively in $\Blin(\TC(\cH))$. Hence we now have
\begin{equation}\nonumber
  \CP(\cH,\cK)
  \subset \Mon(\cH,\cK)
  \subset \HP(\cH,\cK)
  \subset \Blin(\TC(\cH),\TC(\cK)).
\end{equation}
$\Pos(\cH)$ is now ambiguous.
Mostly, for positive trace-class or bounded operators,
we use ``$\TC(\cH)^+$'' or ``$\Blin(\cH)^+$'', but
``$\Pos(\cH)$'' might occur if the context makes clear which is intended.

\subsection{Fate of the Jamio{\l}kowski isomorphism}

The problem with the Jamio{\l}kowski isomorphism is not that
it cannot be defined in the infinite-dimensional setting, it's that
it is not defined on the spaces we would like.
$\HS(\cH,\cK)$ is a Hilbert space, hence so is
$\HS(\HS(\cH,\cK),\HS(\cN,\cM))$, and $\jam$ is an isomorphism
between this Hilbert space and $\HS(\HS(\cH,\cN),\HS(\cK,\cM))$.
In other words, we just decorate everything we did before with
superscripts `2'! Unfortunately, though,
identity operators over infinite-dimensional spaces are not Hilbert-Schmidt.

\section{Beyond norms: topology and systems of seminorms}
\label{sec:seminorms and topology}

This section is more in the nature of background than any other.
Readers comfortable with topology generally and strong-operator
topology in particular need not hesitate to skip it.

\subsection{First bite of the strong-operator topology (SOT)}\label{sec:SOT}

\subsubsection{The SOT seminorms}\label{sec:SOT defined}

The space $\Blin(\sX,\sY)$ of linear operators from one normed space
to another has one natural norm itself, namely
\begin{equation}\nonumber
\|A\| \defeq \sup\setof{\|Ax\|}{\|x\|\le 1}.
\end{equation}
Notice that the three norms occuring in the display are all distinct!
Looking more closely, we realize that
\begin{equation}\label{eq:strong seminorm}
A\mapsto |A|_x \defeq \|Ax\|
\end{equation}
for a fixed $x\colon\sX$
is \textit{almost} a norm.
In fact, it is a \textit{seminorm},
satisfying $|A+B|_x \le |A|_x + |B|_x$ and $|cA|_x = |c| |A|_x$,
but it does not necessarily separate; that is, $|A|_x = 0$ does not imply $A=0$.
However, collectively, the set of seminorms $\setof{|\cdot|_x}{x\in\sX}$
\textit{does} separate.

\paragraph{Sequential convergence in SOT versus norm: a few simple examples.}
\leavevmode

When should we say that a sequence $(A_n)_n$ converges to (say) zero with respect
to the SOT seminorms? Taking our cue from the situation for a norm,
we declare that such convergence means: For each $x$, $|A_n|_x \to 0$.

We give a few examples to give some idea of the difference between
norm and SOT convergence, and for that purpose specialize to the case
$\sX=\sY=\cH$ a separable infinite-dimensional Hilbert space.
Let $e_1,e_2,\ldots$ be an ONB of $\cH$, and consider $A_n \defeq \bpi(e_n)$.
This sequence does not converge in norm ($\|A_n-A_m\|=1$ for $n\neq m$).
However, for $h=\sum c_ne_n$, $A_nh = c_ne_n \to 0$.
As that says $|A_n|_h \to 0$ for every $h$, this sequence converges to zero
in SOT; we write that with fewer words as $A_n \xrightarrow{\mathrm{SOT}} 0$.
Similarly, $B_n\defeq e_1\pam{e_n} \xrightarrow{\mathrm{SOT}} 0$.
The sequence $B_n^\dag = e_n\pam{e_1}$ does not converge even in SOT,
however, since $B_n e_1 = e_n$ does not converge.
Omitting details, this means that the operation of hermitian conjugation is
not SOT-continuous; in contrast, it is norm-continuous.

\subsection{From metric to topology}\label{sec:metric space}

The basic guise of a norm is as a measure of the \textit{size} of a vector,
but one slips, almost without noticing, into using it to measure
\textit{distance} between vectors. The result is a metric space
with distance function $d(x,x') = \|x-x'\|$.
Recall that a metric space is a set $\sX$ equipped with a function
$d\colon \sX\times\sX \rightarrow [0,\infty)$ satisfying
\begin{itemize}
\item $d(x,y) = d(y,x)$ (symmetry)
\item $d(x,z) \le d(x,y) + d(y,z)$ (triangle inequality)
\item $d(x,y) = 0 \Rightarrow x = y$ (separation)  
\end{itemize}
For $\epsilon > 0$, the set $B_\epsilon(x) \defeq\setof{y\in\sX}{d(x,y) < \epsilon}$
is the \textit{open ball} of radius $\epsilon$ about $x$.
A sequence $(y_n)_n$ converges to $x$ iff $d(x,y_n) \to 0$ as $n\to\infty$.
Equivalently, the sequence is eventually in (and stays in) any given
open ball $B_\epsilon(x)$. If we imagine just being supplied with this
set of open balls, \textit{unlabelled}, we could still check the condition.
This indicates that the convergence is really a topological property.

To see what that means, we introduce some more terminology which applies
to topological spaces generally.
If $B_\epsilon(x) \subseteq U$, then $U$ is a \textit{neighborhood} of $x$.
$U$ is \textit{open} if it is a neighborhood of every point it contains.
Open balls are open, so the terminology is consistent, but there are many
more open sets than just those.
Now that we have the open sets for a metric space, we can make the jump
to general topology. A \textit{topology} on a set $\sX$ is a collection $\tau$
of subsets of $\sX$ such that
\begin{enumerate}
\item $\tau$ is closed under arbitrary unions.
\item $\tau$ is closed under finite intersection.
\item $\varnothing,\sX\in\tau$.
\end{enumerate}
Simple, isn't it? But that very austerity makes it difficult to see that
the idea has anything to do with approximation.
Actually, we will only be interested in so-called \textit{Hausdorff}
(or $T_2$) topologies which separate points by open sets, so we add
\begin{enumerate}[resume]
\item For $x\neq y$, there are disjoint $U,V\in\tau$ such that $x\in U$
  and $y\in V$.
\end{enumerate}

We can take this idea of topology backward through the preceding train of thought.
For a topological space $\sX$,
$Y\subseteq \sX$ is a neighborhood of $x$ if $x\in U\subseteq Y$ for
some open set $U$; $U$ is itself an open neighborhood, not just of $x$, but
of every contained point. Coming back to the sequence $(y_n)$, we see that
$y_n\to x$ iff $y_n$ is eventually in each neighborhood of $x$.
This last condition is the definition of convergence of a sequence in
a general topological space.

The topology (i.e., collection of open sets) of a metric space is the
smallest collection satisfying the axioms and containing the open balls.
We say that it is \textit{generated} by the open balls.
In fact, in this particular case, the open sets are just arbitrary unions of
open balls. Generation is a little more complicated in general, as we shall
see for systems of semimetrics.

\paragraph{Closed sets and closure.}\label{sec:closed sets}

\textit{Closed} sets are by definition the complements of the open sets.
We can therefore phrase the axioms of a topology in terms of closed sets
simply by swapping union and intersection.
The \textit{closure} of a set $Y$, denoted $\cl Y$, is the
smallest closed set containing $Y$. That exists because it is the
intersection of all closed sets containing $Y$. 
$x$ is in $\cl Y$ iff every neighborhood of $x$ intersects $Y$.
Heuristically, that means $Y$ contains points which approximate
$x$ arbitrarily well. In a metric space, this means there is a
sequence in $Y$ converging to $x$.
In general, $x$ may have too many neighborhoods for a simple sequence to
do that. 

\paragraph{Many metrics determine the same topology.}

A couple of examples may help the reader appreciate the way in which
the quantitative aspect of a metric is abstracted away in the topology.
It is obvious that simply rescaling $d$ by a constant will not change
the topology. The tricky part in any more drastic change is not to run
afoul of the triangle inequality. One may check that
$(x,y) \mapsto \min(1,d(x,y))$ and $(x,y) \mapsto d(x,y)/(1+d(x,y))$
are both bounded metrics determining the same topology as $d$.
Evidently, boundedness is not a topological property.

\subsection{The topology defined by a system of semimetrics}
\label{sec:semimetric spaces}

Just as a norm provides a metric, a seminorm provides a semimetric.
$d$ is a semimetric if it is symmetric and satisfies the triangle inequality;
it is not required to be separating (but might be!).
Now, if we have a family $\{d_\alpha\}_\alpha$ of semimetrics, 
a topology is generated by the collection of balls
\mbox{ $B_{\alpha,\epsilon}(x) = \setof{y}{d_\alpha(x,y)<\epsilon}$}
  as $x$ and $\alpha$ vary.
  We also call this the topology generated by the semimetrics (or seminorms,
  if the semimetrics come from such).
  If we have a single semimetric $d$, then $U$ is open iff for each $x\in U$,
  some $B_\epsilon(x)$ is contained in $U$ as well.
  With multiple semimetrics, the condition is instead that some
  finite intersection $\cap_{i=1}^n B_{\alpha_i,\epsilon_i}(x)$ is contained
  in $U$. Correspondingly, $y_n$ converges to $x$ with respect to the
  topology generated by a family of semimetrics iff $y_n$ is eventually
  in each such finite intersection.

\subsection{Continuity}\label{sec:continuity}

By definition, a function $f\colon(\sX,\tau)\rightarrow(\sX',\tau')$
from one topological space to another
is \textit{continuous} if $U\in\tau'$ implies that $f^{-1}(U)\in\tau$.
Thus, inverse images (by $f$) of open sets are open. Equivalently, inverse images
of closed sets are closed.

For a loose paraphrase,
let $U$ be an open neighborhood of $f(x)$,
so that $f^{-1}(U)$ is an open neighborhood of $x$.
Thus, we can make $f(y)$ as ``close'' as desired to $f(x)$
by taking $y$ sufficiently close to $x$.
This more closely resembles the definition from elementary calculus.

We often deal with multiple topologies on a set such as $\Blin(\TC(\cH),\TC(\cK))$.
Saying that a map is $\tau/\tau'$ continuous means it is continuous with respect
to topology $\tau$ on the domain and $\tau'$ on the codomain.
If we say that a map is, for instance ``SOT-continuous'', this means with SOT
on both spaces.

\paragraph{Comparing topologies: finer and coarser}

A topology on $\sX$ is a set of subsets of $\sX$, so two of them
might be comparable in terms of containment.
Traditionally, instead of saying ``$\tau\subseteq \tau'$'', we say
``$\tau$ is \textit{coarser} than $\tau'$'' or ``$\tau'$ is \textit{finer}
than $\tau$''.
In this situation, $x_n\to x$ in $\tau$,
also written ``$x_n\xrightarrow{\tau} x$'',
implies that $x_n\xrightarrow{\tau'} x$.
The finer the topology, the ``more difficult'' it is for a sequence 
to converge, but the weaker the topology, the more difficult for a set to be
closed. It is easy to get confused about these. (At least, I think so.)

\paragraph{Topological isomorphism}

$f\colon(\sX,\sigma)\rightarrow(\sY,\tau)$ is a \textit{topological isomorphism}
if it is a continuous bijection with a continuous inverse.
Alternatively, $f$ is a bijection between the sets $\sX$ and $\sY$, and also
between the topologies $\sigma$ and $\tau$.
Actually, people almost universally say \textit{homeomorphic} rather than
\textit{topologically isomorphic}.
Consider the open interval $]-\tfrac{\pi}{2},\tfrac{\pi}{2}[$ and real line
$\Real$ with their usual metric topologies. The mapping
$x \mapsto \tan^{-1} x$ is a topological isomorphism between them.
As metric spaces, they are different (only one is bounded), but as
topological spaces they are the same.

\subsection{Compactness: topological almost-finiteness}

Compactness is a very important topological property, in general and
for our investigation.

\paragraph{Definition.}
\leavevmode
A subset $\sU$ of $\tau$ is an \textit{open covering} of
$\sX$ if $\cup \{U\in\sU\} = \sX$. A \textit{subcovering} of $\sU$ is
a subset of $\sU$ which still covers $\sX$.
$(\sX,\tau)$ is \textit{compact} if every open covering has a finite subcovering.

\paragraph{A way to think about compactness.}
\leavevmode

For an interpretation, consider each member of a covering as
consisting of points we regard as practically indistinguishable.
(Since there are probably overlaps, this means our notion of
\textit{practically indistinguishable} is not transitive, which 
is entirely reasonable.)
If $\sX$ is compact, no matter our notion of practical indistinguishability
(as long as it's represented by open sets), every point falls in one of
a finite number of indistinguishability classes. Picking one point from
each such class, we have a sort of finite approximation to $\sX$.

\subsubsection{Equivalently, a family of closed sets with FIP has nonempty intersection.}

We can also phrase the definition in terms of closed sets.
A set of sets has the \textit{finite intersection property} (FIP) if
every finite subset has nonempty intersection. The space $(\sX,\tau)$
is compact if every family of closed sets with FIP has nonempty
intersection, i.e., there is at least one point which is in every one of
those closed sets.

\subsubsection{A pertinent example}

The closed unit ball of a finite-dimensional space is compact.
Not so for an infinite-dimensional space.
For instance, let $\cH$ be a Hilbert space with ONB $\varphi_1,\varphi_2,\ldots$.
We give two proofs that $\cball \cH$ is not compact.
First, note that $\|\varphi_i - \varphi_j\| = \sqrt{2}$ for every $i\neq j$.
Thus, the cover consisting of open balls of radius $2^{-1/2}$ about each
point of $\cball \cH$ cannot have a finite subcover because no two members
of the ONB can be in the same ball.
Alternatively, let
\mbox{$\cH_n \defeq \setof{h}{\inpr{\varphi_i}{h} = 0,\, i=1,\ldots, n}$}
This is a closed subspace of $\cH$,
hence \mbox{$G_n \defeq \cH_n\cap\setof{h}{\tfrac{1}{2}\le \|h\| \le 1}$} is a
closed subset of $\cball \cH$.
The $G_n$ sequence is nested, hence has FIP, yet the intersection is empty.

\paragraph{Every infinite set in a compact space has accumulation points.}
\leavevmode

$x$ is an \textit{accumulation point} of $S$ if every neighborhood of $x$
intersects $S$ in some point other than $x$ itself.

Suppose, for a contradiction, that
$S$ is an infinite subset of $\sX$ without accumulation point.
For each $x\in\sX$, choose an open neighborhood $U_x$ such that
$U_x\cap S$ is finite. Clearly, $\setof{U_x}{x\in\sX}$ is an open covering,
so by the compactness hypothesis, there is a finite subcovering
$\{U_{x_1},\ldots, U_{x_n}\}$.
Now, $S = \sX\cap S = \cup_{i=1}^n U_n\cap S$, but the final expression is
a finite union of finite subsets, hence finite. This is a contradiction.

\subsubsection{A metric space is compact iff complete and totally bounded}

Recall that a sequence $(x_n)$ in a metric space is \textit{Cauchy}
if the diameters of the tails $\{x_n,x_{n+1},\ldots\}$ tend to zero
as $n\to\infty$. And, a metric space is \textit{complete} if
every Cauchy sequence has a limit (``no holes'').
This should be familiar from calculus.
Alternatively:
a metric space is complete iff every nested sequence
$G_1 \supseteq G_2 \supseteq \cdots$ of closed sets with diameters
$\diam G_n$ decreasing to zero, has nonempty intersection, indeed
consisting of a single point.
For, if this condition holds we can apply it to the closures of the
tails $\{x_n,x_{n+1},\ldots\}$. Conversely, any sequence $x_n\in G_n$
is Cauchy hence has a limit which is in each $G_m$ (because they are closed).
It follows that a subset of a complete metric space is also such
if and only if it is closed.

$M$ is \textit{totally bounded} if it is covered by some finite set of balls
of radius $r$, for any given $r > 0$.
Any subspace of a totally bounded metric space is totally bounded.

Combining the two hereditary properties just mentioned, we see that
in a complete totally bounded metric space, a subset is closed
iff it is complete and totally bounded.
Further, it can be covered by a finite number of such subsets having
diameter not exceeding any given $\epsilon > 0$.

\begin{proof}[Proof of main claim]
  The ``$\Rightarrow$'' direction is easy. We prove ``$\Leftarrow$''.
  For a contradiction, assume $\sX$ has the property
  $\sP\defeq $ `no finite subset of $\sU$ covers'.
By total boundedness, and the observation just preceding the lemma,
there is a finite cover by complete totally bounded subsets with diameter
not exceeding any given $r > 0$. Further since the cover is finite,
one of them must also have property $\sP$.
Iterating,
there is a sequence $\sX \supseteq G_1 \supseteq G_2 \supset \cdots$
of closed subsets of $\sX$ with $\diam G_n \le 2^{-n}$ and property $\sP$.
By completeness,
$\cap G_n = \{x\} \subset B_\epsilon(x) \subseteq U \in \sU$ for some
$x$ and $\epsilon$, since the members of $\sU$ are open.
But then, for $2^{-n} < \epsilon$, $G_n$ is covered by $U$ alone,
which is a contradition. 
\end{proof}

\subsection{Topological products}

\paragraph{Definition.}

The reader is surely familiar with the idea of a Cartesian product set
$\prod_{\alpha\in\sI} \sX_\alpha$ of a family of sets indexed by $\sI$.
Each member of the product is an ``$\sI$-tuple'' $(x_\alpha)_{\alpha\in\sI}$,
or a choice of one element from each factor.
Now, if each set $\sX_\alpha$ in the product is equipped with a topology 
$\tau_\alpha$, there is a standard natural way to produce a topology
for the product, called (what else?) the \textit{product topology}.
It can be defined as follows. A neighborhood of $(x_\alpha)$ is any set
containing a set of form $\prod_\alpha U_\alpha$, where
(i) $x_\alpha \in U_\alpha$ for every $\alpha$ (of course),
(ii) $U_\alpha\in\tau_\alpha$, and
(iii) $U_\alpha = \sX_\alpha$ for all but finitely many indices.
This is actually a very coarse topology.
About the only other one that would make sense as some sort of
general construction is what we would get by dropping (iii).
That usually gives an excessively fine topology, and doesn't even have
a standard name (to my knowledge).
Thus, whenever you see a product of topological spaces, you should just
assume it has the product topology unless the contrary is explicitly stated.

\subsubsection{Tychonoff theorem: A topological product of compact spaces is compact}

This remarkable theorem is a very famous result.
We will use it a few times, but do not give a proof. For that,
see the earlier general references for topology.
One thing to note here is that there is absolutely no restriction
on the size of the index set $\sI$.

\subsubsection{SOT as a product topology}

Now we will see how to think of SOT as a product topology. This is a useful
trick which can be applied to other topologies generated by systems of seminorms.

Consider the enormous product space $\prod_{x\colon\sX} \sY$, each factor
having the norm topology of $\sY$.
We can embed $\Blin(\sX,\sY)$ into this space via 
$A \mapsto \prod_x Ax$, one coordinate for each point on the graph of $A$.
Clearly this is injective, though very far from surjective.
Being sloppy about the distinction between $\Blin(\sX,\sY)$ and its image
in the product, we see that an open neighborhood in either SOT or the product
has the same description, namely
$\setof{B}{Bx_i - Ax_i \in U_i;\, i=1,\ldots,n}$, where the $U_k$ are open
neighborhoods of zero in $\sY$.

\subsection{SOT in more depth}\label{sec:SOT}

\subsubsection{SOT and norm topology compared}\label{sec:SOT and norm}

\paragraph{Lower semicontinuity.}\label{sec:lsc}
A function $f$ on a topological space with values in
\mbox{$\Real\cup\{+\infty\}$}
is \textit{lower semicontinuous} (lsc) if for any $c < f(x)$,
$\setof{y}{f(y) > c}$ is a neighborhood of $x$.

\paragraph{The norm is SOT-lower semicontinuous.}
\label{sec:norm is SOT-lsc}\leavevmode

For the proof, given $c < d < \|A\|$, choose $x$ such that $\|Ax\| > d$.
Then, $\|\cdot\| > c$ on the SOT-neighborhood $\setof{B}{|B-A|_x < d-c}$.

\paragraph{$\cball \Blin(\sX)$ is SOT-closed; its complement is SOT-open.}
\leavevmode

$\oball\Blin(\sX) = \setof{A\in\Blin(\sX)}{\|A\| < 1}$ is the
open unit ball of $\Blin(\sX)$, and
$\cball\Blin(\sX)$ the closed unit ball $\setof{A}{\|A\| \le 1}$.
The claim follows from the previous paragraph because it implies that
the complement of $\cball \Blin(\sX)$ is a SOT-neighborhood of any
$A\not\in \cball \Blin(\sX)$.

\paragraph{norm topology is strictly finer than SOT.}
\leavevmode

Any SOT-semiball contains a norm ball, so the norm topology is finer.
That it is strict follows because any SOT-semiball contains operators
of arbitrarily large norm ($\|Ax\| < \epsilon$ does not constrain
how large $Ay$ can be if $y$ is not proportional to $x$). 

\subsubsection{SOT on (norm) bounded sets}\label{sec:SOT on bdd sets}

If the reader finds this section a strain, it is advisable to
skip over it lightly until it is called on.
In fact, the comments on power series will be used only once,
in \S\ref{sec:cp+ is GMET-closed}.

\paragraph{If $\cl\Span U = \sX$, then $U$-SOT is the same as SOT on  bounded sets.}
\label{sec:smaller set of SOT seminorms}
\leavevmode

We need only show that $U$-SOT is as fine as SOT.
The sets $\setof{|\cdot|_x}{x\in U}$ and
$\setof{|\cdot|_x}{x\in \Span U}$ of seminorms 
generate the same topology.
So, let $\sA$ be a bounded set.
Given $x$ and $\epsilon > 0$, there is $y\in\Span U$ with
$\|y-x\| < \tfrac{\epsilon}{2\cdot\mathrm{ diam } \sA}$,
so that $A,B\in\sA$,
$|A-B|_y < \tfrac{\epsilon}{2} \Rightarrow |A-B|_x < {\epsilon}$.

\paragraph{$A\mapsto A\blank\colon \cball\Blin(\cH)\rightarrow\Blin(\TC(\cH))$
  is SOT/SOT-continuous.}
\leavevmode

The span of $\setof{\bpi(h)}{h\colon\cH}$ is dense in $\TC(\cH)$.
(In contrast, it is \textit{not} dense in $\Blin(\cH)$.)
Therefore, by translation covariance and
\S~\ref{sec:smaller set of SOT seminorms},
it suffices to show that,
given $h\colon\cH$ and $\epsilon > 0$,
there is a SOT neighborhood of zero in $\Blin(\cH)$ on which
$\|A\bpi(h)\|_1 < \epsilon$. This is easy:
\mbox{$\setof{A}{\|Ah\|_1 < \tfrac{\epsilon}{\|h\|_1}}$}
works.

\paragraph{Operator composition is SOT continuous on bounded sets.}
\label{sec:multiplication is SOT-cts}
\leavevmode

$AB-A'B' = (A-A')B + A'(B-B')$.
Therefore, for $A,A'$ constrained as $\|A\|,\|A'\| \le M$,
$|AB-A'B'|_x \le |A-A'|_{Bx} + M |B-B'|_x$.

\paragraph{A power series has the same radius of convergence on
  $\Blin(\sX)$ as on $\Cmplx$.}
\label{sec:radius of convergence}
\leavevmode

Let $f(z) = \sum_{n=0}^\infty c_nz^n$ have radius of convergence $R$
for $z\in\Cmplx$. It has the same radius of convergence on $\Blin(\sX)$
because the relevant bound on the tail just replaces complex modulus
with operator norm. By the same token, the sum converges uniformly
(i.e., in norm) over $r\cdot\cball \Blin(\sX)$ for any $r < R$.

\paragraph{A power series is SOT-continuous for norm strictly less than
  radius of convergence.}
\label{sec:SOT-cts power series}
\leavevmode

By induction on \S~\ref{sec:multiplication is SOT-cts},
\mbox{$A \mapsto \sum_{n=1}^N c_nA^n$} is SOT-continuous 
for any $N < \infty$, and the tail is uniformly small by
\S~\ref{sec:radius of convergence}, so
$|f(A)-f(A')|_x < \epsilon$ for $A'$ in some SOT neighborhood of $A$.

%
%


\section{Ground matrix element topologies  (GMET)}\label{sec:GMET}

We have acquired a good grip on many aspects of $\CP$ maps in finite
dimensions, and it makes sense to leverage that in the separable setting.
After \S~\ref{sec:generators CP evolution}, we will be in 
an analogous situation with respect to continuous-time $\CP$ evolution.
Our method to do that is to lift finite-dimensional
orthoprojections on the base Hilbert space(s) so as to obtain
finite-dimensional approximations. To borrow a possibly more evocative
(and maybe misleading) term from quantum field theory,
we might even call this a \textit{regularization}.
This is the topic of \S~\ref{sec:regularization}.

\S~\ref{sec:BML} pivots to what may at first sight seem unrelated
and develops the \textit{ground matrix element topology}
on $\Blin(\cH,\cK)$, $\TC(\cH)$ and $\Blin(\TC(\cH),\TC(\cK))$.
In the case of $\Blin(\cH,\cK)$, this corresponds to the widely-used
\textit{weak-operator topology} (WOT). Similarly to that case, the
closed unit balls of the other spaces are also compact in GMET
(\S~\ref{sec:GMET-cpt}).

\S~\ref{sec:GMET+projections} shows how GMET fits very naturally
with our regularization scheme. With that, we will be well-equipped
to tackle $\CP$ maps in the separable setting.

\subsection{Regularization via lifted finite-dimensional projections}
\label{sec:regularization}

\subsubsection{Lifting orthoprojections from ground Hilbert spaces}
\label{sec:projections}

\paragraph{The simple idea.}
Let $P$ denote the orthoprojection of $\cH$ onto a finite-dimensional
subspace $P\cH$. There is a natural way to \textit{lift} this projection
to act on $\Blin(\cH)$, namely $A \mapsto PAP$.
With a matrix representation and appropriate choice of basis,
$P$ could correspond to zero-ing all but the first $n$ entries of
a column vector, and the lifting to $\Blin(\cH)$ then corresponds to
zero-ing all but the upper-left $n\times n$ block.
Since $\TC(\cH)$ is a subset of $\Blin(\cH)$, the same lift applies to
it. The idea easily extends to $\Blin(\TC(\cH))$ (and beyond),
but notation becomes a problem.

\paragraph{Lifted polyprojection notation.}\label{sec:lifted polyprojection}
To deal with that problem, we adopt a radically simplified notation wherein
the lift of $P$
to any level is indicated by adding a hat as $\wh{P}$; the specific
operator thus represented is determined by what it is operating on.
For instance, for $A\in\TC(\cH)$, $\wh{P}A = PAP$, and for
$\Lambda\in\Blin(\TC(\cH))$, $\wh{P}\Lambda = \wh{P}\circ\Lambda\circ\wh{P}$,
i.e., $(\wh{P}\Lambda)A = \wh{P}(\Lambda(\wh{P}A))$. There is potential
ambiguity here about whether a particular use of $\wh{P}$ represents
application or composition; parentheses or explicit composition symbols
may occasionally be needed.

Another problem arises when the context involves more that one ground
Hilbert space, e.g., for $\Blin(\TC(\cH),\TC(\cK))$.
In that case, $P$ will be a \textit{polyprojection}, incorporating
a finite-dimensional orthoprojection $P_\cH$ on $\cH$ and another on $\cK$.
If $h\in\cH$, only one of them can possibly apply.
Then, for $\Lambda\in\Blin(\TC(\cH),\TC(\cK))$,
$\wh{P}\Lambda = \wh{P}\circ\Lambda\circ\wh{P}$, the first `$\wh{P}$' on
the RHS is a lift of a projection on $\cK$ and the second a lift of
a projection on $\cH$.

In the following sections, unless explicitly noted otherwise, $P$ and
variants such as $P'$, etc., denote such finite-dimensional polyprojections.

\subsubsection{Convergence of projected objects as $P$ tends to identity}

\paragraph{Ground Hilbert spaces.}

The limit we will always be interested in is one in which each
component tend to the corresponding identity in SOT, i.e.,
$P_\cH\xrightarrow{\mathrm{SOT}}\Id_\cH$ etc. This will be denoted
simply $P\xrightarrow{\mathrm{SOT}}\Id$, or even $P\to\Id$.
Since our Hilbert spaces are separable, this can
be along a sequence, and will be assumed to be so.
For instance, the $n^{\mathrm{th}}$ member of the sequence
might project onto the span of the first $n$ elements $h_1,h_2,\ldots,h_n$
and $k_1,\ldots,k_n$ of some complete orthonormal systems.

\paragraph{In $\Blin(\cH,\cK)$, convergence is in SOT.}\label{sec:P-hat on B(H,K)}
\leavevmode

For $A\colon\Blin(\cH,\cK)$,
\mbox{$\wh{P} A = P \circ A \circ P \xrightarrow{\mathrm{SOT}} A$}
as $P\xrightarrow{\mathrm{SOT}}\Id$
because composition is SOT-continuous on bounded sets
(\S\ref{sec:multiplication is SOT-cts}).

Readers may find it amusing to check that this is
consistent with the previous paragraph when $h\colon\cH$ is
regarded as a member of $\Blin(\Cmplx,\cH)$.

\paragraph{In $\TC(\cH)$, convergence is even in norm.}\leavevmode

The appropriate instantiation of the lifted projection is
here $\wh{P}\colon\Blin(\TC(\cH))$,
and we want to show that this converges in SOT to $\Id_{\TC(\cH)}$.
Now, $\|\wh{P}\|_{1,1} = 1$. (The norm is obviously at least 1, but
$\|\wh{P}\rho\|_1 = \|P\circ\rho \circ P\|_1 \le \|P\|\|\rho\|_1\|P\|$.)
Therefore, we have a bounded sequence,
so \S~\ref{sec:smaller set of SOT seminorms}
applies, and it suffices to show that
$\wh{P}\bpi(h) \to \bpi(h)$. 
But, $\wh{P}\bpi(h) = \bpi(P h)$, and
the trace norm of $\bpi(h) - \bpi(P h)
= h (\pam{h - P h}) + (h-P h)\pam{P h}$
does not exceed $2\|h\|\|h - P h\|\to 0$.

\paragraph{In $\Blin(\TC(\cH),\TC(\cK))$, convergence is in SOT.}
\label{sec:SOT B(T(H),T(K))}
\leavevmode

Given the preceding paragraph, the argument is exactly the same as
in \S~\ref{sec:P-hat on B(H,K)}.
That is, $\wh{P}\Lambda = \wh{P}\circ\Lambda\circ\wh{P}$, where
the operators here are in
$\Blin(\TC(\cK))$,
$\Blin(\TC(\cH),\TC(\cK))$, 
and $\Blin(\TC(\cH))$ respectively.
According to the preceding paragraph, this is a sequence of SOT
convergent operators (the middle one trivially so), therefore is
SOT convergent to $\Id_{\TC(\cK)}\circ\Lambda\circ\Id_{\TC(\cH)}$.
\subsection{Spaces of bounded multilinear functions ($\BML$)}\label{sec:BML}

\subsubsection{The vector space of all functions
  $\cK^*\times\cH\rightarrow \Cmplx$ is topologically
  $\Cmplx^{\cK^*\times\cH}$ } \label{sec:C^{KxH}}

An operator $T$ in $\Blin(\cH,\cK)$ is fully specified by 
its \textit{matrix elements} $\inpr{k}{Th}$ as $k$ and $h$ range over
$\cK$ and $\cH$, respectively.
From this perspective, $T$ belongs to the vector space $\Cmplx^{\cK^*\times\cH}$ of
\textit{all} functions from $\cK^*\times\cH$ to $\Cmplx$. 
This is a product set with one factor of $\Cmplx$ for each argument value
$(\pam{k},h)$. We equip it with the topology generated by the system of seminorms
$\setof{f \mapsto |f(\pam{k},h)|}{h\in\cH,\, \pam{k}\in\cK^*}$.
This is the \textit{product topology}, but in this context it makes more
sense to call it the \textit{topology of pointwise convergence}, or simply
\textit{pointwise topology}.
We will need this ridiculous level of generality (``all functions'')
just until we have reaped the benefits of its topological simplicity.

\subsubsection{The topological subspace $\|\cdot\|_{\mathrm{ML}}\le 1$  is compact}

On $\Cmplx^{\cK^*\times\cH}$ we define also a quasi-norm
($[0,+\infty]$-valued, i.e., possibly infinite) by
\begin{equation}\label{eq:ML norm}
\|f\|_{\mathrm{ML}}\defeq
\inf \setof{M}{\forall \pam{k},h, \; |f(\pam{k},h)| \le M \|k\|\|h\|}.
\end{equation}
The subspace $\|\cdot\|_{\mathrm{ML}} \le 1$ of $\Cmplx^{\cK^*\times\cH}$ 
is topologically isomorphic to the product space
\mbox{$\prod_{\pam{k},h} \|k\|\|h\|{\mathbb D}$}, where
$\mathbb D$ denotes the closed unit disk in $\Cmplx$.
By Tychonoff's theorem, this is a compact space.

\subsubsection{The subspace of bilinear functions is closed}

The condition
$f(\pam{k},{ah+a'h'}) - af(\pam{k},{h}) - a'f(\pam{k},{h'}) = 0$ or
$f(\pam{a}\pam{k}+\pam{a'}\pam{k'},{h})
- {a}f(\pam{k},{h}) - {a'}f(\pam{k'},{h'}) = 0$ for
fixed $a,a',k,k',h,h'$ defines a closed subset of $\Cmplx^{{\cK}^*\times\cH}$.
Hence, the subspace where all such conditions are satisfied is also a
closed subspace.
This subspace consists exactly of the bilinear functions, that is,
functions $f(\pam{k},h)$ which are linear in each variable with the other
held fixed.

\subsubsection{ $\cball ( \BML(\overline{\cK},\cH;\Cmplx))$ is compact}

The subspace of bilinear functions,
bounded with respect to $\|\cdot\|_{\mathrm{ML}}$,
is denoted $\BML(\cK^*,\cH\,;\,\Cmplx)$.
Although supplied with a norm, the intended topology of this space is
the pointwise one.
The closed unit ball
$\setof{f\in\BML(\overline{\cK},\cH;\Cmplx)}{\|f\|_{\mathrm{ML}}\le 1}$ is
the intersection of the compact set \mbox{$\|\cdot\|_{\mathrm{ML}}\le 1$} 
with the closed $\BML(\cK^*,\cH\,;\,\Cmplx)$, hence is compact.

The reader may be relieved to know that,
henceforth, we can confine our attention to $\BML(\overline{\cK},\cH;\Cmplx)$ 
or variants with more arguments, and need not be concerned with
$\Cmplx^{\cK^*\times\cH}$.

\subsubsection{Rephrase in terms of seminorms}

\subsubsection{Generalization to multilinear functions is simple}

In the preceding, we worked with $\cK^*\times\cH$; the multilinear functions
were actually bilinear. The reader may easily check that having more
factors presents no essential difficulty.
To deal with $\Blin(\TC(\cH),\TC(\cK))$, we will need four.

\subsection{Closed balls of $\Blin(\cH,\cK)$, 
  $\TC(\cH)$, and $\Blin(\TC(\cH),\TC(\cK))$ are GMET-compact}\label{sec:GMET-cpt}

\subsubsection{Case $\Blin(\cH,\cK)$}\label{sec:B(H,K)}

\paragraph{$\Blin(\cH,\cK) \cong \BML({\cK}^*,\cH;\Cmplx)$}

We have already (\S~\ref{sec:C^{KxH}})
discussed the injective linear map taking $T$ to the
bilinear function $(\pam{k},h) \mapsto \inpr{k}{Th}$.
Conversely, the Riesz
representation theorem shows that a bounded bilinear function
defines a bounded operator.
Therefore, the correspondence is bijective.
It is clear from (\ref{eq:ML norm}) 
that $\|\cdot\|_{\mathrm{ML}}$ coincides with the ordinary
operator norm $\|\cdot\|$.

\paragraph{GMET = WOT for $\Blin(\cH,\cK)$.}

The topology which $\Blin(\cH,\cK)$ acquires by this identification
is traditionally called weak-operator topology (WOT).
Because we use the same procedure to get topologies on our other two
main classes of spaces, we use the generic term
\textit{ground matrix element topology} (GMET).
The reason for ``ground'' will become clear in \S~\ref{sec:GMET revealed}.

\subsubsection{Case $\TC(\cH)$}

\mbox{$A\mapsto\; [(\pam{k},h)\mapsto \inpr{k}{Ah}]$}
maps $\cball \Blin(\cH,\cK)$ bijectively onto
$\cball \BML(\cK^*,\cH\,;\,\Cmplx)$, so it
is an injection on $\cball(\TC(\cH)) \subset \cball(\Blin(\cH))$.
It follows that
$\cball \Blin(\cH,\cK)$ is GMET-compact if its image in
$\BML(\cK^*,\cH\,;\,\Cmplx)$ is closed.
We proceed to prove that.

\paragraph{The supremum of any collection of lower semicontinuous functions
  is itself lower semicontinuous.}
The definition of lower semicontinuity in \S~\ref{sec:lsc} can be
equivalently phrased: $g$ is lsc if all sublevel sets $\{g\le c\}$ are closed.
An intersection of closed sets being closed, lower semicontinuity of
a supremum as stated follows immediately.

\paragraph{Trace norm is GMET lower semicontinuous.}
  The trace norm is given by
  \begin{equation}\label{eq:trace norm as sup}
    \|T\|_1 =
    \sup
    \setof{\sum_{i=1}^{n}|\langle{h'_i}|{Th_i}\rangle|}{h_1,\ldots,h_n \text{ and }
      h'_1,\ldots, h'_n \text{ orthonormal }}.
  \end{equation}
  But, each
  $A \mapsto \sum_{i=1}^{n}|\langle{h'_i}|{Ah_i}\rangle|$
  is trivially GMET continuous. 

  \paragraph{The image of $\cball \TC(\cH)$ is closed.}
  GMET lower semicontinuity of $\|\cdot\|_1$ implies that
  $\cball \TC(\cH)$ is GMET-closed.
  This is just like in \S~\ref{sec:norm is SOT-lsc}.
  
\subsubsection{Case $\Blin(\TC(\cH),\TC(\cK))$}

\paragraph{GMET fully revealed.}\label{sec:GMET revealed}
For any $A\colon\TC(\cH)$ and $B\colon\Blin(\cK)\cong \TC(\cK)^*$,
$\inpr{B}{\Lambda A}$ is a matrix element of $\Lambda$; the topology
these generate is the ordinary weak-operator topology for this case.
We want to use the much coarser topology generated by the seminorms
associated with just matrix elements of the form
\begin{equation}\label{eq:ground matrix element}
\inpr{k'\pam{k}}{\Lambda\cdot h'\pam{h}}, \quad k,k'\colon\cK,\; h,h'\colon\cH.
\end{equation}
These are \textit{ground} matrix elements, ``{ground}'' indicating that
we go all the way down to the underlying Hilbert spaces.

\paragraph{A $\BML$ space over four factors.}
Via (\ref{eq:ground matrix element}), 
$\Lambda$ in $\Blin(\TC(\cH),\TC(\cK))$ defines
an element of $\BML({\cK}^*,\cK,\cH,{\cH}^*;\Cmplx)$,
a space of complex multilinear functions of four variables.
Essentially everything that we did with $\BML({\cK}^*,\cH;\Cmplx)$
carries over --- in particular compactness of the ($\|\cdot\|_{\mathrm{ML}}$)
unit ball.
The number of factors in the domain is an irrelevant detail.

\paragraph{$\cball \Blin(\TC(\cH),\TC(\cK))$ is GMET-compact.}

The norm of $\Blin(\TC(\cH),\TC(\cK))$ is 
  \begin{equation}\label{eq:1,1 norm}
    \|\Lambda\|_{1,1} = \sup
    \setof{ |\langle{B}|{\Lambda A}\rangle| }{
      A, B \text{ finite-rank },
      \|A\|_1 \le 1, \, \|B\|\le 1 }.
  \end{equation}
  But, $\langle{B}|{\Lambda A}\rangle$ is a sum (i.e., finite) of
  ground matrix elements, hence GMET continuous.
  This shows immediately that
  \begin{itemize}
  \item 
 $\|\cdot\|_{1,1}$ is GMET lower semicontinuous,
\item $\|\cdot\|_{\mathrm{ML}} \le \|\cdot\|_{1,1}$.
  \end{itemize}
  Just as for $\TC(\cH)$, these imply that $\cball\Blin(\TC(\cH),\TC(\cK))$
  is a GMET-closed subset of the unit ball of
$\BML({\cK}^*,\cK,\cH,{\cH}^*;\Cmplx)$.
%
%

\subsection[\phantom{Closed}\dots They are also metrizable]
{\dots They are also metrizable}%
\label{sec:metrizable}

\subsubsection{Just basic matrix elements define the same topology on bounded sets.}

Not all matrix elements $\inpr{k}{Th}$ are needed to identify an operator $T$,
equivalently as we saw in \S~\ref{sec:B(H,K)}, a bilinear function
$\cK^*\times\cH \rightarrow \Cmplx$.
The \textit{basic} matrix elements $\inpr{k_i}{Th_j}$ for orthonormal bases
$k_1,k_2,\ldots$ and $h_1,h_2,\ldots$ suffice.
However, the \textit{topology} defined by the corresponding seminorms is
strictly weaker than GMET. 
Just like what happens for SOT (\S~\ref{sec:SOT on bdd sets}), though
and for the same reason, this otherwise-weaker topology is the same
restricted to a norm bounded set.
[For any $k$, $h$, and $\epsilon > 0$,
there are such finite linear combinations $k'$ and $h'$
such that $\|k-k'\| < \epsilon$, $\|h-h'\| < \epsilon$.]

\paragraph{This works the same way for $\BML$ spaces over more factors.}

\subsubsection{Unit ball of a $\BML$ space over separable Hilbert spaces is metrizable.}

We are always assuming our underlying Hilbert spaces
$\cH,\ldots,\cK$ are separable.
In that case, there are only a countable number of combinations
$(h_i,\ldots,k_j)$ of the complete orthonormal systems.
Consequently the pointwise topology on $\cball\BML(\cH,\ldots,\cK;\Cmplx)$
is actually generated by a countable number of seminorms.
This means that there is a metric which generates the same topology.

\paragraph{The topology defined by a countable collection of seminorms
  is metrizable.}
\leavevmode

If $p_1,p_2,\ldots$ are the seminorms, 
a suitable metric is defined by
\begin{equation}\nonumber
d({x},{y}) \defeq \sum_\alpha 2^{-\alpha}
\max(p_\alpha(x-y),1).
\end{equation}

\paragraph{GMET on closed balls of $\Blin(\cH,\cK)$,
  $\TC(\cH)$ and $\Blin(\TC(\cH),\TC(\cK))$ is metrizable.}
\leavevmode

This is a straightforward consequence of the preceding and the identification
of the spaces in question with subspaces of $\BML$ spaces.

\subsection{GMET and lifted projections are a good match}\label{sec:GMET+projections}

\subsubsection{GMET is the coarsest topology making all $\wh{P}$ continuous}
\label{sec:GMET final for P}

By definition, GMET is the coarsest topology making ground matrix elements
continuous. For instance, a basic open neighborhood of zero in
$\Blin(\TC(\cH),\TC(\cK))$ might be defined by $n$ condidtions 
$\inpr{k_i'\pam{k_i}}{\Lambda\cdot h_j'\pam{h_j}} < \epsilon_{ij}$.
If $P$ projects onto $\Span\setof{k_i,k_i'}{i=1,\ldots,n}$ in $\cK$
and $\Span\setof{h_i,h_i'}{i=1,\ldots,n}$ in $\cH$, then this
open neighborhood admits the alternative description
``$\wh{P}\Lambda$ lies in a [certain] open set of
$\Blin(\TC({P}\cH),\TC(P\cK))$''.

\subsubsection{A simple criterion that can sometimes show GMET-continuity}
\label{sec:GMET-cntty trick}

If, for every $P$, there exists $P'$ and continuous $S$
such that
\begin{equation}\nonumber
  \begin{tikzcd}
    \sX \arrow[r,"T"] \arrow[d,"\wh{P'}"'] & \sY \arrow[d,"\wh{P}"] \\
    \sX \arrow[r,"S"'] & \sY
  \end{tikzcd}
\end{equation}
commutes, then $T$ is GMET-continuous.
\begin{proof}
  The hypothesis says that if $x$ is in the GMET neighborhood
  $(\wh{P'})^{-1}(S^{-1}U)$, then $Tx$ is in the GMET neighborhood $\wh{P}^{-1} U$,
  for any neighborhood $U$ in $\sY$. \S~\ref{sec:GMET final for P} finishes
  the proof.
\end{proof}

\section{Order and topology}
\label{sec:order and top}

To this point, we have been dealing with the $\Pos$, $\Mon$ and $\CP$
convex cones mostly as geometrical objects. Now we consider another
aspect, namely the partial orders they induce on ambient spaces.
To be frank, the major use of this material in these notes is to
SOT convergence of Kraus sums. However, it seems to me that 
\S~\ref{sec:intervals B1(H)} in particular can have significant
application elsewhere.

\S~\ref{sec:order intervals} is trivial in the finite-dimensional setting.

\subsection{Partial orders induced by $\Pos(\cH)$ and $\Mon(\cH,\cK)$}

A \textit{partial order} (or just \textit{order}) on a set is a binary
relation $\le$ with properties
(i) $x\le y$  and $y \le x \Leftrightarrow x=y$ (asymmetry)
(ii) $x\le y \text{ and } y \le z \Rightarrow x\le z$ (transitivity).
It is partial because two given elements may not be comparable at all.
In a vector space,
a pointed convex cone $C$ containing zero induces a partial order
according to $x \le y \;\equiv\; y-x\in C$.
In particular, we are interested in the following orders.
\begin{itemize}
\item $\Blin(\cH)^+$ induces an order on $\Blin(\cH)$.
We saw this already in \S~\ref{sec:pos cone}.
\item $\TC(\cH)^+$ induces an order on $\TC(\cH)$
  This is just the restriction to $\TC(\cH)$ of the previous order.
\item $\Mon(\cH,\cK)$ induces an order on $\Blin(\TC(\cH),\TC(\cK))$.
More explicitly, $\Lambda \le \Gamma$ is equivalent to,
$\forall\rho\colon\Pos(\cH),\; \Lambda\rho \le \Gamma\rho$.
The two occurences of ``$\le$'' here refer to orders on different spaces.
\end{itemize}
$x < y$ means $x\le y$ and $x\neq y$;
we may write ``$y\ge x$'' instead of ``$x\le y$'' etc.
$0 \le x$ simply means that $x$ belongs to the cone $C$.
If $\dim\cH < \infty$, then $\Mon(\cH)\subset\Blin(\Blin(\cH))$
consists of operators on a Hilbert space, hence ``$\le$'' is
potentially ambiguous in this case.
In practice, this is not a problem; whenever the $\Mon$-induced order
is an option, that is the one intended.

\subsection{Order and projection}

A function $f$ on an ordered set is \textit{decreasing} if $f(x) \le x$.

\paragraph{Projections lifted to $\TC(\cH)$ are not necessarily decreasing
  on $\Pos(\cH)$.}
\leavevmode

This means that, if $Q$ is a (not necessarily finite-dimensional)
orthoprojection on $\cH$ and $0\le\rho$, then $\wh{Q}\rho \le \rho$ does not
always hold; it depends on $Q$ and $\rho$.
This is related to the fact that two polarizing filters can be used to
rotate polarization arbitrarily, albeit with intensity loss.

\paragraph{Projections lifted to $\Blin(\TC(\cH),\TC(\cK))$ are decreasing
  on $\Mon(\cH,\cK)$.}\label{sec:projections decrease}
\leavevmode

In light of the preceding, this may seem surprising at first.
The reason is that monotonicity is stable under composition
and lifted projections (i.e., $\btheta(Q)$) are monotone. 
Suppose $0 \le \Lambda$. Then, 
$\wh{Q}\Lambda
= \wh{Q}\circ\Lambda\circ\wh{Q}
= \btheta(Q)\circ\Lambda\circ\btheta(Q)$ is monotone as a composition
of such.
In addition,
{$\Lambda - \wh{Q}\circ\Lambda\circ\wh{Q}
= \btheta(Q^\perp)\circ\Lambda\circ\btheta(Q^\perp)
+ \btheta(Q)\circ\Lambda\circ\btheta(Q^\perp)
+ \btheta(Q^\perp)\circ\Lambda\circ\btheta(Q)$}
is monotone as a sum of compositions of monotone operators.

\subsection{Order and norm}\label{sec:order and norm}

Generally, a function from one ordered set to another is called
\textit{monotone} if it preserves order. For the orders we are
considering, the norm turns out to be a monotone function from
the defining cone to the set $[0,\infty)$ with its usual
(in fact, total) order.
\begin{itemize}
\item $\|\cdot\|$ is monotone on $\Blin(\cH)^+$
\item $\|\cdot\|_1$ is monotone on $\TC(\cH)^+$
\item $\|\cdot\|_{1,1}$ is monotone on $\Mon(\cH,\cK)$
\end{itemize}
The first two are simple. Here is a proof of the last.
  Choose $\rho \ge 0$ with $\|\rho\|_1 = \Tr \rho = 1$
  such that $\|\Lambda\rho\|_1 =\Tr \Lambda\rho > \|\Lambda\|_{1,1}-\epsilon$.
  Then, $\|\Gamma\|_{1,1} \ge \Tr \Gamma\rho \ge \Tr \Lambda \rho$.

\subsection{Compactness of order intervals}\label{sec:order intervals}

Now we come to the most interesting aspect of our orders.
On an ordered set, we can define an \textit{order interval}
as $[x,z]_\le = \setof{y}{x \le y \le z}$. On the linearly ordered
set $\Real$, this is just an ordinary interval.
We show that such order intervals in our operator spaces are
compact with respect to useful topologies.

\paragraph{Closed bounded $\le$-intervals in $\Blin(\cH)$ are WOT (=GMET) compact.}
\label{sec:intervals B(H)}
\leavevmode

For purposes of proving this, we may as well assume that we
deal with an interval of the form $[0,A]_\le$ with $0 < A$
because translating so that the lower bound is zero is a topological
isomorphism and if $0\nleq A$ then the interval is empty hence
trivially conpact. This remark applies also to the following two cases.

The definition of $0\le B \le A$ is that
for each $\psi\in\cH$, $\inpr{\psi}{B\psi}$ lies in the compact interval
$[0,\inpr{\psi}{A \psi}]$.
But, WOT is a product topology, and this says that $0\le B \le A$ is
a product of compact factors. Hence $\setof{B}{0\le B \le A}$ is compact.

\paragraph{Closed bounded $\le$-intervals in $\TC(\cH)$ are GMET-compact.}
\label{sec:intervals B1(H) trivial}
\leavevmode

If $\rho$ is trace-class, then $[0,\rho]_\le$, regarded \textit{a priori}
as a subset of $\Blin(\cH)$, is actually in $\TC(\cH)$.
Therefore, the compactness is such an immediate corollary of the previous
paragraph that it hardly bears mentioning.
We do so merely to point out that we can do much better, as shown next.

\paragraph{Closed bounded $\le$-intervals in $\TC(\cH)$ are norm-compact.}
\label{sec:intervals B1(H)}
\leavevmode

Now we are working in a metric space.
Therefore, since $[0,\rho]_\le$ is clearly closed,
we need demonstrate only total boundedness.

We can choose a finite-dimensional orthoprojector $P$ such that
$\rho = P\rho P + P^\perp \rho P^\perp$ and 
$\|P^\perp \rho P^\perp\|_1$ is as small as desired.
For instance, take the range of $P$ to be the direct sum of
the eigenspaces for the largest $n$ eigenvalues of $\rho$.

Effectively, $P\rho P$ is in the finite-dimensional space $\TC(P\cH)^+$
and $X \defeq \setof{\rho'\in\TC(P\cH)}{0\le\rho'\le P\rho P}$ is a 
totally bounded subset thereof.
Now, consider $0 \le \sigma \le \rho$. Then,
\begin{equation}\nonumber
\sigma = P\sigma P + P\sigma P^\perp + P^\perp\sigma P + P^\perp \sigma P^\perp,
\end{equation}
with $\| P\sigma P \|_1 \le \| P\rho P \|_1 \le \|\rho\|_1$,
$\| P^\perp \sigma P^\perp \|_1 \le \| P^\perp \rho P^\perp \|_1$,
and, by the Lemma below, $\| P \sigma P^\perp \|_1 \le \|\rho \|_1^{1/2}
\| P^\perp \rho P^\perp \|_1^{1/2}$. Thence, $P\sigma P$ is in the
totally bounded set $X$ and
\begin{equation}\nonumber
\| \sigma - P\sigma P \|_1 \le 
2 \|\rho\|_1^{1/2} \| P^\perp \rho P^\perp \|_1^{1/2} + \| P^\perp \rho P^\perp \|_1.
\end{equation}
As noted at the outset, the bound here as small as desired
by choice of $P$, so the order interval $[0,\rho]_\le$ is totally bounded
by the following lemma.

\begin{lem*}
  If $\sigma \in \TC(\cH)^+$, then
  $\Blin(\cH)\times\Blin(\cH) \ni (A,B)\mapsto \Tr \sigma A^\dag B$
  is a well-defined, possibly degenerate, inner product, hence
  satisfies the Cauchy-Schwarz inequality
  $|\Tr \sigma A^\dag B| \le (\Tr \sigma A^\dag A) (\Tr \sigma B^\dag B)$
\end{lem*}
\begin{proof}
  Consider the basic case $\sigma=\bpi(h)$.
  $\Tr \bpi(h) A^\dag B$ is linear in $B$, conjugate-linear in $A$ and
  non-negative if $A=B$, as required.
The only real question then about the general case $\sigma = \sum\bpi(h_n)$ with
$\sum \|h_n\|^2 < \infty$, is whether it is well-defined.
However, $B \bpi(h) A^\dag = (B h)\pam{(A h)}$ has trace-norm
bounded by $\|A\|\|B\|\|h\|^2$, so all is well.
\end{proof}

\paragraph{Closed bounded $\le$-intervals in $\Blin(\TC(\cH),\TC(\cK))$ are SOT-compact.}
\label{sec:SOT-compact intervals}
\leavevmode

This proof is exactly parallel to that in \S~\ref{sec:intervals B(H)}.
SOT is a product topology and $\Gamma \in [0,\Lambda]_\le$ is
equivalent to $\Gamma\rho \in [0,\Lambda\rho]_\le$ in $\TC(\cK)$ for each
$\rho \in \TC(\cH)^+$.
But the order interval $[0,\Lambda\rho]_\le$ in $\TC(\cK)$ is
compact by \S~\ref{sec:intervals B1(H)}.

\newpage
\section{$\CP$ maps in separable setting}\label{sec:CP separable}

\paragraph*{Highlights.}\leavevmode

\begin{enumerate}
\item No revision of the definition of $\CP$ is required.
 (We still need only finite-dimensional auxiliary identities.)
\item Both $\CP(\cH,\cK)$ and the set of its extreme points are GMET-closed.
\item Extreme $\CP$ maps and $\btheta$-maps are the same thing.
\item An algorithm is given for a SOT-convergent
  Kraus decomposition of a $\CP$ map $\Lambda$,
yielding decompositions of a sequence of chosen projections $\wh{P}\Lambda$
  at intermediate stages.
\end{enumerate}

\paragraph*{Isn't it obvious that we should just replace finite Kraus sums by infinite ones?}
No, it isn't. First, we should remember than an ``infinite sum'' isn't really
a sum at all, but a limit, so we always must ask in what sense it converges.
Second, we should recall the phenomenon of continuous spectrum.
A self-adjoint operator on a finite-dimensional space has a spectral
decomposition with number of terms not exceeding the dimension.
Yet spectral representation of the position operator on the
separable Hilbert space $L^2[0,1]$, for instance, requires an \textit{integral}.
The fact that we do get a sum for the Kraus decomposition appears to be
linked to the relatively strong finite approximability of trace-class operators,
as compared to general bounded operators.

\subsection{Our approach to Kraus decomposition vs. traditional approach}

\paragraph*{Traditional recipe.}
The traditional proof of existence of Kraus decompositions
rests on two ingredients: Stinespring's theorem, and a knowledge of the
representations of the $C^*$-algebra $\Blin(\cH)$.
I am not going to explain either of these.
Expositions of Stinespring's theorem can be found in
34.7 of Ref. \cite{Conway-ACOT}, Thm. IX.4.2 of Ref. \cite{Davidson},
Thm. IV.3.6 of Ref. \cite{Takesaki-I}, or \S 9.2 of Ref. \cite{Davies-76}.
Theorem 3.3 of \cite{Kraus-71} or Chapter 3 of \cite{Kraus-83}
explain how to cook the result from these ingredients.

\paragraph*{Criticism.}
That demonstration of Kraus decomposition strikes me as sort of a magic
trick. One is amazed, but does not know what to think of it or
what to do with it. It seems hermetically sealed.

\paragraph*{Present approach.}
We will build on the finite-dimensional theory to obtain
an algorithm which produces Kraus decompositions for the restriction
of a $\CP$ map to larger and larger finite-dimensional spaces.
Also, the result at any stage is exactly reproduced by restricting a
later-stage result.
The proof that this algorithm converges in SOT involves auxiliary
results which are interesting in their own right.

\paragraph*{On the other hand\dots}
Of course, the ingredients on which the traditional proof rests are
themselves important for other things. See, for example, an entire chapter
on dilation theory in \cite{BLPY}. Some people for whom Kraus decomposition
is important will therefore need to understand those things as well, but
I think it is relatively small.

\subsection{$\CP$-induced partial order ($\precsim$) on $\HP(\cH,\cK)$}\label{sec:precsim}

Just like $\Mon$, the $\CP$ cone induces an order on
$\HP(\cH,\cK)$ or $\Blin(\TC(\cH),\TC(\cK))$, which we will
denote by $\precsim$ (and the usual variants).
Note that $0 \prec \Lambda$ means merely that $\Lambda$ is a nonzero $\CP$ map.
Since $\CP(\cH,\cK) \subset \Mon(\cH,\cK)$, $\precsim$ is a
\textit{suborder} of $\le$, i.e.,
\begin{equation}\nonumber
\Lambda \precsim \Gamma \quad\Rightarrow\quad \Lambda \le \Gamma.
\end{equation}

Since $\precsim$ is a suborder of $\le$, it immediately follows from
\S~\ref{sec:order and norm} that the norm $\|\cdot\|_{1,1}$ is monotone on $\CP(\cH,\cK)$.
Similarly, lifted projections are decreasing with respect to $\precsim$.
That is, if $0\precsim\Lambda$, then $0 \precsim \wh{Q}\Lambda \precsim \Lambda$.
Using the fact that $\btheta(Q)$ is $\CP$, 
the argument in \S~\ref{sec:projections decrease} goes through
\textit{mutatis mutandis}.

\subsection{GMET-closedness}
\label{sec:GMET closure}

\subsubsection{$\Mon(\cH,\cK)$ is GMET-closed}
\label{sec:Mon GMET-closed}

Suppose $\Lambda$ is not monotone.
Then $\inpr{k}{(\Lambda\rho)k}\not\in [0,\infty)$ for
some $\rho \in\TC(\cH)^+$ and $k\in\cK$.
Since that matrix element is a norm-continuous function of $\rho$, and
$\rho$ is the norm limit of finite-rank operators, we can assume that
$\rho$ has finite rank. But then, 
$\setof{\Gamma}{\inpr{k}{(\Gamma\rho)k}\not\in [0,\infty)}$ is
a GMET neighborhood of $\Lambda$.

\subsubsection{$\CP(\cH,\cK)$ is GMET-closed}
\label{sec:CP is GMET-closed}

By definition,
\begin{equation}\nonumber
\CP(\cH,\cK) =
\bigcap_N \Big(\text{ -- }\otimes \Id_{\Blin(\Cmplx^N)}\Big)^{-1}
\Mon(\cH\otimes{\Cmplx^N},\cK\otimes{\Cmplx^N}).
\end{equation}
But,
$\Mon(\cH\otimes{\Cmplx^N},\cK\otimes{\Cmplx^N})$ is GMET-closed by
Lemma \ref{sec:Mon GMET-closed}, hence so is each inverse image above,
by \S~\ref{sec:otimes-1 GMET-cts}.

\paragraph{Lemma.}
\label{sec:otimes-1 GMET-cts}
The map $\text{ -- }\otimes\Id_{\Blin(\Cmplx^N)}$,
i.e., $\Gamma \mapsto \Gamma\otimes\Id_{\Blin(\Cmplx^N)}$, 
from $\Blin(\TC(\cH),\TC(\cK))$ to \\
\mbox{$\Blin(\TC(\cH)\otimes\Cmplx^N,\TC(\cK)\otimes\Cmplx^N)$}
is both WOT and GMET continuous.
In fact, it is an isomorphism onto its image in either topology.
\begin{proof}
Use the realization of $B\colon\Blin(\cK\otimes\Cmplx^N)$
as an $N\times N$ matrix with entries in $\Blin(\cK)$, and similarly for
$\rho\colon\TC(\cK\otimes\Cmplx^N)$.
 
Then $\Lambda \otimes \Id_{\Blin(\Cmplx^N)}$ applies $\Lambda$ to each entry,
i.e.,
$
  \langle{B}\mid{(\Lambda \otimes \Id_{\Blin(\Cmplx^N)}) \rho}\rangle
 = \sum_{i,j} \langle{B_{ij}}\mid{\Lambda \phi_{ij}}\rangle
 $.
This immediately proves WOT continuity.
Since the ranks of the entries $B_{ij}$ do not exceed that of $B$,
GMET continuity follows similarly.
\end{proof}
\subsubsection{$\graph(\precsim)$ is GMET$\times$GMET-closed}
\label{sec:graph(le_CP) is GMET-closed}
  $\graph(\precsim)$ is the inverse image of the GMET-closed set
  $\CP(\cH,\cK)$ under the GMET-continuous map
\mbox{$(\Lambda,\Gamma) \mapsto \Gamma - \Lambda$}.

\subsection{Monoidal functor \usebox{\boxt}, redux}
\label{sec:theta functor separable}

\subsubsection{Trouble?}

\S~\ref{sec:theta functor} showed that --- for finite-dimensional Hilbert
spaces --- $\Hilb$ and $\CP$ are
monoidal categories under tensor product and $\btheta$ is a monoidal
functor $\Hilb\rightarrow \CP$.
There is a potential difficulty in extending this to separable Hilbert
spaces.
According to Def.~\ref{sec:CP def},
$\Lambda$ is CP if $\Lambda\otimes\Id_{\Cmplx^N}$ is monotone for any
natural number $N$. In the finite-dimensional setting, this was enough
to guarantee (Prop.~\ref{prop:CP is OK}) that $\Lambda\otimes\Gamma$ is
CP as soon as $\Lambda$ and
$\Gamma$ are so, because they operate on finite-dimensional spaces.
Is it necessary to strengthen the defintion of CP in the
separable setting? In fact no, as \S~\ref{sec:tensor separable} shows.
\S~\ref{sec:theta props} establishes some useful properties of $\btheta$.

\subsubsection{No worries, the old CP definition works}
\label{sec:tensor separable}

First, we must identify
the space on which $\Lambda\otimes\Gamma$ operates, assuming
$\Lambda\colon\Blin(\TC(\cH))$ and $\Gamma\colon\Blin(\TC(\cK))$.
In contrast to Hilbert spaces, the choice of norm on an algebraic
tensor product of Banach spaces, hence its completion,
is often not canonical\cite{Ryan-02,Kubrusly-23}. Fortunately, in our
case, we can identify $\TC(\cH)\otimes\TC(\cK)$ (more precisely,
its completion) with $\TC(\cH\otimes\cK)$, the tensor product
in this last expression being the ordinary Hilbert one.

\paragraph{
$\Lambda\in\CP(\cH,\cK)$
implies that $\Lambda\otimes\Id_{\TC(\cN)}$ is $\CP$,
if $\cN$ is separable.}
\label{sec:tensor product separable}\leavevmode

\begin{equation}\nonumber
\wh{P} (\Lambda) \otimes \wh{P} (\Id_{\TC(\cN)})
=\wh{P} (\Lambda\otimes \Id_{\TC(\cN)})
\xrightarrow{\mathrm{SOT}} 
\Lambda \otimes\Id_{\TC(\cN)}.
\end{equation}
But, $\wh{P} (\Id_{\TC(\cN)})$  can be identified with $\Id_{\TC(\wh{P} \cN)}$,
the identity on a finite-dimensional space, and $\wh{P}\Lambda$ is $\CP$,
so 
\mbox{$\wh{P} (\Lambda) \otimes \wh{P} (\Id_{\TC(\cN)})$} is
in $\CP(\cH\otimes\cN,\cK\otimes\cN)$, which is SOT-closed by
\S~\ref{sec:CP is GMET-closed}.

\subsubsection{ \usebox{\boxt} is  GMET-continuous and commutes with lifted projections}
\label{sec:theta props}

\paragraph{$\|\btheta(A)\| = \|A\|^2$.}\leavevmode

On the one hand, $\|A\rho A^\dag\|_1\le \|A\| \|\rho A\|_1\le \|A\|^2 \|\rho\|_1$.
On the other hand, let $h$ be a unit vector nearly saturating the norm of
$A$: $\|Ah\| \simeq \|A\|$. Then
$\|A\bpi(h)A^\dag\|_1 = \|\bpi(Ah)\|_1 = \|Ah\|^2 \simeq \|A\|^2$.
(The reader can translate `$\simeq$' into something involving $\epsilon$'s.)
\paragraph{$\btheta$ commutes with lifted projection: $\wh{P}\circ\btheta = \btheta\circ\wh{P}$.}
\label{sec:theta P commute}\leavevmode

$ \wh{P}(\btheta(A))\cdot \rho
= (\wh{P}\circ\btheta(A)\circ\wh{P})\rho
= (\wh{P}\circ\btheta(A))\cdot\wh{P}(\rho)
= {P}A{P}\rho {P} A^\dag {P}
= ({P} A {P}) \rho ({P} A^\dag {P})
= \btheta({P} A {P}) \cdot \rho
= \btheta(\wh{P}(A)) \cdot \rho. $ 
\paragraph{$\btheta$ is norm and GMET continuous,
  and SOT continuous on bounded sets.}\leavevmode

$D\btheta(A)\cdot B = B\blank A^\dag + A\blank B^\dag$ is
a bounded $\Real$-linear operator. So, $\btheta$ is not only norm-continuous,
but differentiable.
Then, $\btheta$ is GMET-continuous by \S\S~\ref{sec:theta P commute},
\ref{sec:GMET-cntty trick}.
Finally, 
by \S~\ref{sec:smaller set of SOT seminorms}, SOT continuity on
\mbox{$\cball(\Blin(\cH,\cK))$} follows immediately from
$\|\btheta(A) \bpi(h)\|_1 = \|A h\|^2$.

\subsection{Extreme $\CP$ maps}\label{sec:extreme CP}

\subsubsection{\usebox{\boxt}-maps are extreme}\label{sec:theta is extreme}

$\wh{P}(\btheta(A)) = \btheta(\wh{P}(A))$ (\S~\ref{sec:theta P commute})
is extreme by the finite-dimensional theory. 
Thus, $\btheta(A)$ is extreme by \S~\ref{sec:nonextreme shows itself}
immediately following.

\paragraph{
 $\Lambda\not\in\extr\CP \;\Rightarrow$
 eventually $\wh{P}\Lambda\not\in\extr\CP$
 along any sequence $P \xrightarrow{\mathrm{SOT}} \Id$
 .}
\label{sec:nonextreme shows itself}
\leavevmode

By hypothesis, there is a decomposition $\Lambda = \Gamma+\Gamma'$ into
linearly independent $\CP$ maps.
If $\wh{P}$ is not extreme, then
$c_P\wh{P}\Gamma + c_P'\wh{P}\Gamma' = 0$ for 
some $c_P,c'_P$ which can be scaled so that
$(c_P,c_P')\in[-1,1]^2$. That square being compact, there must be a
subsequence along which $(c_P,c_P') \to (c,c')$. But, that would
imply that $c\Gamma + c'\Gamma' = 0$, a contradiction.

\subsubsection{Extreme $\CP$ maps are \usebox{\boxt}-maps}
\label{sec:extreme CP maps are theta-maps}

\S~\ref{sec:reduction still works} below (no danger of circularity)
shows that if $0 \prec \Lambda$, then $\btheta(A)\precsim\Lambda$ for some
$A\neq 0$. Therefore, $\Lambda$ cannot be extreme unless it is
a $\btheta$-map.

\subsubsection{$\extr \CP(\cH,\cK)$ is GMET-closed}
\label{sec:extr CP GMET closed}

By \S~\ref{sec:nonextreme shows itself},
to show that $\Lambda$ is a pure $\CP$ map, it
suffices to show that $\wh{P}\Lambda$ is such for every
finite-dimensional $P$.

\paragraph{$\Lambda\in\cl_{\mathrm{GMET}}(\extr\CP(\cH,\cK))
  \;\Rightarrow\; \wh{P}\Lambda \text{ is pure } \CP$.}
\leavevmode

Since $\extr \CP(\cH,\cK) = \btheta(\Blin(\cH,\cK))$, 
by \S~\ref{sec:GMET final for P}, the hypothesis is equivalent to
\mbox{$\wh{P}\Lambda \in \cl \wh{P}\circ\btheta(\Blin(\cH,\cK))$}.
But, since $\btheta$ commutes with lifted projection
(\S~\ref{sec:theta P commute}) and by closedness of pure $\CP$ maps in
the finite-dimensional case (\S~\ref{sec:extr CP closed}),
$\cl \wh{P}\circ\btheta(\Blin(\cH,\cK))
= \cl \btheta(\Blin(P\cH,P\cK))
= \btheta(\Blin(P\cH,P\cK))
= \extr\CP(P\cH,P\cK)$.

\subsection{Construction of a SOT-convergent Kraus decomposition}
\label{sec:Kraus construction 2}

Now we will show that the reduction algorithm for constructing a Kraus decomposition,
developed in \S~\ref{sec:Kraus construction}, works over a separable space.
Recall the formula 
for $(A,\Lambda') = \reduce(k,h,\Lambda)$, namely
\begin{equation}
  \tag{\ref{eq:CP reduction}}
  \inpr{k'}{A h'} =
\left\{\begin{array}{ll}
0, & c=0 \\    
c^{-1/2} \inpr{k'\pam{k}}{\Lambda\cdot h'\pam{h}}, & \text{otherwise}
 \end{array}\right\}
\quad \text{where }
c = {\inpr{\bpi(k)}{\Lambda\cdot \bpi(h)}}.
\end{equation}
This is clearly well-defined, even over an infinite dimensional Hilbert space.

\subsubsection{Reduction still works: the remainder is $\CP$}
\label{sec:reduction still works}

In \S~\ref{sec:Kraus construction},
the conclusion that $\Lambda'= \Lambda - \btheta(A)$ is $\CP$
relied on the Jamio{\l}kowski transform. 
Hence, we have more work to secure this in the separable case.
It follows immediately from (\ref{eq:CP reduction}) that
\begin{equation}\label{eq:projected reduction}
k,h \in \Ran P \quad\Rightarrow\quad 
\reduce(k,h,\wh{P}\Lambda)=(\wh{P}A,*),
\end{equation}
so, $\btheta(\wh{P}A)\precsim \wh{P}\Lambda$.
Since $\mathrm{Graph}(\precsim)$ is GMET$\times$GMET closed
(\S \ref{sec:graph(le_CP) is GMET-closed}), 
${P}\xrightarrow{\mathrm{SOT}} \Id$ yields
\mbox{$\btheta(A)\precsim \Lambda$}, as required.

\subsubsection{Iterated reduction}

The iterative algorithm is not essentially different from the
finite-dimensional one presented in \S~\ref{sec:Kraus construction},
but the discussion will be facilitated by some extra notation.
Let
\begin{itemize}
\item  
$\sB_\cH = h_1,h_2,\ldots$ be a complete orthonormal system in
$\cH$ and $\cH_n = \Span\{h_1,\ldots,h_n\}$,
\item $\sB_\cK$ and $\cK_m$ be defined similarly for $\cK$,
\item $n\mapsto(k_{l(n)},h_{r(n)})$ be an enumeration of all pairs in
  $\sB_\cK\times\sB_\cH$,
\item $N(i,j)$ be the index of the last member of this enumeration in
$\cK_i\times\cH_j$,
\item $P_{n,m}$ be the polyprojection of $\cK$ onto $\cK_n$ and $\cH$ onto
  $\cH_m$.
\end{itemize}
The iterative procedure is the same as before:
From 
$\Lambda_0 \defeq \Lambda$ and 
$(A_m,\Lambda_m) \defeq \reduce(k_{l(m)}\pam{h_{r(m)}},\Lambda_{m-1})$,
we obtain
\begin{equation}\label{eq:reduction n again}
\Lambda_0 = \sum_{i=1}^{n} \btheta(A_i) + \Lambda_n.
\end{equation}

From (\ref{eq:projected reduction}), it follows that
\begin{equation}
\wh{P_{k,j}}\Lambda = \sum_{i=1}^{N(k,j)} \btheta(\wh{P_{k,j}}A_i),  
\end{equation}
and therefore $\wh{P_{k,j}}\Lambda_m = 0$ for $N(k,j) \le m$.
Eq. (\ref{eq:projected reduction}) also shows very clearly that only
a finite number of matrix elements of $\Lambda$ and the $A_i$'s are needed
at any stage of the reduction. Assuming that individual matrix elements
are available, the algorithm is computationally feasible.

\subsubsection{The series converges to $\Lambda_0$}

\paragraph{\dots in GMET.}
$\Lambda_{n+1} \precsim \Lambda_n$, implies that
the norms $\|\Lambda_{n}\|_{1,1}$ are bounded (\S~\ref{sec:precsim}).
Therefore, the observation that \mbox{$\wh{P_{i,j}}\Lambda_n \to 0$}
for every $(i,j)$ implies that $\Lambda_n \to 0$ in {GMET}.

\paragraph{\dots and even in SOT.}\label{sec:Kraus SOT convergent}

The preceding demonstration of GMET convergence is not used in the
following and was given because of its simplicity. 

Let $S_n \defeq \sum_{i=1}^n \btheta(A_i)$ be the partial sums of the series,
$T_n \defeq \{S_n,S_{n+1},\ldots\}$ the tails, and $\overline{T}_n$ the
SOT-closure of $T_n$. If $\cap \overline{T}_n$ is a singleton, it is the
SOT-limit of the series, hence the GMET-limit, $\Lambda$.

Now, $\overline{T}_n \subseteq [S_n,\Gamma]_\le$ for $\Gamma$ any upper bound
for $T_1$; there is at least one such, namely $\Lambda$.
The set $\sI$ of such intervals
evidently has FIP and its elements are SOT-compact by
\S~\ref{sec:SOT-compact intervals}, hence $\cap\sI$ is nonempty.
But, any $\Psi$ in $\cap\sI$ satisfies $S_n \le \Psi \le \Gamma$ for
every $n$ and upper bound $\Gamma$, hence $\Psi$ must be \textit{the}
least upper bound of the partial sums.
Finally, observe that $\cap \overline{T}_n$ is contained in $\cap \sI$,
yet nonempty by the same argument as for $\cap\sI$.

\vspace*{0.8cm}
\centerline{\rule{0.5\textwidth}{0.8pt}}
\vspace*{15pt}
\centerline{\huge{Continuous-time evolution}}
\vspace*{10pt}
\centerline{\rule{0.5\textwidth}{0.8pt}}
\vspace*{0.8cm}
\addcontentsline{toc}{part}{Continuous-time evolution}

\section{(Two-real-parameter) evolution families}\label{sec:evolution}

One learns in a first course on quantum mechanics that
the Schr\"{o}dinger equation $\tfrac{d\psi}{dt} = -iH\psi$ with constant
Hamiltonian has solution $\psi(t) = e^{-itH}\psi(0)$.
Abstracting away the state vector, we find that the propagator
$U(t,0) = e^{-itH}$ also satisfies the equation $\tfrac{d}{dt}U(t,0) = -iH U(t,0)$.
This is nice because the propagator can be used over and over with distinct
initial state vectors.

But matters become much more interesting with a time-dependent Hamiltonian.
Then we really need both arguments in the propagator.
The solution $U(t,s)$ of $\tfrac{d}{dt}U(t,s) = -iH(t)U(t,s)$ with initial
condition $U(s,s) = \Id_\cH$ propagates (better --- \textit{evolves}, as a
transitive verb)
any state vector at time $s$ to time $t$.
In this section, we generalize from the state vector in a Hilbert space
to a vector in a Banach space, such as a density operator is.

\paragraph*{For readers following the finite-dimensional track.}
\leavevmode

A finite-dimensional Banach space is just a normed space.
From this point on, there may occasionaly be adjectives or qualifiers
which are irrelevant to the finite-dimensional setting.
E.g., norm continuity is simply continuity and $\TC(\cH)$ can be
considered simply $\Blin(\cH)$.
Sections specific to the separable setting are mixed in, and
frequently flagged explicitly, but anything about GMET or SOT
should be skipped.

\paragraph*{On the integration theory.}
\leavevmode

Techically, the integration theory involved here is Bochner-Lebesgue.
For a first-rate exposition of integration incorporating Banach-space
valued integrands from the very beginning (with no complication),
Chapter 11 of Lang's analysis text\cite{Lang-Analysis} is recommended.

\paragraph{Highlights.}\leavevmode

\begin{enumerate}
\item An evolution family $\cE$ which is norm continuous in each variable
  is invertible.
\item The generator need be only locally integrable.
  \begin{itemize}
  \item This corresponds to weak differentiability of $\cE$.
  \item An integral equation formulation is convenient.
  \item An iterative procedure converges to the solutin of the integral equation.
  \end{itemize}
\end{enumerate}

\subsection{Linear evolution}\label{sec:linear evolution}

\subsubsection{The basic `algebraic'  idea}\label{sec:algebraic E}
  A \textit{\paren{two-parameter} evolution family}
  $\cE$ on a Banach space $\sX$ is a map
  \begin{equation}\nonumber
  \setof{(t,s)}{-\infty < s \le t < \infty}\rightarrow \Blin(\sX)
  \end{equation}
  obeying the conditions
\begin{subequations}
\begin{align}
 \cE(t,t) &= \Id_{\sX}  \label{eq:E Id} \\ 
  \cE(u,s) &= \cE(u,t)\cE(t,s), \quad
             \text{if }
             u\ge t\ge s. \label{eq:E divisibility} 
\end{align}
\end{subequations}
The condition (\ref{eq:E divisibility}) is often called \textit{divisibility},
and we will adopt this terminology.

\subsubsection{ If $\cE$ is norm continuous in \textit{each}  argument,
  then it is contained in $\GL(\sX)$.
}\label{sec:cts E}

$\GL(\sX)$
denotes the subset of invertible linear operators in $\Blin(\sX)$
(``$\GL$'' stands for ``general linear (group)'').
The important point about this set is that it is norm open, or at least
that the identity is in the norm interior of $\GL(\sX)$.
\begin{proof}
 Define a relation $R$ by: if $t \ge s$, `$tRs$' means $\cE(t,s)\in\GL(\sX)$
 and if $t < s$, `$tRs$' means $sRt$. Evidently, $R$ is transitive
[$\cE(t,r)^{-1} = \cE(s,r)^{-1}\cE(t,s)^{-1}$ if the RHS inverses exist],
hence an equivalence relation. By hypothesis, the $R$-equivalence class of
$t$ includes some open interval containing $t$. Therefore, equivalence classes
are open. And also closed, because the union of all but one is open.
The only subsets of $\Real$ with this property (`clopen') are $\varnothing$ and
$\Real$.
\end{proof}

Due to divisibility (\ref{eq:E divisibility}), continuity follows already from:
$\forall t$, $\cE(t+\delta,t) \to \Id_\sX$ and 
\mbox{$\cE(t,t-\delta) \to \Id_\sX$} as \mbox{$\delta \downarrow 0$}.

\paragraph{Remark.}
It was important in the proof 
that $\cE(t,s)$ be continuous in each argument.  Had we only 
continuity in $t$, we would have been guaranteed only an interval
$[t,t+\epsilon[$ in $t$'s equivalence class.
We shall see that differentiability in $t$ ensures continuity also in $s$.

\subsubsection{(Weakly) differentiable evolution families}\label{sec:diff'ble E}

Now we will be more demanding, and suppose that $\cE$ is differentiable in its
first argument. In particular,
\begin{equation}\nonumber
  \cE(s+\epsilon,s)
  = \Id_\sX + \epsilon A(s) + o(\epsilon), \quad \epsilon > 0
\end{equation}
and therefore
$\cE(t+\epsilon,s) = \Big( \Id_{\sX}+ \epsilon A(t) + o(\epsilon) \Big) \cE(t,s)$,
  or equivalently
\begin{equation}\label{eq:dE/dt}
\frac{\partial}{\partial t}\cE(t,s) = A(t)\cE(t,s).
\end{equation}

\paragraph{Integral equation formulation.}\leavevmode

We were a little imprecise there because
a reformulation is preferable, specifically as
an integral fixed-point equation, which incorporates the boundary condition:
\begin{tcolorbox}[standard jigsaw,opacityback=0]
\begin{equation}\label{eq:E integral eq}
\cE(t,s) = \Id_{\sX} + \int_s^t A(u) \cE(u,s)\, du.
  \end{equation}
\end{tcolorbox}
The solution is given, formally at least, by the infinite series
\begin{equation}\label{eq:E series}
\cE(t,s) \defeq \sum_{n=0}^\infty I_n(t,s),
\end{equation}
where the terms are defined recursively as
\begin{equation}\label{eq:I recursion}
  \begin{split}
  I_0(t,s) & \defeq \Id_\sX, \\
  I_{n+1}(t,s) &\defeq \int_s^t A(u) I_n(u,s)\, du. 
  \end{split}
\end{equation}

But, when does the series (\ref{eq:E series}) converge (and in what sense)?

\paragraph{Local integrability, $L^1_{\mathrm{loc}}$.}\leavevmode

For a measurable function $F\colon\Real \rightarrow \sX$, $\sX$ a Banach space,
we define quasi-seminorms for $s < t$ by
\begin{equation}\label{eq:L1-loc}
\|F\|_{L^1_{\mathrm{loc}}(s,t)} = \int_s^t \|F(u)\| \, du,  
\end{equation}
where $\|F(u)\|$ is the norm of $F(u)$ as a member of $\sX$.
$F$ is \textit{locally integrable} if every
$\|F\|_{L^1_{\mathrm{loc}}(s,t)}$ is finite.
$L^1_{\mathrm{loc}}(\Real;\sX)$ (even just $L^1_{\mathrm{loc}}$ if unambiguous)
denotes the vector space of locally integrable functions valued in $\sX$,
equipped with the system of seminorms (\ref{eq:L1-loc}).
Thus, ``$F_n \to F$ in $L^1_{\mathrm{loc}}$'' means that every
$\|F-F_n\|_{L^1_{\mathrm{loc}}}$ tends to zero.
\paragraph{
  If $A$ is locally integrable ($L^1_{\mathrm{loc}}$),   the series (\ref{eq:E series}) converges.
}\label{sec:E bounds}\leavevmode

In fact, with the change of measure
\begin{equation}\label{eq:lem change of var}
d\tau(u) \defeq \|A(u)\| \, du  
\end{equation}
we have bounds
\begin{subequations}
\begin{align}
  \|I_n(t,s)\| & \le \tfrac{1}{n!}[\tau(t)-\tau(s)]^n
  \label{eq:lem In bound} \\
  \|\cE_A(t,s)\| & \le e^{\tau_A(t)-\tau_A(s)}
                         \label{eq:lem E bound}
\end{align}
\end{subequations}

\begin{proof}
  (\ref{eq:lem In bound}) is proven by induction. The case $n=0$ is clear.
  For the induction step,
\begin{equation}\nonumber
  \begin{split}
\|I_{n+1}(t,s)\|
    & \le \int_s^{t} \|I_{n}(u,s)\| \, d\tau(u)
    \quad\qquad \why{(\ref{eq:I recursion}) \text{ and } (\ref{eq:lem change of var})} \\
& \le \int_{\tau(s)}^{\tau(t)} \tfrac{1}{n!}[\tau - \tau(s)]^n d\tau
    \qquad \why{\text{induction hypothesis}} \\
& = \int_{\tau(s)}^{\tau(t)} d\Big(\tfrac{1}{(n+1)!}[\tau - \tau(s)]^{n+1}\Big) 
 =  \tfrac{1}{(n+1)!}[\tau(t)-\tau(s)]^{n+1}.
  \end{split}
\end{equation}
(\ref{eq:lem E bound}) follows by the power series for $\exp$.
\end{proof}

\subsubsection{Differentiability still holds almost everywhere}

By moving to an integral equation formulation, we secured some
welcome flexibility. But in what sense does the differential
equation (\ref{eq:dE/dt}) hold, then?
The answer is given by a version of the Lebesgue differentiability theorem
for Banach-space valued integrands: $\tfrac{\partial}{\partial t}\cE(t,s)$
exists almost everywhere with respect to Lebesgue measure, and equals
$A(t)$ in the $L^1_{\mathrm{loc}}$ sense. See Thm. 2 of \S{V.5} of
Yosida\cite{Yosida-80} for this.
But, why do we even care? Because it matters for translating restrictions
on an evolution family to restrictions on its generator. This will be
important in \S\ref{sec:generators CP evolution}. 

\subsubsection{Change of perspective: $\cE$ maps $L^1_{\mathrm{loc}}$ generators to
evolution families}\label{sec:E as map}

To this point, we have been dealing with only an individual evolution family,
which we have written simply as $\cE$. It is determined by a generator function
$A$. Shortly, we will have two generators $A$ and $A'$; to keep
the associated evolution families straight, it is natural to subscript them as
$\cE_A$ and $\cE_{A'}$. An even better perspective, though, is to
regard $\cE$ as a \textit{map} from $L^1_{\mathrm{loc}}(\Real,\Blin(\sX))$ into
the set of evolution families.
Continuity of this map then becomes an obvious
subject of inquiry; we turn to that next.

\subsubsection{Formula for a perturbed evolution family}\label{sec:perturbing E}

\begin{tcolorbox}[standard jigsaw,opacityback=0]
\begin{equation}
\label{eq:E perturbation}
  \cE_{A'}(t,s)
  =
  \cE_{A}(t,s)
+  \int_s^t \cE_{A'}(t,u)\, [A'(u)-A(u)] \,
  \cE_{A}(u,s)\, du
\end{equation}
\end{tcolorbox}

\begin{proof}
\begin{equation}\nonumber
  \begin{split}
\int_s^t \cE_{A'}(t,u)A'(u)\cE_{A}(u,s)\, du 
&=
\int_s^t \cE_{A'}(t,u)A'(u)
\Big(\Id + \int_s^u A(u')\cE_{A}(u',s)\, du'\Big) \, du \\
&=
\cE_{A'}(t,s)
    + \int_s^t \int_s^{u}
    \cE_{A'}(t,u)
    A'(u)
    A(u')
    \cE_{A}(u',s) \, du'\, du.
  \end{split}
\end{equation}
Each equality here is a use of the integral equation (\ref{eq:E integral eq}),
in different ways.
A parallel calculation with the r\^{o}les of $A$ and $A'$ swapped yields
\begin{equation}\nonumber
  \begin{split}
\int_s^t \cE_{A'}(t,u)A(u)\cE_{A}(u,s)\, du 
=
\cE_{A}(t,s)
    + \int_s^t \int_s^{u}
    \cE_{A'}(t,u)
    A'(u)
    A(u')
    \cE_{A}(u',s) \, du'\, du.
  \end{split}
\end{equation}
To obtain (\ref{eq:E perturbation}), simply
subtract this second display from the previous one.
\end{proof}

\subsubsection{ $L_{\mathrm{loc}}^1$ convergence of the generator implies
locally uniform convergence of  $\cE$}\label{sec:E cts}

Bounding the norm of the integral in (\ref{eq:E perturbation}) as
\mbox{$\int_s^t \|\cE_{A'}(t,u)\|\|A'(u)-A(u)\|
  \|\cE_{A}(u,s)\|\, du$} and applying the bound (\ref{eq:lem E bound})
yields
\begin{tcolorbox}[standard jigsaw,opacityback=0]
\begin{equation}  \label{eq:E perturbation bound}
  \|\cE_{A'}(t,s) - \cE_{A}(t,s)\|
  \le
e^{\|A\|_{L^1(s,t)} + \|A'\|_{L^1(s,t)}}
\|A'-A\|_{L^1(s,t)}.
\end{equation}
\end{tcolorbox}

\subsubsection{A differentiable evolution family is invertible}

According to \S~\ref{sec:cts E},
$\cE_A$ lies in $\GL(\sX)$ if it is continuous in each
argument, and as noted there, this requires only
that $\cE(t+\delta,t)-\Id_\sX$ and $\cE(t,t-\delta,t)-\Id_\sX$ tend
to zero as $\delta\downarrow 0$.
However, from the bound (\ref{eq:E perturbation bound}), these
are controlled by 
$\|A\|_{L^1(t-\delta,t)}$ or $\|A\|_{L^1(t,t+\delta)}$.

\subsection{Unitary evolution as prototype evolution in a monoid}
\label{sec:unitary evolution}

This section determines the modifications to the general linear evolution if
we consider only evolution families which consist entirely of unitary operators
on $\Blin(\cH)$, i.e.,
$\cE_A(t,s)\in\Uni(\cH)$.
We will go through some simple things very carefully as a warm-up
exercise for the $\CP$ case.

\subsubsection{The generator must be forward tangent to $\Uni(\cH)$ at the identity}

If all the evolution operators are unitary, this is true in particular of
$\cE(t+\epsilon,t)$, that is 
$\Id_\cH + \epsilon A(t) + o(\epsilon) \in \Uni(\cH)$.
What the notation here means is that \textit{some} $o(\epsilon)$
added to $\Id_\cH + \epsilon A(t)$ will result in an operator
in $\Uni(\cH)$, not that that is true for \textit{any} $o(\epsilon)$ terms.
[Logically, it might be better to write
\mbox{$\Id_\cH + \epsilon A(t) \in \Uni(\cH)+o(\epsilon)$}, but I think
that looks strange.]
In terms of the following definition, this says that $A(t)$ is
\textit{forward tangent} to $\Uni(\cH)$ at $\Id_\cH$.
\paragraph{Definition: forward-tangent, tangent cone.}\label{sec:tangent def}
\leavevmode

If $x$ is a point in a subset $\sS$ of a normed space $\sX$,
 the vector $v\colon\sX$ is
\textit{forward-tangent} to $\sS$ at $x$ precisely when
\begin{equation}
  x + \epsilon v + o(\epsilon) \in \cC, \quad \text{as } \epsilon\downarrow 0.    
\end{equation}
Equivalently, the distance between $x+\epsilon v$ and $\cC$ is $o(\epsilon)$ as 
$\epsilon\downarrow 0$.    

The set of vectors forward-tangent to $\sS$ at $x$ in this sense is a
closed cone, called the \textit{tangent cone} to $\sS$ at $x$.
The set of vectors $v$ such that both $v$ and $-v$ are forward-tangent
to $\sS$ at $x$ is the \textit{tangent space} at $x$.

\paragraph{Remarks.}


The reader may be puzzled by this notion. Why not just say ``tangent''?
Because $\Uni(\cH)$ is a smooth manifold in $\Blin(\cH)$, a vector forward
tangent to it is also backward tangent (with obvious defintion).
However, this is not so for $\CP(\cH)$, as we shall see.

The reason we only need forward tangency is because $\cE(t,s)$ is defined
for $t\ge s$, but not for $t < s$.

\subsubsection{Such forward tangency amounts to skew-adjointness}

This is simple computation. Unitarity requires that
$(\Id_\cH + \epsilon A(t) + o(\epsilon))^\dag
(\Id_\cH + \epsilon A(t) + o(\epsilon))
= \Id_\cH + \epsilon( A(t)+ A(t)^\dag) + o(\epsilon)$.
equal $\Id_\cH$. For that to hold to $o(\epsilon)$ requires
$A(t) = - A(t)^\dag$. Equivalently,
$A(t) = -iH \in i\cdot\SA(\cH)$.

\subsubsection{Any constant skew-adjoint operator generates a unitary evolution}
\label{sec:skew-adjoint constant}

So far we have established only that skew-adjointness is \textit{necessary}.
But, we easily verify directly that if $A(t) = -iH$ is constant,
then $\cE_ A(t,s) = e^{-i(t-s)H}\in\Uni(\cH)$.

\subsubsection{Skew-adjoint step maps generate unitary evolution}
\label{sec:skew-adjoint step maps}

\paragraph{Step maps.}
A \textit{step map} is a function $A$ such that
for any bounded interval $[a,b]$, there are \textit{finitely many} breakpoints
$a = t_0 < t_1 < \cdots < t_{n-1} < t_n = b$ such that $A$
takes constant value $A_i$ on $]t_i,t_{i+1}[$ (values at endpoints are irrelevant).

\paragraph{}
For instance, if $A$ is a step map with breakpoints as above
and $t_1 < s < t_2$, then
\mbox{$\cE(s,t_0) = \exp[(s-t_1) A_1]\exp[(t_1-t_0) A_0]$}.
This is
unitary because the two factors are so (this was the previous step), and
$\Uni(\cH)$ is closed under multiplication; that is, it is a \textit{monoid}.
[To algebraists and computer scientists, a monoid is a set equipped with
an everywhere-defined binary operation and an identity element for said operation.
We used a form of this word already in the discussion of monoidal categories.
According to this classification, a quantum dynamical semigroup should be called
a quantum dynamical monoid.]
Actually, of course, $\Uni(\cH)$ is not just a monoid, but a group.
However, existence of inverses is not relevant here, which is fortunate since
they are scarce in $\CP(\cH)$ (\S~\ref{sec:Iso(CP) example}).

\subsubsection{Any  locally integrable
  $A\colon\Real \rightarrow i\cdot\SA(\cH)$
  generates a  unitary evolution family}
\label{sec:skew-adjoint L1_loc}

The final property of $\Uni(\cH)$ which comes into play is that it is closed.
This is so because $A\mapsto A^\dag A$ is continuous and $\Uni(\cH)$ is the
inverse image of $\Id_\cH$ under this map.
$\cE_ A(t,s)$ is in $\Uni(\cH)$ for any skew-adjoint step map, and it is
a continuous function of $L^1_{\mathrm{loc}}(\Real,\Blin(\sX))$, according to
\S~\ref{sec:E cts}. Therefore, $\cE(t,s)\in\Uni(\cH)$ for any $A$ in the
$L^1_{\mathrm{loc}}$ closure of the step maps. But this is \textit{all}
of $L^1_{\mathrm{loc}}$.
On this denseness, see \S{11.6} of Lang\cite{Lang-Analysis}, for instance.

\subsubsection{All of $\Uni(\cH)$ is accessible}

$\cE(t,s)$ for $t > s$ can be anything in $\Uni(\cH)$ for appropriate generator.
In fact, for $t_0 < t_1 < \cdots < t_n$, any sequence of $\cE(t_k,t_0)$
is realizable. This is related to the fact that right multiplication,
$A \mapsto AU$, is an isomorphism of the tangent space at $\Id_\cH$
with the tangent space at $U$.
A very simple way to realize any $U$ is to let $iA$ be a logarithm of it,
so the curve $t\mapsto e^{itA}$ goes from $\Id_\cH$ to $U$ in unit time.

\section{Synthesizing generators of $\CP$ evolution}\label{sec:generators CP evolution}

Recall the notions of tangent cone/space, Definition~\ref{sec:tangent def}.
The following notation will be heavily used in the rest of these notes.
\begin{notation}
$\cp^+(\cH)$ and $\cp(\cH)$
denote the tangent cone and tangent space to $\CP(\cH)$ at $\Id_{\Blin(\cH)}$, respectively.
\end{notation}

\paragraph*{Highlights.}\leavevmode

\begin{enumerate}
\item The generator of a $\CP$ evolution family must be valued
  in $\cp^+(\cH)$ (almost everywhere).
  Subject to that, the only restriction on time-variation is the
  unavoidable integrability condition of \S\ref{sec:E bounds}.
\item $\cp^+(\cH)$ and $\cp(\cH)$ are norm-closed.
\item Evolution families with $\cp(\cH)$-valued generators are
  precisely those lifted from $\Blin(\cH)$.
\item Any operator in $\cp^+(\cH)$ can be written
  in the form $L(\Psi,K) = \Psi + K\blank + \blank K^\dag$ for
  (nonunique) $\Psi\in\CP(\cH)$ and $K\in\Blin(\cH)$.
  We call the pair $(\Psi,K)$ a \textit{Lindblad parametrization}.
\end{enumerate}

\subsection{Lifting general evolution families from $\Blin(\cH)$ to
  $\Blin(\TC(\cH))$}

Suppose $\cE$ is a weakly differentiable evolution family on $\cH$
with generator $A(t)$;
its constituent operators take values in $\GL(\cH)$ according
to \S\ref{sec:cts E}.
It follows immediately from the functorial nature of $\btheta$ that
\mbox{$\hat{\cE}(t,s) \defeq \btheta(\cE(t,s))$} is a weakly differentiable
evolution family on $\TC(\cH)$.
Since $\cE(s,s)=\Id_\cH$, the chain rule gives
$
\tfrac{\partial}{\partial t}\hat{\cE}(t,s)|_{t=s}
= D\btheta(\Id_{\cH})\cdot\tfrac{\partial}{\partial t}\cE(t,s)|_{t=s}
= D\btheta(\Id_{\cH})\cdot A(s)
$
almost everywhere.

Now, the derivative of $\btheta$ is
\begin{equation}\label{eq:Dtheta}
D\btheta(S) =  T \mapsto S\blank T^\dag + T\blank S^\dag.
\end{equation}
\begin{proof}
For \textit{real} $t$, $\btheta(S+tT) = \btheta(S) + 
t(S\blank T^\dag + T\blank S^\dag) + t^2\btheta(T)$.
\end{proof}
Therefore, the generator
$D\btheta(\Id_\cH)\cdot A$ of $\hat{\cE}$ has the more explicit expression
$A\blank\Id_\cH+\Id_\cH\blank A^\dag$.
\paragraph{Dropping redundant $\Id$'s.}
If we understand $K\blank\Id_\cH$ as a purely syntactic device
(``put the argument here''), there is no problem shortening it to
$K\blank$, and we shall do this whenever convenient.

\subsection{The Lindblad map $L$}\label{sec:Lindblad map}


Motivated by the discussion of lifting from $\Blin(\cH)$ and the
observation that $\Id_\cH + tK$ and $\exp tK$ are equal to $\mathcal{O}(t)$,
we consider the smooth nonlinear map
\begin{equation}\label{eq:gamma}
  \begin{array}{ccccl}
    \HP({\cH}) &\times &\Blin(\cH)
    & \xrightarrow{\gamma} & \HP({\cH}) \\
    (\Psi&,&K) & \mapsto & \Psi + \btheta(\Id_\cH+K).
  \end{array}
\end{equation}

The $\btheta$ term is always $\CP$, so $\gamma$ maps
$\CP({\cH}) \times \Blin(\cH)$ into $\CP(\cH)$, and
for $\Psi\in\CP(\cH)$, $\gamma(t\Psi,tK)$ is $\CP$ for $t\ge 0$.

\subsubsection{Definition}

\begin{equation}
L \defeq D\gamma(0).
\end{equation}
Using (\ref{eq:Dtheta}) unpack this as
\begin{equation}\nonumber
  \begin{array}{ccccl}
 \HP({\cH}) &\times &\Blin(\cH) &\xrightarrow{L}& \HP({\cH}) \\
    (\Psi&,&K) & \mapsto & \Psi +
                           K\blank + \blank K^\dag.
  \end{array}
\end{equation}

\subsubsection{$L$ maps $\CP(\cH)\times\Blin(\cH)$, but nothing else, into $\cp^+(\cH)$}
\leavevmode

If $\Psi\in\CP(\cH)$, then 
$\gamma(t\Psi,tK)$ is in $\CP(\cH)$ for $t > 0$, so
$L(\Psi,K)\in\cp^+(\cH)$. But, $\Psi=0$ works,
so $L(0,-K)\in\cp^+(\cH)$, and
$L(\Psi,K)\in\cp^+(\cH)$ implies
$\Psi = L(\Psi,0)\in\cp^+(\cH)$.

\subsection{$L$ maps onto $\cp^+(\cH)$ --- finite-dimensional case}
\label{sec:L onto in finite dim}


Recall that $\jam$ is an isometry between
$\CP(\cH)\subset\HP(\cH)$
and $\Pos(\Blin(\cH))\subset\SA(\Blin(\cH))$,
with $\Id_{\Blin(\cH)} = \btheta(\Id_\cH) \bijam \bpi(\Id_\cH)$.
also between the tangent cone $\cp^+(\cH)$
of $\CP(\cH)$ at $\Id_{\Blin(\cH)}=\btheta(\Id_{\cH})$ and
the tangent cone $C^+$
of $\Pos(\Blin(\cH))$ at $\bpi(\Id_\cH)$,
with
\begin{equation}\label{eq:C+}
L(\Psi,K) = \Psi+ K\blank + \blank K^\dag \bijam
\jam\Psi + K\pam{\Id_\cH} + \Id_\cH\pam{K}.
\end{equation}

\subsubsection{$\SA(\Blin(\cH))\cong\SA(\Blin(\cH)^0)\times \Blin(\cH)/i\Real$ }
\label{sec:SA=...}

If, by $\Blin(\cH)^0$,
we denote the subspace of trace-zero operators (the orthogonal complement of
$\Cmplx\Id_\cH$),
then relative to the orthogonal decomposition
$\Blin(\cH) = \Cmplx \Id_{\cH} \oplus \Blin(\cH)^0$,
{any} $\Gamma$ in $\SA(\Blin(\cH))$ has unique block-matrix representation
\begin{equation}\label{eq:block matrix}
\left(  \begin{array}{c|c}
          c\colon\Real & \pam{B} \\
          \hline
    B\colon\Blin(\cH) & \Phi\colon\SA(\Blin(\cH)^0)
        \end{array}
      \right).
\end{equation}
Flattened, this says
\begin{equation}
\Gamma = \Phi + K\pam{\Id_\cH} + \Id_\cH\pam{K},
      \;\;
      \text{where  } K = B+\tfrac{c}{2}\Id_\cH.
    \end{equation}
    If we add a real multiple of $i\Id_\cH$ to $K$, it makes
    no difference to $\Gamma$. Thus, we consider $K$ to be
    in the quotient space (over $\Real$) $\Blin(\cH)/i\Real\Id_\cH$,
    but suppress `$\Id_\cH$' in the notation.
    
\subsubsection{$C^+ \cong \Pos(\Blin(\cH)^0)\times\Blin(\cH)/i\Real$}
\label{sec:tangent cone to Pos}
    
Now, if $A$ is trace-zero, then
$\inpr{A}{(\bpi(\Id_\cH)+t\Gamma)A} = t\inpr{A}{\Phi A}$.
Therefore, if $\Gamma\in C^+$, $\Phi$ is positive on the trace-zero operators.
Any such $\Phi$ is $\jam\Psi$ for some $\Psi\in\CP(\cH)$.
Therefore, every $\Gamma$ in $C^+$ takes the form (RHS \ref{eq:C+}),
with $\Psi\in\CP(\cH)$.
This means that $\jam$ maps $L(\CP(\cH)\times\Blin(\cH))$ \textit{onto} $C^+$,
and therefore $L(\CP(\cH)\times\Blin(\cH))$ is the entirety of $\cp^+$,
as announced.

\subsection{$L$ maps onto $\cp^+(\cH)$ --- separable case}
\label{sec:parametrizations separable}

\subsubsection{$L$ is GMET-continuous}\label{sec:L is GMET-cts}

\paragraph{$\wh{P}(LZ) = L(\wh{P}(Z)) \circ \wh{P}$.}\label{sec:PLZ etc}
\leavevmode



If $Z=(\Psi,0)$, this is trivial.
If $Z = (0,K)$, $LZ = K\blank + \blank K^\dag$.
Concentrating on the first term,
$  (\wh{P}(K\blank))B
=  (\wh{P}\circ(K\blank)\circ\wh{P})B
=  {P}K{P}B P
=  (\wh{P}K)(\wh{P}B)
=  (((\wh{P}K)\blank)\circ\wh{P})B$.
Discarding the arbitrary $B$, 
$  \wh{P}(K\blank) =  ((\wh{P}K)\blank)\circ\wh{P}$.
The computation for the other term is similar.

\paragraph{}

$L(\wh{P}(Z)) \circ \wh{P}$
can be read as
the composition $(\blank\wh{P})\circ L\circ \wh{P}$
applied to $Z$; $\blank\wh{P}$ is the linear operator
``precompose with $\wh{P}$'', as usual.
Therefore, it immediately follows from 
\S~\ref{sec:GMET-cntty trick} that $L$ is GMET-continuous.

\subsubsection{Convergent sequences of  finite-dimensional parametrizations}

\paragraph{Promotion.}\leavevmode

Recall the discussion of projection and promotion in \S\ref{sec:projections}.
${P}$ can be considered a map from $\cH$ to itself, or as a map into
the distinct Hilbert space $P\cH$. In the latter view, the adjoint of
$P$ is the promotion $\iota$. Both of these can be lifted so that,
for example, $\wh{\iota}$ may inject $\Blin(\TC(P\cH))$ into 
$\Blin(\TC(\cH))$; it is usually harmless to regard $\iota$ as
inclusion, but occasionaly a little care is needed.
Suppose
$Z\in\HP(P\cH)\times\Blin(P\cH)$ is a Lindblad parametrization of
$A\colon(\TC(P\cH))$. Then, 
$L(\wh{\iota} Z) \neq \wh{\iota}(LZ) = \wh{\iota}\Lambda$.
For, if $Z=(\Psi,K)$, then $(K\blank)\rho = K(P\rho P) + K(P\rho P^\perp)$,
and the second term carries $P^\perp \cH$ into $P\cH$.
The same consideration shows, however, that
$(L(\wh{\iota} Z))\circ\wh{P}$, $\wh{\iota}\Lambda$, and
$\wh{P}(L(\wh{\iota} Z))$ are all equal.
Also, if $\Psi\in\CP(P\cH)$, then $\wh{\iota}\Psi\in\CP(\cH)$.
We now go back to suppressing $\iota$.

%
%
%
%

\paragraph{Setup: an increasing exhaustive sequence $\sS$ of orthoprojections in $\cH$}
\leavevmode

Now, given $\Lambda\colon\HP(\cH)$, the finite-dimensional theory
(\S\ref{sec:L onto in finite dim})
for $P\cH$ says that there is $Z_P$ in \mbox{$\HP(P\cH)\times\Blin(P\cH)$} such that
$LZ_P = \wh{P}\Lambda$.
For the promoted operators, we have
\begin{equation}\label{eq:promoted LZP}
\wh{P}(LZ_P) = (LZ_P)\circ\wh{P} = \wh{P}(\Lambda).  
\end{equation}
according to the previous paragraph.
In the following, we assume that we have such a $Z_P$ for each $P$
in an \textit{increasing} sequence $\sS$ of projections which tends to
$\Id_\cH$ in SOT.

\paragraph{If $(LZ_P)_{P\in\sS}$ has GMET accumulation points, it converges to $\Lambda$}
\label{sec:Lambda only limit of LZ_P}
\leavevmode

If $P'\le P \in\sS$, (\ref{eq:promoted LZP}) implies that
$\wh{P'}(LZ_P - \Lambda) = \wh{P'}\circ\wh{P}(LZ_P - \Lambda) = 0$.
This is not enough to guarantee GMET convergence (due to the restriction on $P'$),
but, if $LZ_P\xrightarrow{\mathrm{GMET}} \Lambda'$ along some subsequence
of $\sS$, it says that $\wh{P'}(\Lambda' - \Lambda)=0$.
And, therefore, $\Lambda' = \Lambda$, since $P'$ is arbitrary in $\sS$.

\paragraph{GMET accumulation points of  $(Z_P)_P$ are Lindblad parametrizers of $\Lambda$.}
\leavevmode

Suppose $Z_P\xrightarrow{\mathrm{GMET}} Z$ along some subsequence $\sS'\subseteq \sS$.
Since $L$ is GMET-continuous (\S\ref{sec:L is GMET-cts}),
$L Z_P \xrightarrow{\mathrm{GMET}} LZ$. But then, 
\S\ref{sec:Lambda only limit of LZ_P} implies that $LZ=\Lambda$.
If $\Lambda$ is $\cp^+$, $Z$ must have a $\CP$ first component in
order to be a Lindblad parametrization of $\Lambda$. But this follows from
the facts that
every $Z_P$ has that property and $\CP(\cH)$ is GMET-closed.

\paragraph{If $(Z_P)$ is bounded, then it has GMET accumulation points.}\label{sec:if Z bdd}
\leavevmode

This is an immediate consequence of GMET-compactness of closed balls
in $\HP(\cH)$ and $\Blin(\cH)$.

\paragraph{The finite-dimensional theory provides a  bounded sequence $(Z_P)$ as above.}
\leavevmode

\S\ref{sec:unitary trick} shows that the minimal Lindblad parametrizers
$\deltamin_P$ (see \S\ref{sec:minimal}) are uniformly bounded, hence so is
the (promoted) sequence $(\deltamin_P\Lambda: P\in\sS)$.

Readers worried about a potential circularity in citing a later result
should note that the finite-dimensional development is independent of
the separable, hence we are free to cite any part of the former, wherever
it may be, without fear.

%
%
%
\subsection{Integration}\label{sec:integration}

\subsubsection{$exp$ maps $\cp^+(\cH)$ into $\CP(\cH)$}

\begin{proof}
Take $Z\in\CP(\cH)\times\Blin(\cH)$.
Then $\gamma(Z/n)$ is $\CP$, 
as is $\gamma(Z/n)^n$ for $n=1,2,\ldots$, because $\CP(\cH)$ is a monoid.
Since $\CP(\cH)$ is norm-closed,
even the $n\to\infty$ limit, if it exists, is $\CP$.
But, by an elementary formula for exponential,
\begin{equation}\nonumber
\gamma\left(\tfrac{Z}{n}\right)^n =   
\left(\Id_{\Blin(\cH)} + \tfrac{L(Z) + o(1)}{n}\right)^n
\xrightarrow[n\to\infty]{\mathrm{norm}} \exp L(Z).
\end{equation}
\end{proof}

%
\subsubsection{Extension}\label{sec:L1 generators for CP}


The final steps are very similar to those in the unitary case in
\S~\ref{sec:skew-adjoint constant} -- \S~\ref{sec:skew-adjoint L1_loc}.

\paragraph{$\cp^+(\cH)$-valued step maps  generate $\CP$ evolution.}\leavevmode

As for $\Uni(\cH)$, this follows because $\CP(\cH)$ is a monoid and
we can splice segments of exponential curves together.

\paragraph{$\cp^+(\cH)$-valued  locally integrable  maps generate $\CP$ evolution.}
\leavevmode

$\CP(\cH)$ is norm closed. 

\subsubsection{Not all of $\CP(\cH)$ is accessible}

According to \S\ref{sec:cts E}, a continuous $\CP$ evolution family 
necessarily lies in $\GL(\TC(\cH))$.
$\CP$ maps which are not invertible in $\Blin(\TC(\cH))$,
and possibly others, are therefore
inaccessible by this sort of evolution. Lifted orhtoprojections are an example.
This is a limitation of continuous $\CP$ evolution families.

\subsection{$\cp^+(\cH)$ and $\cp(\cH)$ are norm-closed}

This section will show that $\cp^+(\cH)$ and $\cp(\cH)$ are
GMET closed, hence norm closed. As we discussed in connection
with $\CP(\cH)$, such closedness is generally interesting.
Norm closedness, however, suffices for the
particular technical need we have later in this section 
(\S\ref{sec:L is iso separable}).

\subsubsection{From $\cp^+$ to $\cp$}

If $\cp^+(\cH)$ is closed, then so is 
$\cp(\cH) = \cp^+(\cH) \cap (-\cp^+(\cH))$, as the intersection
of two closed sets. This works for any topology.

In finite dimension, one can argue more directly.
Any subspace is closed.

\subsubsection{Norm-closedness}\label{sec:cp+ is norm closed}

Assume $\Lambda$ is in the \textit{norm} closure of $\cp^+(\cH)$.
To show that $\Lambda\in\cp^+(\cH)$, we need to show that, given $\epsilon > 0$,
$\Id_{\TC(\cH)} + t \Lambda$ is within $\epsilon t$ of $\CP(\cH)$ 
for $t$ small enough. We work with $\TC(\cH)$ rather than $\Blin(\cH)$
to include the separable case. In finite dimension, the difference is
just a numerically different, yet equivalent, norm.

Now, take $\Gamma\in\cp^+(\cH)$ with
\mbox{$\|\Lambda - \Gamma\|_{1,1} < \tfrac{\epsilon}{2}$}.
Then,
$\Id_{\TC(\cH)} + t \Gamma$
is within $\tfrac{\epsilon}{2}t$ of $\CP(\cH)$ for small enough $t$,
so
\begin{equation}\nonumber
\Id_{\TC(\cH)} + t \Lambda  
= (\Id_{\TC(\cH)} + t \Gamma) + t(\Lambda - \Gamma),
\end{equation}
is within $\epsilon t$ of $\CP(\cH)$ for small enough $t$.

In finite dimension, there is nothing more to be said because there is
only one topology compatible with the linear structure.
The stronger claim of GMET-closedness will be established in
\S\ref{sec:cp+ is GMET-closed}.

\subsection{Conventional forms}\label{sec:Lindblad conventional}

We unpack $L(\Psi,K) = \Psi + K\blank + \blank K^\dag$ into a
more conventional form.
$\Psi$ is $\CP$, hence has a Kraus decomposition
$\Psi = \sum_i \btheta(A_i)$.  
Occasionally, as here, we want to decompose $K$ into hermitian and antihermitian
parts, and adopt a naming convention according to which that decomposition is
\begin{equation}
K = -G -iH.
\end{equation}
Thus,
\begin{equation}\nonumber
K\blank + \blank K^\dag = -[G,\;\cdot\;]_{+}-[iH,\;\cdot\;],
\end{equation}
and
\begin{equation}\label{eq:L(Psi,K) conventional}
L(\Psi,K) =   
\sum_i \btheta(A_i) -[G,\;\cdot\;]_{+}-[iH,\;\cdot\;].
\end{equation}

\subsubsection{Trace-preserving case}

$L(\Psi,K)$ generates a trace-preserving evolution
is equivalent to $L(\Psi,K)^\dag\Id_{\cH} = 0$.
Now, from (\ref{eq:L(Psi,K) conventional})
\begin{equation}\nonumber
  L(\Psi,K)^\dag\Id_{\cH} = \sum_i A_i^\dag A_i - 2G.
\end{equation}
So, a generator of a trace-preserving evolution can be written as
\begin{equation}\nonumber
L(\Psi,K) =   
\sum_i\Big( \btheta(A_i) -\tfrac{1}{2}[A_i^\dag A_i,\;\cdot\;]_{+}\Big)
- [iH,\;\cdot\;].
\end{equation}
This is a very common form.

\section{Lindblad parametrizers}\label{sec:parametrizers}

The previous section showed that $\cp^+(\cH)$ is nonuniquely parametrized
by $\CP(\cH)\times\Blin(\cH)/i\Real$, via $L$.
That is, $L$ constructs a $cp^+$ operator from a parametrization.
Suppose, conversely, we are given a $\cp^+$ operator.
How can we --- theoretically, at least --- obtain a parametrization?

\paragraph{Definition.}\label{sec:parametrizer def}
\leavevmode

A \textit{Lindblad parametrizer} is a bounded linear map
\mbox{$\Delta\colon\HP(\cH) \rightarrow$}
  \mbox{$\HP(\cH)\times\Blin(\cH)/i\Real$}
  which
\begin{enumerate}[label=\textnormal{(}\roman*\textnormal{)}]
  \item 
 is right-inverse to $L$, i.e., $L\circ\Delta$ is identity,
\item
  maps $\cp^+(\cH)$ into $\CP(\cH)\times\Blin(\cH)/i\Real$.
\end{enumerate}
  Since the trivial map
  $\Lambda \mapsto (\Lambda,0)$ is right-inverse to $L$,
  clause (ii) is the only reason this definition amounts to anything.
  It ensures that, if $\Lambda\in\cp^+(\cH)$, then $\Delta\Lambda$
  is the kind of parametrization existence of which we proved in
  the previous section.
  Should add the requirement that $\Delta\Lambda$ has the
  form $(0,K)$ of a lifted generator when that is possible?
  That turns out to be automatic.

  \paragraph{Highlights.}\leavevmode
  \begin{enumerate}
  \item Lindblad parametrizers exist. For finite dimension, one is given explcitly.
  \item $L$ is a (Banach) isomorphism between $0\times\Blin(\cH)/i\Real$
    and $\cp(\cH)$. Restricted to $\cp(\cH)$, any parametrizer is thus
    the inverse of that.
  \item Parametrizers are in one-to-one correspondence with
    complementary subspaces to $\cp(\cH)$ satisfying (\ref{eq:cp+ cap M}),
    their ``fixed-spaces'', on which they are essentially identity.
  \end{enumerate}

\subsection{Lindblad parametrizers exist --- finite-dimensional case}
\label{sec:parametrizers, finite dim}

\subsubsection{The minimal parametrizer $\deltamin$}\label{sec:minimal}

Let us expose the \textit{minimal parametrizer} we have already found
in an explicit form, bypassing the Jamio{\l}kowski transformation.
Let $A_0=\tfrac{1}{\sqrt{\dim\cH}}\Id_\cH,A_1,\ldots,A_{(\dim\cH)^2-1}$
be an orthonormal basis of $\Blin(\cH)$.
Then, $\Lambda\colon\HP(\cH)$ is decomposed on the basis $A_i\blank A_j^\dag$ 
(\S\ref{sec:ONB}) as
\begin{equation}
 c A_0\blank A_0 
 + \sum_{i \ge 1}
 \left( d_i A_i \blank A_0 + \overline{d_i} A_0 \blank A_i^\dag \right) 
 +  \sum_{i,j \ge 1} h_{ij} A_i\blank A_j^\dag,
\end{equation}
where $h$ is a hermitian {matrix}. Reduce it to the form
$h = \sum_\alpha \epsilon_\alpha b_\alpha b_\alpha^\dag$ with the $b_\alpha$'s
orthogonal column vectors and $\epsilon_\alpha = \pm 1$.
Then, with the definitions $B_\alpha = \sum (b_\alpha)_i A_i$ and
$K= \tfrac{c}{2\dim\cH}\Id_\cH + \tfrac{1}{\sqrt{\dim\cH}}\sum d_iA_i$,
\begin{equation}
  \Lambda =
  \sum_\alpha \epsilon_\alpha \btheta(B_\alpha)
  + (K\blank + \blank K^\dag).
\end{equation}
$\Lambda$ is in $\cp^+(\cH)$
iff $\Psi = \sum \epsilon_\alpha \btheta(B_\alpha)$ is $\CP$,
iff every $\epsilon_\alpha$ is $+1$.

\subsubsection{$\|\deltamin\|$ is bounded independently of dimension}
\label{sec:unitary trick}

This section concerns the finite-dimensional setting, but the only
use we will make of the bound proven is to show existence of parametrizations
in the separable setting.

\paragraph{Averaging over the unitary group.}
\leavevmode

The proof method borrows a trick from Thm. 5 of Lindblad\cite{Lindblad-76}.
We consider functions $f(U)$ of a unitary variable $U$, and
average over the unitary group $\Uni(\cH)$, an operation denoted
$\langle f(U) \rangle$. In other words $\langle \cdot \rangle$ is
integration with respect to a measure on $\Uni(\cH)$ which is
normalized ($\langle A \rangle = A$ if $A$ does not
depend on $U$) and invariant ($\langle f(VU)\rangle = \langle f(U) \rangle$
for fixed unitary $V$).
An explicit construction will not be needed.

\paragraph{Schur.}
\leavevmode

Consider, in particular, a function of form $\btheta(U)A$ for
$A\colon\Blin(\cH)$. $\langle\btheta(U)A\rangle$ commutes with
all unitaries, since
\mbox{$V\langle\btheta(U)A\rangle V^\dag = \langle \btheta(VU)A\rangle$},
and therefore $\langle \btheta(U)\rangle$ is proportional to
orthoprojection onto $\Id_\cH$ [a form of Schur's Lemma].
Since $\btheta(U)\Id_\cH= \Id_\cH$, the proportionality constant is one:
\begin{equation}\nonumber
  \langle \btheta(U)\rangle
  = {\Id_\cH} \tfrac{\Tr}{\dim \cH}
  = \hat\bpi(\Id_\cH).
\end{equation}

\paragraph{Proof of the bound.}
\leavevmode

  Take $\Lambda:\HP(\cH)$, with $\deltamin(\Lambda) =(\Psi,-G-iH)$.
\begin{equation}
\begin{split}
\langle \Lambda(U)\cdot U^{-1} \rangle  
= &
\langle \Psi(U)\cdot U^{-1} \rangle  
- \langle [G,U]_{+}\cdot U^{-1} \rangle  
- \langle [iH,U]\cdot U^{-1} \rangle  \\
  =
& 
- \langle [G,U]_{+}\cdot U^{-1} \rangle  
  - \langle [iH,U]\cdot U^{-1} \rangle
\quad \why{ \Tr A = 0 \Rightarrow \langle \btheta(A)U\cdot U^{-1} \rangle  
= A \langle UA^\dag U^{-1} \rangle =  0} \\
= &  
-\left(G + \tfrac{\Tr G}{\dim\cH}\Id_\cH \right) - iH.
\end{split}
\end{equation}
$\|\langle \Lambda(U)\cdot U^{-1} \rangle\| \le \|\Lambda\|$ because
we are averaging over $U$ and $\|U\|=1$.
Since the norm of any operator is at least as large as
that of its hermitian or antihermitian part,
\begin{equation}\nonumber
\|G + \tfrac{\Tr G}{\dim\cH} \Id_\cH \|,\, \|H\| \le \|\Lambda\|.
\end{equation}
Now, assume $\|G\|$ is an eigenvalue of $G$ (otherwise, $-\|G\|$ is one, and
we work with $-G$), and abbreviate $c = \tfrac{1}{\dim\cH} \Tr G$.
If $c > 0$, then clearly $\|G\| \le \|G+c\Id_\cH\|$.
If $c \le 0$, then $G$ must have a non-positive eigenvalue, so
the largest eigenvalue of $G - |c|\Id_\cH$ is $\|G\|-|c|$ and the smallest
is no greater than $-|c|$.
No matter the value of $c$, one of these is at least $\tfrac{\|G\|}{2}$
in absolute value. Therefore, $\|G\| \le 2\|G + c\Id_\cH\| \le 2 \|\Lambda\|$.
Finally,
\begin{equation}\nonumber
\|\Psi\|
\le
\|\Lambda\| + \|[G,\;\cdot\;]_{+}\| + \|[iH,\;\cdot\;]\|  
\le \|\Lambda\| + 4 \|\Lambda\| + 2\|\Lambda\|.
\end{equation}

\subsection{Lindblad parametrizers exist --- separable case}
\label{sec:parametrizers separable}

This section shows, again nonconstructively, the existence of Lindblad
parametrizers, relying on results of \S\ref{sec:parametrizations separable}.

\subsubsection{A variation on the Banach-Alaoglu theorem}
\label{sec:B-A variation}

The Banach-Alaoglu theorem says that if $\sX$ is a Banach space, then
the closed unit ball of $\sX^*$ is weak-* compact. We will use the
following variation on that classical theorem.

\begin{prop*}
Let $\sX$ and $\sY$ be Banach spaces,
and $\tau$ a topology under which $\cball\sY$ is compact and Hausdorff.
Then, $\cball\Blin(\sX,\sY)$ is compact under the topology $\hat{\tau}$
generated by the sets $\setof{T}{Tx\in \mathcal{O}}$ as $x$ ranges over
$\sX$ and $\mathcal{O}$ over $\tau$.
\end{prop*}
The important points about $\Cmplx$ in the proof of the Banach-Alaoglu theorem
are that it is normed, so that $\sX^*$ has a norm, and that the unit disk
$\mathbb D$ of $\Cmplx$ is compact. These r\^{o}les can be separated.
Write $\mathbb B$ for $\cball\sY$ with the $\tau$ topology, under
which it is compact.
We simply mimic the Banach-Alaoglu proof with $\mathbb D$ replaced by $\mathbb B$.
Readers who remember the proof can probably see how this works. Here are
details. There is considerable overlap with ideas from \S\ref{sec:BML} on BML spaces.
\begin{proof}
It suffices to show that we get a topological isomorphism from
$\cball\Blin(\sX,\sY)$ with
topology $\hat\tau$ to a closed subset of the compact space
$\sZ \defeq \prod_{x\in\sX}\|x\|\mathbb{B}$ via
the map $f$ given by $\cball\Blin(\sX,\sY)\ni T \mapsto \prod_{x\in\sX} Tx$.
All $f$ is doing is just taking its argument to an enormous table of
values. Clearly, $f$ is injective. Linearity of $T$ is embodied in a 
collection of conditions like $Tx + Tx' - T(x+x') = 0$ or $a(Tx)-T(ax)=0$.
If a point in $\sZ$ passes all these linearity tests,
then it is obviously the image by $f$ of something in $\cball\Blin(\sX,\sY)$,
so any point $t=\prod_{x\in\sX} t_x$ not in that image must fail one of those tests.
All that we need to do is show that some neighborhood of $t$
also fails that test. For instance, if $t_x+t_{x'}-t_{x+x'} = c > 0$, 
then by definition of the product topology
$\setof{s}{s_x+s_{x'}-s_{x+x'} > \tfrac{c}{2}}$ is such a neighborhood of $t$.
\end{proof}

\subsubsection{Application: parametrizers exist}

In Prop. \ref{sec:B-A variation}, take $\sX = \Blin(\TC(\cH))$,
$\sY=\HP(\cH)\times\Blin(\cH)/i\Real$, and $\tau=$ GMET.
Then, the sequence $(\deltamin_P)$ has a $\hat{\tau}$ accumulation point $\Delta$.
We must verify that it is a parametrizer. By definition of $\hat{\tau}$,
for each $\Lambda\in\Blin(\TC(\cH))$, $\Delta\Lambda$ is a
GMET accumulation point of $(\deltamin_P\Lambda)$, so is a Lindblad
parametrization of $\Lambda$ by \S\ref{sec:parametrizations separable}.

\subsection{$L$ and $\Delta$ effect a (Banach space) isomorphism
 $0\times\Blin(\cH)/i\Real \leftrightarrow \cp(\cH)$}\label{sec:L,D iso on cp}

The situation
is summed up by the following diagram, where $L\circ\Delta$ is identity
across each row, ``$\rightarrowtail$'' and
``$\twoheadrightarrow$'' denote linear injection and surjection, respectively.
The two top rows essentially paraphrase of the clauses of the definition
\ref{sec:parametrizer def}.
We proceed to justify the bottom row.
\begin{equation}\label{eq:parametrizer diagram}
  \begin{tikzcd}
\HP(\cH)
   \arrow[r,rightarrowtail,"\Delta"]
   &
   \HP(\cH)\times\Blin(\cH)/i\Real 
   \arrow[r,twoheadrightarrow,"L"]
   &
   \HP(\cH)
   \\
   \cp^+(\cH)
   \arrow[r,rightarrowtail,"\Delta"]
  \arrow[u,hookrightarrow,"\subset"']
   &
   \CP(\cH)\times\Blin(\cH)/i\Real 
   \arrow[r,twoheadrightarrow,"L"]
   \arrow[u,hookrightarrow,"\subset"']
   &
   \cp^+(\cH)
   \arrow[u,hookrightarrow,"\subset"']
   \\
   \cp(\cH)
   \arrow[r,rightarrow,"\Delta","\cong"']
   \arrow[u,hookrightarrow,"\subset"']
   &
   0\times\Blin(\cH)/i\Real
   \arrow[u,hookrightarrow,"\subset"']
   \arrow[r,rightarrow,"L","\cong"']
   &
   \cp(\cH)
   \arrow[u,hookrightarrow,"\subset"']
   \\
  \end{tikzcd}
\end{equation}


\paragraph{The indicated codomains in the bottom row are okay.}
\leavevmode

$\cp(\cH)$, respectively \mbox{$0\times\Blin(\cH)/i\Real$},
is the maximal linear space contained in $\cp^+(\cH)$, 
respectively \mbox{$\CP(\cH)\times\Blin(\cH)/i\Real$}.
Hence, since $\Delta$ and $L$ are linear maps,
the middle row of (\ref{eq:parametrizer diagram}) implies
that $\Delta$ maps $\cp(\cH)$ into $0\times\Blin(\cH)/i\Real$,
and $L$ the other way around.

In the following paragraphs, $\Delta$ and $L$ refer to those
restrictions.

\paragraph{$L$ is injective on $0\times\Blin(\cH)/i\Real$.}
\leavevmode

Suppose $L(0,K) = K\blank + \blank K^\dag$ is zero.
Application to $\Id_\cH$ shows
  that $K = -iH$ is antihermitian, so $L(0,K) = -[iH,\;\cdot\;]$.
  But, the only operators commuting with all of $\Blin(\cH)$ are multiples
  of the identity, so $K \in i\Real\Id_\cH$, which is zero in the
  quotient space.

  In infinite dimensions, $\Id_\cH$ is not in the domain $\TC(\cH)$ of
  $L(0,K)$, so we should work instead with its adjoint
  $K^\dag\blank + \blank K$, with the same conclusion.
  
\paragraph{$L$ is an isomorphism $0\times\Blin(\cH)/i\Real\xrightarrow{\cong}\cp(\cH)$.}
\leavevmode

It only remains to see that $L$ is surjective. But this is immediate from
$L\circ\Delta = \Id$ on $\cp(\cH)$.

\paragraph{$\Delta$ is a \textit{fixed} isomorphism
  $\cp(\cH) \xrightarrow{\cong}  0\times\Blin(\cH)/i\Real $.}
\leavevmode

By the previous paragraph and $L\circ\Delta=\Id$ on $\cp(\cH)$,
it follows that $\Delta$ is the inverse of $L$, as a linear map.
That is, all parametrizers have the same restriction to $\cp(\cH)$.

\paragraph{A final step for the separable case.}\label{sec:L is iso separable}
\leavevmode

\S\ref{sec:cp+ is norm closed} proved that $\cp(\cH)$ 
is norm closed. That is, it is a Banach space.
Above, we showed that
$L\colon 0\times\Blin(\cH)/i\Real\rightarrow\cp(\cH)$ is an
algebraic isomorphism. But $L$ is also a bounded operator, so
it is a Banach space isomorphism. Hence, so is its inverse,
the restriction of $\Delta$ to $\cp(\cH)$.

\subsection{Parametrizer fixed-space}

Define the fixed-space of $\Delta$ by
\begin{equation}
\sF(\Delta) \defeq \ker (p_2\circ\Delta),
\end{equation}
where $p_2$ is projection on the second factor, i.e., $p_2(\Psi,K) = K$.
If $\Psi$ is in $\sF(\Delta)$ with $\Delta\Psi = (\Psi',0)$,
then $L\circ\Delta = \Id_{\Blin(\TC(\cH))}$ implies that $\Psi'=\Psi$.
That is
\begin{equation}\label{eq:Delta on M}
\Delta|_{\sF(\Delta)} = (\Id, 0).  
\end{equation}
Up to a trivial modification, $\Delta|_{\sF(\Delta)}$ is identity,
explaining the name ``fixed-space''.

\paragraph{Separable case detail.}
$\sF(\Delta)$ is
closed because it is the kernel of a continuous linear map.

%
%
%
%
\subsubsection{$\sF$ is a parametrizer fixed-space iff the following two conditions hold}
    \leavevmode
    
\begin{align}
&  \sF  \oplus  \cp(\cH)
  = \Blin(\TC(\cH))  \label{eq:cp+M}
      \\
&  \sF \cap \cp^+(\cH)
  \subseteq \CP(\cH)  \label{eq:cp+ cap M}
\end{align}

\paragraph{Proof of necessity.}
    \leavevmode
    
\begin{proof}[Proof of (\ref{eq:cp+M})]
$\Delta\Lambda = (\Psi,K)$ implies that $\Lambda = \Psi + L(0,K)$
  and therefore $\Delta\Psi=(\Psi,0)$, i.e., $\Psi\in\sF(\Delta)$.
  Hence, $\Blin(\TC(\cH)) = \cp(\cH) + \sF(\Delta)$.
  Since $\cp(\cH)\cap \sF(\Delta) = \{0\}$
(because $\Delta\cp(\cH)\cap \Delta\sF(\Delta) = \{0\}$),
the sum is actually direct.
\end{proof}


\begin{proof}[Proof of (\ref{eq:cp+ cap M})]
  This is an immediate consequence of Clause (ii) of
  Def.~\ref{sec:parametrizer def} and (\ref{eq:Delta on M}).
\end{proof}


\paragraph{Proof of sufficiency.}
\leavevmode

If $\sF$ is complementary to $\cp(\cH)$, 
any $\Gamma$ is uniquely decomposed into a sum of
$\Psi\in\sF$ and $L(0,K)\in\cp(\cH)$.
Then defining $\Delta\Gamma = (\Psi,K)$, $\Delta$
is evidently a parametrizer
as long as (\ref{eq:cp+ cap M}) is satisfied.

\subsubsection{A particular case we already know}

\begin{equation}
\sF(\deltamin) = \cp(\cH)^\perp.
\end{equation}
This is immediately verified from \S\ref{sec:minimal}, or even \S\ref{sec:SA=...}.

\section{Structure of the tangent cone}\label{sec:tangent cone}

This section demonstrates the following claims.
\begin{tcolorbox}[standard jigsaw,opacityback=0]
\begin{align}
  & \cp^+(\cH) \text{ is GMET-closed} \nonumber \\
&  {\cp(\cH)} \, +\, \CP(\cH) =  \cp^+(\cH) \nonumber \\
  & \cp(\cH) \cap \CP(\cH) = \Real^+ \Id_{\Blin(\cH)} \nonumber 
\end{align}
\end{tcolorbox}

\subsubsection{$\cp^+(\cH)$ is GMET-closed}\label{sec:cp+ is GMET-closed}

Suppose $\Lambda \in \cl \cp^+(\cH)$ (closure now taken in GMET).
Then, $\wh{P}\Lambda \in \cl \cp^+(P\cH)$, and since $P\cH$ is
finite-dimensional, $\cp^+(P\cH)$ is closed by the preceding subsection,
so $\wh{P}\Lambda \in \cp^+(P\cH)$.
Therefore, 
\mbox{$\exp (\wh{P}\Lambda) = $}
\mbox{$\Id_{\Blin(P)}+\wh{P}\Lambda + \cdots\in\CP(P\cH)$}.
Since $\CP$ is closed under composition, this implies
\mbox{$\exp (\wh{P}\Lambda) \circ \wh{P}\in\CP(\cH)$}, where now
exponentiation is over $\cH$ (first term $\Id_\cH$).
Now, as \mbox{$\wh{P}\xrightarrow{\mathrm{SOT}} \Id_{\Blin(\cH)}$},
\mbox{$\wh{P}\Lambda\xrightarrow{\mathrm{SOT}} \Lambda$}, and 
therefore
\mbox{$\CP(\cH)\ni\exp(\wh{P}\Lambda)\xrightarrow{\mathrm{SOT}} \exp \Lambda$},
according to \S\ref{sec:SOT-cts power series}.
Thus, \mbox{$\exp\Lambda\in\CP(\cH)$} because $\CP(\cH)$ is SOT-closed.
Since the same is true for $t\Lambda$ when $t\ge 0$, 
$\Lambda\in\cp^+(\cH)$.

\subsubsection{$\cp(\cH) \, +\, \CP(\cH) =  \cp^+(\cH)$}

\noindent
Direction $\subseteq$:
This we have known for some time. Since $\cp(\cH)$ and $\CP(\cH)$
  are both in the convex cone $\cp^+(\cH)$, so is their set sum.

\noindent Direction $\supseteq$:
  This direction is equivalent to the statement that $L$ maps
  \textit{onto} $\cp^+(\cH)$, or that parametrizations always exist.
 This was proven for the finite-dimensional
 case in \S\ref{sec:L onto in finite dim}, and for the separable case
 will be proven in \S\ref{sec:parametrizations separable}.

\subsubsection{$\cp(\cH) \cap \CP(\cH) = \Real^+ \Id_\cH$}

\paragraph{finite-dimensional case.}
The following proof is limited to finite-dimensions because it relies on $\jam$.

As we know from \S\ref{sec:L,D iso on cp},
  \begin{equation}\nonumber
  \cp(\cH) = L(0\times\Blin(\cH))
 =\setof{K\blank + \blank K^\dag}{K\in\Blin(\cH)}.
  \end{equation}
Now, it is clear that $\Id_\cH\blank + \blank\Id_\cH = 2\Id_{\TC(\cH)}$ is
$\CP$, so suppose
$K$ is not a multiple of $\Id_\cH$, yet $K\blank + \blank K^\dag$ is $\CP$.
Applying $\jam$, $K\pam{\Id_\cH} + \Id_\cH\pam{K} \in \Pos(\Blin(\cH))$.
This is equivalent to \mbox{$\,\mathrm{Re} \inpr{A}{K}\inpr{\Id_\cH}{A} \ge 0$},
\mbox{$\forall A\in\Blin(\cH)$}. Since $K\not\propto \Id_\cH$, there is some
$N\perp K$ such that $\Tr N < 0$. Take $A=K+N$ for a contradiction.
\hfill\qedsymbol

\paragraph{separable case.}
  
Suppose $\Gamma\in\cp(\cH)\cap\CP(\cH)$.
Applying a lifted finite-dimensional projection,
$\wh{P}\Gamma \in\cp(P\cH)\cap\CP(P\cH)$.
Now we know
\begin{enumerate}[label=\textnormal{(}\roman*\textnormal{)} ]
\item 
$\wh{P}\Gamma = c_P \wh{P}$, by the
finite-dimensional case, 
\item
$\wh{P}\Gamma \xrightarrow{\mathrm{SOT}} \Gamma$
and $\wh{P} \xrightarrow{\mathrm{SOT}} \Id_{\Blin(\TC(\cH))}$, as
$P\xrightarrow{\mathrm{SOT}} \Id_{\cH}$.
\end{enumerate}

Together, these imply that $\Gamma\propto\Id_{\Blin(\TC(\cH))}$.

\section*{acknowledgements}
This work was supported by NSF MRSEC DMR-2011839.
The author is grateful to an open quantum systems discussion group at
Penn State for stimulation.
\vfill\eject
\appendix
\section*{Symbols}\label{sec:glossary}
\addcontentsline{toc}{section}{Symbols}

\begin{center}
\begin{tabular}{ll}
  $\jam$
  &  Jamio{\l}kowski transform,  \S \ref{sec:jam defined}
  \\
  $\blank$ & $A\blank B = \rho \mapsto A\rho B$, Def.~\ref{def:box} 
  \\
  $\bpi$ & $\bpi(h) = h \pam{h}$, Def.~\ref{def:pi} 
  \\
  $\btheta$ & $\btheta(A) = A\blank A^\dag$, Def.~\ref{def:theta-map} 
  \\
  $\SA(\cH)$ & hermitian operators, \S\ref{sec:OQS world}
  \\
  $\Pos(\cH)$ & positive operators, \S\ref{sec:OQS world}
  \\
  $\HP(\cH)$  & hermiticity-preserving superoperators, \S~\ref{sec:HP Mon} 
  \\
  $\Mon(\cH)$  & monotone superoperators, \S~\ref{sec:HP Mon}  
  \\
  $\CP$ & completely positive, Def.~\ref{sec:CP def} 
  \\
  $\extr$ & set of extreme vectors or points, Def.~\ref{sec:extremality defs}
  \\
  $\cp^+(\cH)$, $\cp(\cH)$ & tangent cone/space to $\CP(\cH)$ at $\Id_{\Blin(\cH)}$,
                        \S\ref{sec:tangent def} 
  \\
  $\deltamin$ & minimal Lindblad parametrizer, \S\ref{sec:minimal}
  \\
  $\TC(\cH)$ & trace-class operators, \S~\ref{sec:TC}
  \\
  $\|\cdot\|_{1,1}$ & operator norm on $\Blin(\TC(\cH),\TC(\cK))$,
                      \S\ref{sec:B(B1(H),B1(K))}
  \\
  SOT & strong-operator topology, \S~\ref{sec:SOT} 
  \\
  GMET & ground matrix element topology, \S~\ref{sec:GMET} 
  \\
  $P$ & generic finite-dimensional polyprojection, \S\ref{sec:lifted polyprojection}
  \\
  $\wh{P}$ & lifting of $P$, \S\ref{sec:lifted polyprojection}
  \\
  $\precsim$ & order induced by $\CP(\cH)$, \S~\ref{sec:precsim}
  \\
  $\Blin(\cdot,\cdot)$  & bounded linear operators with operator norm \\
  $x\colon X$ & ``$x$ belongs to $X$'', as a given, not a claim to be checked \\
  $(x_n)$ or $(x_n)_n$ & sequence, indexed by $n$ \\
  $\cl$ & closure with respect to some topology, \S\ref{sec:closed sets}
          \\
  $\tau/\tau'$ continuous & continuous with topology $\tau$ ($\tau'$)
                            on domain (codomain), \S\ref{sec:continuity} \\
  $\oball(\sX)$ & open unit ball $\setof{x\colon\sX}{\|x\| < 1}$ of a normed space \\
  $\cball(\sX)$ & closed unit ball $\setof{x\colon\sX}{\|x\| \le 1}$ \\
  $\GL(\sX)$ & general linear group on a normed space, \S\ref{sec:cts E} \\
\end{tabular}
\end{center}

\bibliography{gks}
\bibliographystyle{amsplain}
\end{document}

%% file: macros.tex
\newcommand{\paren}[1]{{\normalfont{(}}{#1}{\normalfont{)}}}

\newcommand{\defeq}{ {\kern 0.2em}:{\kern -0.5em}={\kern 0.2 em} }  
\newcommand{\eqdef}{ {\kern 0.2em}={\kern -0.5em}:{\kern 0.2 em} }  
\newcommand{\setof}[2]{\left\{ #1 \,\colon\, #2 \right\}}  

\newcommand{\wh}{\widehat}

\DeclareMathOperator{\Ran}{ran}  
\DeclareMathOperator{\rk}{rk}  
\DeclareMathOperator{\cl}{Cl}  
\DeclareMathOperator{\Id}{\mathbbm{1}} 
\DeclareMathOperator{\diam}{\mathrm{diam}}
\newcommand{\Real}{{\mathbb R}}    
\newcommand{\Cmplx}{{\mathbb C}}   

\newcommand{\graph}{\mathrm{Graph}\,}

\DeclareMathOperator{\Span}{\mathrm{span}}

\newcommand{\inpr}[2]{\left\langle {#1} \middle| {#2} \right\rangle}    
\newcommand{\ilinpr}[2]{\langle {#1} \,|\, {#2} \rangle}   
\newcommand{\outpr}[2]{\left| {#1} \middle\rangle\middle\langle {#2} \right|}   
\newcommand{\Dket}[1]{\left| {#1} \right\rangle}                
\newcommand{\Dbra}[1]{\left\langle {#1} \right|}                
\DeclareMathOperator{\Tr}{\mathrm{Tr}}

\DeclareMathOperator{\oball}{\mathrm{ball}}
\DeclareMathOperator{\cball}{\overline{\mathrm{ball}}}

\newcommand{\cC}{\mathcal{C}}
\newcommand{\cE}{\mathcal{E}}
\newcommand{\cH}{\mathcal{H}}
\newcommand{\cK}{\mathcal{K}}

\newcommand{\cV}{\mathcal{V}}

\newcommand{\cO}{\mathcal{O}}
\newcommand{\cM}{\mathcal{M}}
\newcommand{\cN}{\mathcal{N}}

\newcommand{\linfont}{\mathfrak}
\newcommand{\GL}{\mathsf{GL}} 
\newcommand{\Blin}{\linfont{B}} 
 
\newcommand{\TC}{\Blin^1}
\newcommand{\HS}{\Blin^2}
\newcommand{\Uni}{\mathsf{U}}
\newcommand{\HP}{\mathsf{HP}}
\newcommand{\Mon}{\mathsf{Mon}}
\newcommand{\SA}{\mathrm{SA}}
\newcommand{\Pos}{\mathrm{Pos}}

\newcommand{\Hilb}{\ensuremath{\mathfrak{B}}}
\newcommand{\CP}{\ensuremath{\mathsf{CP}}}
\newcommand{\BML}{\mathrm{BML}}
\newcommand{\LLin}[2]{\left({#2}\leftarrow {#1}\right)} 

\newcommand{\sA}{\mathscr{A}}
\newcommand{\sB}{\mathscr{B}}

\newcommand{\sF}{\mathscr{F}}
\newcommand{\sI}{\mathscr{I}}

\newcommand{\sP}{\mathscr{P}}

\newcommand{\sS}{\mathscr{S}}
\newcommand{\sU}{\mathscr{U}}

\newcommand{\sX}{\mathscr{X}}
\newcommand{\sY}{\mathscr{Y}}
\newcommand{\sZ}{\mathscr{Z}}

\newcommand{\jam}{\mathfrak{J}}
\newcommand{\pam}{\overline}
\newcommand{\bijam}{ \;\xleftrightarrow{\;\jam\;}\; } 

\newcommand{\why}[1]{{\bm\{}{#1}{\bm\}}}

\newcommand{\blank}{\,\square\,}

\newcommand{\deltamin}{\Delta^{\scriptstyle{min}}}
\newcommand{\bpi}{{\bm{\pi}}}
\newcommand{\btheta}{{\bm{\theta}}}
\newcommand{\reduce}{\mathsf{red}}
\DeclareMathOperator{\extr}{\mathrm{extr}}

%% file: theorems.tex
\theoremstyle{plain}
\newtheorem{thm}{Theorem}[section]

\newtheorem{lem}[thm]{Lemma}

\newtheorem*{lem*}{Lemma}
\newtheorem*{cor*}{Corollary}
\newtheorem*{thm*}{Theorem}
\newtheorem*{prop*}{Proposition}
\theoremstyle{definition}
\newtheorem{bareDefinition}{Definition}[section]
\newtheorem*{bareDefinition*}{Definition}
\newenvironment{definition}
    {\begin{bareDefinition}}
    {\hfill$\dashv$ \end{bareDefinition}}

\newenvironment{definition*}
    {\begin{bareDefinition*}}
    {\hfill$\dashv$ \end{bareDefinition*}}

\newtheorem{bareExample}{Example}[section]

\newtheorem{notation}{Notation}[section]

\theoremstyle{remark}
\newtheorem{remark}{Remark}[section]
\newtheorem*{remark*}{Remark}